\documentclass[final,1p,times,authoryear]{elsarticle}

\usepackage{amsmath,amssymb,mathtools}
\usepackage{graphicx}
\usepackage[hidelinks,hypertexnames=false]{hyperref}
\hypersetup{pdftitle={Mollification Regularization and Asymptotic Correction for Analytical Expansion of BSDEs with Non-Smooth Terminal Conditions. I. Theory},pdfauthor={Liangliang Zhang}}

\newtheorem{theorem}{Theorem}[section]
\newtheorem{lemma}[theorem]{Lemma}
\newtheorem{proposition}[theorem]{Proposition}
\newtheorem{corollary}[theorem]{Corollary}
\newdefinition{definition}[theorem]{Definition}
\newdefinition{assumption}[theorem]{Assumption}
\newdefinition{example}[theorem]{Example}
\newdefinition{remark}[theorem]{Remark}
\newproof{proof}{Proof}

\makeatletter
\def\@opargbegintheorem#1#2#3{\trivlist
  \let\baselinestretch\@blstr
  \item[\hskip\labelsep{\bfseries #1\ #2}]{\bfseries\itshape (#3)}\ \itshape}
\def\@opargbegindefinition#1#2#3{\trivlist
  \item[\hskip\labelsep{\bfseries #1\ #2}]{\bfseries\upshape (#3)}\ \upshape}
\makeatother

\makeatletter
\renewcommand\appendix{\par
  \setcounter{section}{0}
  \setcounter{subsection}{0}
  \setcounter{equation}{0}
  \gdef\thefigure{\@Alph\c@section.\arabic{figure}}
  \gdef\thetable{\@Alph\c@section.\arabic{table}}
  \gdef\thesection{\@Alph\c@section}
  \@addtoreset{equation}{section}
  \gdef\theequation{\@Alph\c@section.\arabic{equation}}
  \gdef\@seccntformat##1{\appendixname~\csname the##1\endcsname\@seccntDot\hskip 0.5em}
}
\makeatother

\journal{Stochastic Processes and their Applications}

\usepackage{bm}
\usepackage{dsfont}
\usepackage{enumitem}
\usepackage{booktabs,longtable,array,tabularx}
\usepackage{float}
\usepackage{algorithm,algorithmic}
\usepackage[protrusion=true,expansion=true]{microtype}

\numberwithin{equation}{section}

\DeclareMathOperator{\supp}{supp}
\newcommand{\R}{\mathbb{R}}
\newcommand{\E}{\mathbb{E}}
\newcommand{\N}{\mathbb{N}}
\newcommand{\Prob}{\mathbb{P}}
\newcommand{\diff}{\mathrm{d}}

\begin{document}

\begin{frontmatter}

\title{Mollification Regularization and Asymptotic Correction
for Analytical Expansion of BSDEs
with Non-Smooth Terminal Conditions. \\
I. Theory}

\author{Liangliang Zhang\corref{cor1}}
\ead{liangliangzhang81@qq.com}
\cortext[cor1]{Corresponding author}
\affiliation{organization={Shanghai Kingstar Fintech Co., Ltd.}, city={Shanghai}, country={China}}

\begin{abstract}
The analytical expansion method for BSDEs of Zhang (2017) breaks down when the terminal condition~$\psi$ has kinks or discontinuities.  We replace~$\psi$ by $\psi_\varepsilon=\psi*\rho_\varepsilon$ and derive a joint error decomposition whose expansion constants explicitly track the derivative blow-up as $\varepsilon\to0$.  For symmetric mollifiers the leading weak bias cancels.  The associated balanced rates are conditional on an explicitly stated uniform local-to-global Taylor-propagation assumption: under that assumption the benchmark $(l,m,\alpha)=(0,2,1)$ gives $\varepsilon^*\sim\delta_t^{1/3}$ and weak rate $2/3$.  Without it, the rigorous additive bound gives no decay when $m-2l=2$; an unconditional positive rate is obtained only for $m-2l>2$.  Asymptotic corrections cancel the weak bias through order $\varepsilon^{2J}$; their faster exponents are formal targets unless the auxiliary expansion errors and the Picard defect satisfy additional rate conditions, which we state explicitly.  For the driver-free fixed-center scheme we identify an $n$-independent terminal-representation error and show that it depends on the expansion center and on the terminal representative; change-of-basis invariance does not imply centering independence.  A Gaussian moment-compression layer combines whitening, a Stein recursion, and knapsack truncation with a jet-conditioned error certificate.  This Part~I contains the theory, with all proofs self-contained; numerical evidence and implementation measurements are reported separately in the companion paper (Part~II).
\end{abstract}


\end{frontmatter}

\section{Introduction}

\subsection{Background on Forward-Backward Stochastic Differential Equations}

Forward-backward stochastic differential equations (FBSDEs) are a central framework in stochastic analysis and mathematical finance \citep{PardouxPeng1990, Antonelli1993, ElKarouiPengQuenez1997, MaYong1999}. The decoupled FBSDE system considered throughout this paper takes the form
\begin{equation}\label{eq:FBSDE}
\begin{cases}
\diff X_t = \mu(t, X_t)\,\diff t + \sigma(t, X_t)\,\diff W_t, & X_0 = x \in \R^d, \\[6pt]
\diff Y_t = -f(t, X_t, Y_t, Z_t)\,\diff t + Z_t\,\diff W_t, & Y_T = \psi(X_T),
\end{cases}
\end{equation}
where $W$ is a $d$-dimensional Brownian motion on a filtered probability space $(\Omega, \mathcal{F}, (\mathcal{F}_t)_{0 \leq t \leq T}, \Prob)$, the coefficients $\mu$, $\sigma$ govern the forward dynamics, $f$ is the BSDE driver, and $\psi$ is the terminal condition. Under Lipschitz and square-integrability conditions, the system admits a unique adapted solution $(Y, Z) \in \mathcal{S}^2(0,T;\R) \times \mathcal{H}^2(0,T;\R^d)$ (the usual spaces of square-integrable adapted processes, with $\mathcal{S}^2$ endowed with the supremum norm in time) \citep{PardouxPeng1990}, with the existence proof based on a Picard iteration contraction argument that is central to the expansion method below. The BSDE framework encompasses derivative pricing, hedging, stochastic control, and risk measurement \citep{DuffieEpstein1992, ElKarouiPengQuenez1997, Pham2009, Bichuch2018}, motivating the need for efficient solvers---particularly in high-dimensional and non-smooth settings.

A key theoretical tool is the nonlinear Feynman--Kac formula \citep{PardouxPeng1992, MaYong1999}. Under suitable regularity, the BSDE solution satisfies
\begin{equation}
Y_t = u(t, X_t), \qquad Z_t = \nabla_x u(t, X_t) \cdot \sigma(t, X_t),
\end{equation}
where $u: [0,T] \times \R^d \to \R$ solves the semilinear parabolic PDE
\begin{equation}\label{eq:PDE}
\partial_t u(t,x) + \mathcal{A} u(t,x) + f\bigl(t, x, u(t,x), \nabla_x u(t,x) \cdot \sigma(t,x)\bigr) = 0, \quad u(T, x) = \psi(x),
\end{equation}
with the infinitesimal generator
\begin{equation}\label{eq:generator}
\mathcal{A} = \sum_{i=1}^{d} \mu_i(t,x)\,\frac{\partial}{\partial x_i} + \frac{1}{2} \sum_{i,j=1}^{d} a_{ij}(t,x)\,\frac{\partial^2}{\partial x_i \partial x_j},
\end{equation}
where $a(t,x) = \sigma(t,x)\,\sigma(t,x)^\top$. This BSDE--PDE correspondence transforms the stochastic problem into a deterministic one, and it is precisely this transformation that the analytical expansion method of Zhang (\citeyear{Zhang2017}) exploits.

\subsection{The Analytical Expansion Method}\label{sec:intro-expansion}

The analytical expansion method for BSDEs, developed by Zhang (\citeyear{Zhang2017}) building on the PDE expansion framework of Lorig, Pagliarani, and Pascucci (\citeyear{PagliaraniPascucci2012, Lorig2017}), constructs explicit approximate solutions to the FBSDE system \eqref{eq:FBSDE} through five orchestrated steps, each introducing a controlled level of approximation (formal definitions and convergence results are given in Section~\ref{sec:prelim}):
\begin{enumerate}[label=\textbf{Step \arabic*.}, leftmargin=4.5em, itemsep=1pt]
\item \emph{Picard iteration.} The nonlinear BSDE is linearized by a Picard scheme in which each iterate solves a \emph{linear} BSDE with the driver evaluated at the previous iterate; the iterates converge geometrically in the $(\mathcal{S}^2, \mathcal{H}^2)$ norm (Definition~\ref{def:Picard}, Proposition~\ref{prop:Picard-convergence}), so that $k = 2$ or $3$ typically suffices.
\item \emph{PDE coefficient expansion.} The generator $\mathcal{A}$ is Taylor expanded about a reference point $\bar{x}$, yielding a constant-coefficient leading operator $\mathcal{A}_0^{\bar{x}}$---which generates a Gaussian semigroup---plus polynomial perturbations; the solution is expanded accordingly as $u^{(k)} = \sum_{l\ge 0} u_l^{(k),\bar{x}}$ (Section~\ref{sec:prelim-coeff}).
\item \emph{Taylor expansion of the terminal condition.} The terminal condition $\psi$ and source terms are replaced by their order-$m$ Taylor polynomials about $\bar{x}$, reducing the hierarchy to Gaussian moment computations (Section~\ref{sec:prelim-Taylor}). This step requires derivatives $\partial_x^\beta \psi(\bar{x})$ up to order $|\beta| = m$, and it is the step that breaks down for non-smooth $\psi$.
\item \emph{Diagonal evaluation.} Setting $\bar{x} = x$ captures the local behavior optimally at each spatial point.
\item \emph{Time discretization.} The interval $[t, T]$ is divided into $n$ sub-intervals, and the expansion is applied in a backward recursion (Section~\ref{sec:prelim-time}).
\end{enumerate}
The combined error bound for the $(k, l, m, n)$-th order approximation is given by Theorem~\ref{thm:convergence}:
\begin{equation}\label{eq:error-bound}
\begin{aligned}
\|Y - Y^{(k,l,m,n)}\|_{\mathcal{S}^2}^2 + \|Z - Z^{(k,l,m,n)}\|_{\mathcal{H}^2}^2
&\leq K\,\frac{(2\delta)^k}{1 - 2\delta} + C\!\left(\frac{T-t}{n}\right)^{\!l+2} \\
&\quad + C\,n^{2l+2}\!\left(\frac{T-t}{n}\right)^{\!m - 2l},
\end{aligned}
\end{equation}
where the constant $C$ depends crucially on the supremum norms of derivatives of $\psi$ up to a certain order $\chi$. It is this dependence on the smoothness of $\psi$ that constitutes the fundamental limitation of the method in its original form.

\subsection{The Challenge of Non-Smooth Terminal Conditions}

The error bound \eqref{eq:error-bound} reveals a fundamental tension between the mathematical requirements of the analytical expansion method and the practical demands of financial applications. The method achieves its efficiency---explicit, closed-form approximations without Monte Carlo simulation or PDE grid discretization---by Taylor expanding all quantities involved, which requires at minimum that $\psi \in C_b^\chi(\R^d)$ with $\chi$ typically satisfying $\chi \geq m + d + 2$.

In mathematical finance, however, the terminal conditions that arise in practice are generically \emph{non-smooth}: European calls and puts $(x-K)^+$, $(K-x)^+$ have a kink at the strike $K$; digital payoffs $\mathds{1}_{\{x \geq K\}}$ have a jump; barrier and lookback payoffs inherit non-smoothness from extremal functionals; spread and basket payoffs $(x_1 - x_2 - K)^+$ have non-smooth loci forming hyperplanes in $\R^d$; and credit derivatives involve default indicators $\mathds{1}_{\{\tau \leq T\}}$. These structures are not exotic exceptions but the standard building blocks of the derivatives market, so any practically relevant BSDE solver must handle them effectively.

When the expansion method is applied naively to such data, three pathologies arise. First, the Taylor operator $\mathcal{T}_m^{\bar{x}} \psi$ (formally defined in Section~\ref{sec:prelim-Taylor}) is undefined wherever $\psi$ fails to be $m$-times differentiable, and Step~4 evaluates at every spatial point, so the approximation breaks down on the entire non-smooth locus. Second, near a kink or jump the Taylor remainder involves divergent intermediate derivatives, producing Gibbs-type oscillations whose amplitude grows with $m$. Third, the constants $C$ in \eqref{eq:error-bound} depend on $\|\partial^\beta \psi\|_\infty$ for $|\beta| \leq \chi$, which are infinite for non-smooth $\psi$: the bound becomes vacuous and the convergence guarantee evaporates.

Zhang (\citeyear{Zhang2017}) acknowledged this difficulty in the numerical experiments of Chapter~4, replacing the put payoff $(K - S)^+$ by the smooth Black--Scholes put price at a small time-to-maturity $\varepsilon$. This ad hoc substitution worked for the specific example considered, but it relies on a known smooth approximation of the particular payoff, does not generalize, and lacks an error analysis relating $\varepsilon$ to the expansion error. A systematic, theoretically grounded framework for non-smooth terminal conditions is therefore needed.

\subsection{Literature Review: Numerical Methods for BSDEs}

Before describing our contributions, we briefly survey the existing landscape of numerical methods for BSDEs, with particular attention to how each method handles non-smooth terminal conditions. For a comprehensive modern survey we refer to Chessari, Kawai, Shinozaki, and Yamada (\citeyear{ChessariKawaiShinozakiYamada2023}).

\paragraph{Time discretization and regression-based Monte Carlo methods.}
The numerical analysis of BSDE time discretization was initiated by Zhang (\citeyear{ZhangJ2004}), who established an $L^2$ convergence rate of $O(n^{-1/2})$ for the Euler-type backward scheme, and by Bouchard and Touzi (\citeyear{BouchardTouzi2004}), who analyzed a discrete-time approximation in which the conditional expectations are computed by Monte Carlo simulation combined with regression; the $O(n^{-1/2})$ rate for merely Lipschitz terminal data, which serves as a benchmark throughout this paper, originates from these works. Gobet, Lemor, and Warin (\citeyear{GobetLemor2005, LemorGobetWarin2006}) introduced regression-based Monte Carlo schemes that approximate the conditional expectations in the backward recursion by least-squares projection onto basis functions, with a complete error analysis accounting for the regression error. Gobet and Turkedjiev (\citeyear{GobetTurkedjiev2016a, GobetTurkedjiev2016b}) subsequently refined this approach through adaptive local regression and Malliavin-weight representations, improving both the approximation architecture and the convergence rates. Fahim, Touzi, and Warin (\citeyear{FahimTouziWarin2011}) proposed a probabilistic numerical scheme based on the PDE interpretation of the BSDE, replacing the conditional expectation by the solution of an associated linear PDE and obtaining convergence without ellipticity assumptions. These methods are naturally suited to handle non-smooth terminal conditions, since the forward simulation of $X_T$ and the evaluation of $\psi(X_T)$ do not require any smoothness of $\psi$; the non-smoothness is ``averaged out'' by the expectation operator. However, Monte Carlo methods suffer from the well-known curse of dimensionality in regression (the number of basis functions required for accurate regression grows exponentially with dimension), slow convergence rates ($O(N^{-1/2})$ in the number of simulated paths), and high computational cost for achieving the level of accuracy that analytical methods provide in the smooth case.

\paragraph{Deep learning methods.}
The deep BSDE method of Han, Jentzen, and E \citep{HanJentzenE2018} reformulates the BSDE as a stochastic optimization problem and uses deep neural networks to approximate the solution processes $Y_0$ and $Z_t$. The approach has demonstrated remarkable success in high-dimensional problems (with $d$ up to several hundred), where traditional methods are infeasible due to the curse of dimensionality. The deep BSDE method, like Monte Carlo methods, does not require smoothness of $\psi$, since the terminal condition appears only through the evaluation $\psi(X_T)$ at simulated terminal values. Extensions include the deep backward dynamic programming method of Hur\'e, Pham, and Warin (\citeyear{HurePhamWarin2020}); Germain, Pham, and Warin (\citeyear{GermainPhamWarin2023}) provide a survey of these neural-network-based algorithms for stochastic control problems and BSDEs, covering algorithmic variants, approximation results, and open questions. Henry-Labord\`ere and Touzi (\citeyear{HenryLabordereTouzi2021}) developed a branching diffusion representation for semilinear Cauchy problems that yields a Monte Carlo method free of regression steps. However, deep learning methods generally face challenges related to training stability, hyperparameter sensitivity, lack of convergence guarantees, and difficulty in certifying the accuracy of the computed solution.

\paragraph{High-order probabilistic schemes.}
A distinct line of work constructs high-order discretizations of BSDEs by replacing the Euler step with cubature or higher-order weak approximations. Crisan and Manolarakis (\citeyear{CrisanManolarakis2012, CrisanManolarakis2014}) developed second-order discretizations of BSDEs based on cubature on Wiener space, achieving rates beyond $O(n^{-1/2})$ under smoothness assumptions on the coefficients and terminal data. Chassagneux (\citeyear{Chassagneux2014}) designed a class of multidimensional numerical schemes for BSDEs using high-order transient integrators, and Chassagneux and Crisan (\citeyear{ChassagneuxCrisan2014}) combined these ideas with cubature to obtain run-time-efficient high-order schemes with rigorous error bounds. These works demonstrate that high-order probabilistic schemes are available for BSDEs, but their error estimates require smooth terminal data; they are the natural benchmarks against which the present high-order expansion method should be compared, and our contribution can be viewed as restoring such high-order behavior when the terminal condition is non-smooth.

\paragraph{Perturbation and asymptotic expansion methods.}
Several authors have developed perturbation-based approaches to BSDEs that share some features with the analytical expansion method. Fujii and Takahashi (\citeyear{FujiiTakahashi2012a}) proposed an analytical approximation for nonlinear FBSDEs using a perturbation parameter that measures the strength of the nonlinearity, combined with Malliavin calculus techniques to express the expansion coefficients. Takahashi and Yamada (\citeyear{TakahashiYamada2016}) developed an asymptotic expansion for forward-backward SDEs using a Malliavin calculus approach that provides explicit representations of the expansion terms; Naito and Yamada (\citeyear{NaitoYamada2019}) subsequently extended this line of work to obtain a second-order discretization using Malliavin-calculus-based computation of Greeks, thereby improving the accuracy of the expansion approximation. Fujii, Takahashi, and Takahashi (\citeyear{FujiiTakahashiTakahashi2019}) used an asymptotic expansion as a prior within a deep-learning solver for high-dimensional BSDEs, illustrating how expansion information can reduce the difficulty of network training. Gobet and Pagliarani (\citeyear{GobetPagliarani2015}) studied analytical approximations of BSDEs with non-smooth drivers, addressing the case where the nonlinearity $f$ is non-smooth but the terminal condition $\psi$ is smooth---a complementary problem to the one considered here. The PDE expansion framework of Lorig, Pagliarani, and Pascucci (\citeyear{PagliaraniPascucci2012, LorigPagliaraniPascucci2015, LorigPagliaraniPascucci2015b, Lorig2017}) provides the analytical foundation for coefficient expansions and has been applied successfully to option pricing in local and stochastic volatility models, including L\'evy-type processes. Bompis and Gobet (\citeyear{BompisGobet2013}) developed an analytical approximation approach for diffusion models with non-smooth payoffs, in which the transition density (rather than the payoff) is expanded and the resulting approximation carries a priori error estimates that remain valid for kinked terminal functions.

\paragraph{Error expansions, extrapolation, and payoff smoothing.}
The idea of expanding a numerical error in powers of a small parameter and cancelling its leading terms is classical: Talay and Tubaro (\citeyear{TalayTubaro1990}) established an error expansion for the Euler scheme for weak approximation of SDEs and used it to justify Richardson--Romberg extrapolation (see \citep{Milstein2004} for a comprehensive treatment of weak SDE approximation).  Weak error estimates under limited regularity remain an active topic: for rough volatility models, where the diffusion coefficient is merely H\"older in time, weak rates have recently been obtained in \citep{FrizSalkeldWagenhofer2025}. For BSDEs, Gobet and Labart (\citeyear{GobetLabart2007}) proved an error expansion for the discrete-time approximation of BSDEs with smooth terminal data and derived a Romberg-type correction that raises the convergence order. Our asymptotic correction terms (Section~\ref{sec:asymptotic}) share this expand-and-correct philosophy, but the object being expanded is different: Gobet and Labart expand the \emph{time-discretization} error and compute the correction by simulation, whereas we expand the \emph{mollification bias} in the regularization bandwidth $\varepsilon$ and compute the correction functions analytically through auxiliary linear PDEs with distributional terminal data within the same expansion framework. On the payoff smoothing side, Bayer, Siebenmorgen, and Tempone (\citeyear{BayerSiebenmorgenTempone2018}) showed that smoothing the payoff of a basket option restores the high-order convergence of quadrature-based pricing methods, and Bayer, Ben Hammouda, and Tempone (\citeyear{BayerBenHammoudaTempone2023}) analyzed the optimal choice of the smoothing parameter by balancing the smoothing bias against the quadrature error---a balancing principle structurally analogous to our coupled choice $\varepsilon^*(n)$, though developed for integration-based option pricing rather than for BSDE solvers. None of these works addresses the analytical expansion method of \citep{Zhang2017} or provides joint estimates for the regularization--expansion combination considered here.

\paragraph{Wiener chaos expansion.}
Briand and Labart (\citeyear{BriandLabart2014}) proposed a method for simulating BSDEs based on the Wiener chaos expansion of the solution processes. The method projects the BSDE solution onto the orthogonal basis of Wiener chaos (Hermite polynomials of iterated stochastic integrals) and truncates the expansion at a finite order. While elegant, the Wiener chaos method faces computational challenges in high dimensions and requires careful handling of the terminal condition through its chaos expansion coefficients.

\paragraph{Finite difference and finite element methods.}
For low-dimensional problems, classical PDE discretization methods (finite differences, finite elements) can be applied to the semilinear PDE \eqref{eq:PDE}. These methods can handle non-smooth terminal conditions through appropriate mesh refinement near the non-smooth locus, but they suffer from the curse of dimensionality and are typically limited to problems with $d \leq 3$ or $4$.

\paragraph{Methods specifically addressing non-smooth terminal conditions.}
Several works in the BSDE literature have directly confronted the challenge of non-smooth or irregular terminal data, and a clear comparison with these contributions is essential for positioning the present work. Bally and Pag\`es \citep{BallyPages2003} developed a quantization-based numerical scheme for reflected BSDEs (equivalently, optimal stopping/obstacle problems), in which the forward process is approximated by an optimally quantized discrete counterpart and non-smooth payoffs are handled without any smoothness requirement on $\psi$. However, the quantization approach incurs substantial preprocessing costs for computing optimal quantization grids, and the achievable accuracy depends on the quantization level, which grows combinatorially with dimension. Gobet and Makhlouf (\citeyear{GobetMakhlouf2010}) established sharp $L^2$ time-regularity results for the $Z$-component of BSDEs with irregular terminal conditions and used them to quantify the time-discretization error; notably, their proofs already employ smooth approximations of the terminal function as an analytical device, a methodological overlap with the present work that we acknowledge. Geiss, Geiss, and Gobet (\citeyear{GeissGeissGobet2012}) placed these estimates in a fractional smoothness (Besov/interpolation space) framework, showing that the discretization rate is governed by the Besov regularity of $\psi$ relative to the smoothing scale of the forward diffusion, rather than by classical pointwise differentiability. Geiss, Labart, and Luoto (\citeyear{GeissLabartLuoto2020}) established mean-square convergence rates for random walk approximations of BSDEs under merely H\"older terminal conditions, obtaining rates that degrade with the H\"older exponent (of order $n^{-\alpha/2}$ for $\alpha$-H\"older data); their results provide a direct rate benchmark for the low-regularity regime. For context, the classical benchmark rate $O(n^{-1/2})$ for Lipschitz terminal data is due to Zhang (\citeyear{ZhangJ2004}) and Bouchard and Touzi (\citeyear{BouchardTouzi2004}) (see above); while this rate cannot be improved for the bare Euler scheme without additional smoothness, our mollification-expansion approach achieves a \emph{tunable} rate that depends on the expansion orders $(l, m)$ and the balanced coupling $\varepsilon^*(n)$, and approaches first order as the correction order increases. Our approach is complementary to the fractional smoothness analyses above: while Geiss et~al.\ exploit the probabilistic smoothing of $X_T$ to sharpen convergence rates of discrete schemes, we employ deterministic mollification of $\psi$ itself to restore the applicability of the analytical expansion framework, yielding explicit closed-form approximations rather than discrete numerical solutions. We also note the work of Heston and Zhou (\citeyear{HestonZhou2000}), who determined the rate of convergence of discrete-time (e.g.\ binomial-tree) approximations of contingent claims and showed that the rate deteriorates for non-smooth payoffs, and the multilevel Monte Carlo method of Giles (\citeyear{GilesMultilevel2008}), which provides an alternative strategy for variance reduction when pricing non-smooth payoffs.

\paragraph{Positioning of the present work.}
The survey above reveals the following gap. Within analytical expansion methods one must distinguish two routes: the \emph{density/operator expansion} route of Lorig, Pagliarani, and Pascucci (\citeyear{PagliaraniPascucci2012, LorigPagliaraniPascucci2015}), which expands the transition density (or the pricing operator) and then integrates it against the terminal payoff, and therefore tolerates kinked or discontinuous $\psi$ provided it is integrable against the approximating densities (see also Bompis and Gobet (\citeyear{BompisGobet2013})); and the \emph{payoff expansion} route of Zhang (\citeyear{Zhang2017}), which Taylor expands the terminal condition $\psi$ itself at each step of a backward recursion and thereby produces fully explicit, closed-form approximations for both components $(Y, Z)$ of the BSDE with nonlinear drivers, but is restricted to smooth data. The present paper addresses the payoff expansion route: to the best of our knowledge, no systematic, theoretically grounded extension of the Zhang-type expansion method to non-smooth terminal conditions---with joint regularization--expansion error estimates, a balanced parameter coupling, and asymptotic corrections---has previously been available. Methods based on simulation, regression, quantization, or neural networks offer different compromises among dimensionality, statistical error, implementation complexity, and deterministic representation; no uniform efficiency ranking is made here.

\subsection{Contributions of This Paper}

The main contributions of this paper are fivefold:

\begin{enumerate}[label=(\arabic*), leftmargin=3em, itemsep=4pt]
\item \textbf{Mollification framework with joint error estimates.} We introduce a systematic mollification regularization approach that replaces the non-smooth terminal condition $\psi$ by its convolution $\psi_\varepsilon = \psi * \rho_\varepsilon$ with a standard mollifier of bandwidth $\varepsilon > 0$. The total approximation error is decomposed into a regularization error $E_1(\varepsilon)$ and an expansion error $E_2(\varepsilon, l, m, n)$. We derive estimates for both components: $E_1(\varepsilon) = O(\varepsilon^\alpha)$ in the supremum norm for H\"older data, where $\alpha$ is the H\"older exponent of $\psi$ (with $\alpha = 1$ for Lipschitz payoffs; for BV payoffs $E_1(\varepsilon) = O(\varepsilon^{1/2})$ holds in the $L^2(\Prob)$ sense at fixed $t<T$ through parabolic regularization effects, Corollary~\ref{cor:E1-disc}), and
\begin{equation}
\begin{aligned}
E_2(\varepsilon, l, m, n) &\;=\; O\!\left(\varepsilon^{-(m - 2l - \alpha)}\,\delta_t^{(m-2l)/2}\right) + O\!\left(\delta_t^{\lceil(l+1)/2\rceil}\right), \\
&\qquad \delta_t = (T-t)/n,
\end{aligned}
\end{equation}
established by tracking the derivative blow-up rates $\|\partial^\beta \psi_\varepsilon\|_\infty \leq C\,\varepsilon^{-(|\beta| - \alpha)}$ through the expansion hierarchy (Appendix~\ref{app:exp-error}); the complete bound of Theorem~\ref{thm:E2} contains an additional terminal-step term $O(\delta_t^{l+1}\varepsilon^{-(2l+2-\alpha)})$, omitted here for readability, which is of no lower order than the displayed terms whenever $m\le 4l+2$ (cf.\ Remark~\ref{rem:terminal-step-impact}).  Here $l$ and $m$ are the coefficient- and terminal-data expansion orders introduced in Steps~2--3 above, and $n$ is the number of time sub-intervals, so that $\delta_t=(T-t)/n$ is the sub-interval length.

\item \textbf{Balanced coupling under an explicit propagation assumption.}  Balancing $E_1=O(\varepsilon^\alpha)$ with the conditional expansion bound of Theorem~\ref{thm:E2} gives the crude coupling $\varepsilon^*=\delta_t^{1/2}$ and rate $O(\delta_t^{\alpha/2})$.  For symmetric kernels and piecewise smooth payoffs, the weak bias is $O(\varepsilon^2)$; conditional on Assumption~\ref{ass:telescoping-gain}, the same balance gives
\begin{equation}
\varepsilon^*=\delta_t^{q/(2(q+2-\alpha))},\qquad
\text{weak error }O\!\left(\delta_t^{q/(q+2-\alpha)}\right),
\qquad q=m-2l,
\end{equation}
provided the terminal-step term is of no lower order.  For $(l,m,\alpha)=(0,2,1)$ these formal values are $\delta_t^{1/3}$ and $2/3$.  They are not unconditional theorems: the additive propagation bound gives no decay at $q=2$.  Proposition~\ref{prop:additive-E2} instead supplies an unconditional bound and a positive rate for $q>2$.  The BV/digital rates $\delta_t^{1/5}$ (crude) and $\delta_t^{1/2}$ (weak) likewise inherit Assumption~\ref{ass:telescoping-gain}; the faster $\alpha_{\mathrm{eff}}=1/2$ values remain heuristic.

\item \textbf{Asymptotic correction terms and sufficient rate conditions.}  For drivers independent of $(y,z)$, or linear in $(y,z)$ with sufficiently smooth bounded coefficients, we construct correction functions through auxiliary linear PDEs with Dirac-type terminal data.  After $J$ corrections the remaining pointwise ($t<T$) and weak bias is $O(\varepsilon^{2(J+1)})$.  Balancing this bias with the conditional expansion bound defines
\begin{equation}
\varepsilon^*_J=\delta_t^{q/(2(q+2(J+1)-\alpha))},\qquad
r_J=\frac{(J+1)q}{q+2(J+1)-\alpha},\qquad q=m-2l.
\end{equation}
The exponent $r_J$ is a formal target, not an unconditional total-error rate.  Theorem~\ref{thm:improved-coupling} states the additional requirements on the auxiliary expansion errors, the Picard defect, the coefficient layer, and the terminal step under which the total weak error is $O(\delta_t^{r_J})$.  The numerical correction tables use Richardson extrapolation in $\varepsilon$ and therefore do not validate an auxiliary-$v_j$ solver or these end-to-end exponents.

\item \textbf{Driver-free analysis of the fixed-center scheme.}  We extend the mollification framework to the second expansion scheme of~\citep{Zhang2017}, which evaluates Taylor coefficients at the fixed point $x_0$ and eliminates diagonal evaluation.  For $f\equiv0$, Theorem~\ref{thm:E2-second} decomposes the error into an $n$-independent terminal-representation term and the $n$-dependent coefficient-expansion terms.  The floor is not invariant under a change of center: it is the transported defect of the chosen terminal representative.  Diagonal Taylor centering, $L^2(\nu_T)$ terminal projection, and $p$-refinement are three distinct mechanisms for reducing it.  Theorem~\ref{thm:E2-second} does not assert the same terminal-step concentration for a general $(y,z)$-dependent driver, and no convergence rate is claimed for digital data under this scheme.

\item \textbf{Gaussian moment compression with a jet-conditioned certificate.}  We separate the monomial layer, whose $\binom{m+d}{d}$ coefficients are not claimed to be compressible, from the Gaussian moment layer.  Whitening factorizes the moment table, a Stein recursion computes it with constant work per entry even for singular covariances, and a separable weight turns truncation into a knapsack problem.  Once the current jet is available, Proposition~\ref{prop:budget} certifies the discarded moment mass and its linear multistep accumulation.  This is a jet-conditioned, a posteriori certificate, not a bound computed from model constants alone before the jet exists.  Basis invariance of the truncated conditional-expectation matrix is proved separately from terminal-representation accuracy.  Measured tests cover genuinely $(y,z)$-dependent drivers in five dimensions and extensions to ten and twenty dimensions.

\end{enumerate}

The numerical study is reported separately in the companion paper (Part~II) \citep{Zhang2026II}.  It covers European put, digital, and differential-funding examples, Gaussian moment-compression testbeds in dimensions five to twenty, and author-reported $n$-step diagnostics.  Its evidential scope is deliberately narrower than the theoretical targets of the present paper: direct-integration experiments measure only the mollification bias along prescribed bandwidth schedules, the reported secant exponents concern the bias law $O(\varepsilon^2)$ rather than an end-to-end rate, and no reported number is used to validate the conditional exponents of Sections~\ref{sec:error}--\ref{sec:asymptotic}.

\paragraph{Relationship to the doctoral dissertation \citep{Zhang2017}.}
This paper is based in part on Chapter~3 of the doctoral dissertation \citep{Zhang2017}, which developed two analytical expansion schemes for FBSDEs with \emph{smooth} terminal conditions and established convergence in the $(k,l,m,n)$-parameter space. The original Picard iteration framework, the PDE coefficient expansion hierarchy, the time-discretization backward recursion, and the convergence theorem (Theorem~\ref{thm:convergence}) all originate from~\citep{Zhang2017}. The present paper contributes the following \emph{new} results beyond the thesis: (i)~the entire mollification regularization framework (Section~\ref{sec:mollification}), including the derivative estimates for mollified non-smooth terminal conditions (Theorem~\ref{thm:deriv-est}); (ii)~the joint error decomposition and the $\varepsilon$-dependent expansion error bounds (Theorem~\ref{thm:E2}), which require tracking derivative blow-up rates through the expansion hierarchy; (iii)~the balanced coupling analysis and explicit formulas (Theorems~\ref{thm:opt-lip}--\ref{thm:opt-bv}); (iv)~the asymptotic correction theory for smooth drivers, and for linear drivers with sufficiently regular coefficients (Theorems~\ref{thm:first-corr}--\ref{thm:J-corr} and the corrected error bounds in Theorem~\ref{thm:corrected-bound}); and (v)~the Gaussian moment-compression layer of Section~\ref{sec:moment-compression}---whitening with moment factorization, the Stein-recursion moment table, knapsack truncation with a jet-conditioned budget, and the basis-invariance result.  The numerical experiments involving non-smooth terminal conditions are likewise new relative to the dissertation and are reported in Part~II \citep{Zhang2026II}. In summary, while the smooth-data expansion framework is inherited from~\citep{Zhang2017}, the non-smooth extension---comprising the mollification theory, error analysis, parameter coupling, asymptotic corrections, and moment compression---constitutes the entirety of the new theoretical contribution in the present work.

\subsection{Organization of the Paper}

The remainder of this Part~I is organized as follows. Section~\ref{sec:prelim} fixes the problem formulation and standing assumptions, recalls the convergence theorem of Zhang (\citeyear{Zhang2017}), and analyzes the failure modes of the method under non-smooth terminal data. Section~\ref{sec:mollification} develops the mollification regularization framework and derives the regularization and derivative blow-up estimates. Section~\ref{sec:error} carries out the joint error analysis and derives the balanced coupling between the regularization bandwidth and the discretization parameters. Section~\ref{sec:asymptotic} constructs the asymptotic correction terms and derives formal corrected couplings under the disclosed assumptions. Section~\ref{sec:extension} extends the theory to the fixed-center second expansion scheme. Section~\ref{sec:moment-compression} develops the Gaussian moment-compression layer. Section~\ref{sec:conclusion} concludes with limitations and an outlook. Appendix~\ref{app:semigroup} collects properties of the Gaussian semigroup operator; Appendices~\ref{app:reg-error} and~\ref{app:exp-error} give the proofs of the regularization and expansion error estimates; Appendix~\ref{app:gaussian} records the Gaussian integral formulas; Appendix~\ref{app:aux} collects auxiliary lemmas; and Appendix~\ref{app:expansion-details} gives the derivation details of the analytical expansion framework. The numerical study---direct-integration bias measurements, Richardson corrections, sensitivity analyses, author-reported $n$-step diagnostics, and moment-compression testbeds---is reported in the companion paper (Part~II) \citep{Zhang2026II}.

\textbf{Notation.} We consolidate the key notation used throughout the paper. Vectors are denoted by boldface or plain lowercase ($x \in \R^d$); matrices by uppercase ($A$). Multi-indices: $\beta \in \N_0^d$ with $|\beta| = \sum_i \beta_i$. We write $\delta_t = (T-t)/n$ for the sub-interval length. The Picard contraction parameter is denoted by $\delta$ and equals $L_f$ times the length of the interval on which the iteration is performed: $\delta = L_f (T-t)$ on the full interval (Proposition~\ref{prop:Picard-convergence}) and $\delta = L_f\,\delta_t$ on a sub-interval of the time-discretized scheme (Theorem~\ref{thm:convergence}). All constants $C$ are generic and may change from line to line; dependencies are stated explicitly where they matter. We write $\Prob$ and $\E$ for the underlying probability measure and expectation, $L^p(\Prob)$ for the associated Lebesgue spaces, and $\mathds{1}_A$ for the indicator function of a set $A$. The mollification bandwidth is denoted by $\varepsilon>0$; $\rho_\varepsilon(\cdot)=\varepsilon^{-d}\rho(\cdot/\varepsilon)$ is the scaled mollifier kernel and $\psi_\varepsilon=\psi*\rho_\varepsilon$ the mollified terminal condition (Section~\ref{sec:mollification}). The asymptotic correction order is denoted by $J\in\N_0$ (Section~\ref{sec:asymptotic}), and the coupling exponent $\gamma$ is defined through $\varepsilon=C(\delta_t)^{\gamma}$ (Section~\ref{subsec:optimal}). The symbol $K$ denotes the option strike in payoff examples and a generic constant in error estimates; the context always disambiguates. The point $x_0$ denotes the fixed initial point of the forward diffusion used as the expansion center of the second scheme (Section~\ref{sec:extension}). See also the norm conventions in Section~\ref{subsec:error-decomp}.

\section{Preliminaries}\label{sec:prelim}

This section fixes the problem formulation and standing assumptions, recalls the convergence theorem of the analytical expansion method of Zhang (\citeyear{Zhang2017}) and Lorig, Pagliarani, and Pascucci (\citeyear{Lorig2017}), and analyzes the difficulties arising from non-smooth terminal conditions.  The derivation details of the expansion framework---the coefficient expansion hierarchy, the Gaussian semigroup representations, the moment identities, and the full algorithm---are collected in Appendix~\ref{app:expansion-details}.

\subsection{The FBSDE Framework}

Let $T > 0$ be a fixed time horizon and $(\Omega, \mathcal{F}, \Prob)$ a complete probability space carrying a $d$-dimensional Brownian motion $W = (W_t)_{0 \leq t \leq T}$ with augmented natural filtration $(\mathcal{F}_t)_{0 \leq t \leq T}$ satisfying the usual conditions. For $p \geq 1$, let $\mathcal{S}^p(t, T; \R)$ and $\mathcal{H}^p(t, T; \R^d)$ denote the standard spaces of continuous adapted processes and progressively measurable processes, respectively, with
\begin{equation}
\|Y\|_{\mathcal{S}^p}^p := \E\!\left[\sup_{t \leq s \leq T} |Y_s|^p\right] < \infty,
\qquad
\|Z\|_{\mathcal{H}^p}^p := \E\!\left[\left(\int_t^T |Z_s|^2\,\diff s\right)^{p/2}\right] < \infty.
\end{equation}
The product space $\mathcal{B}^p(t, T) := \mathcal{S}^p \times \mathcal{H}^p$ is a Banach space; we abbreviate $\mathcal{S}^2 = \mathcal{S}^2(0, T; \R)$, $\mathcal{H}^2 = \mathcal{H}^2(0, T; \R^d)$, and $\mathcal{B}^2 = \mathcal{B}^2(0, T)$. We use standard multi-index notation: for $\beta \in \N_0^d$, $|\beta| = \sum_i \beta_i$, $x^\beta = \prod_i x_i^{\beta_i}$, $\partial_x^\beta = \prod_i \partial_{x_i}^{\beta_i}$. The space $C_b^k(\R^d)$ consists of functions with bounded continuous partial derivatives up to order $k$, and $C^{0,\alpha}(\R^d)$ denotes the H\"older space with seminorm $[g]_{C^{0,\alpha}} := \sup_{x \neq y} |g(x) - g(y)|/|x - y|^\alpha$.

We now state the standing assumptions on the coefficients of the FBSDE \eqref{eq:FBSDE}.

\begin{assumption}[Forward SDE coefficients]\label{ass:forward}
The drift $\mu: [0,T] \times \R^d \to \R^d$ and diffusion $\sigma: [0,T] \times \R^d \to \R^{d \times d}$ satisfy:
\begin{enumerate}[label=(\alph*), leftmargin=2.5em]
\item \textbf{Lipschitz continuity}: There exists a constant $L_\mu > 0$ such that for all $t \in [0,T]$ and $x, x' \in \R^d$,
\begin{equation}
|\mu(t,x) - \mu(t,x')| + |\sigma(t,x) - \sigma(t,x')| \leq L_\mu\,|x - x'|.
\end{equation}

\item \textbf{Linear growth}: There exists a constant $K_\mu > 0$ such that for all $(t,x) \in [0,T] \times \R^d$,
\begin{equation}
|\mu(t,x)| + |\sigma(t,x)| \leq K_\mu\,(1 + |x|).
\end{equation}

\item \textbf{Uniform ellipticity}: The diffusion matrix $a(t,x) = \sigma(t,x)\,\sigma(t,x)^\top$ satisfies, for some constant $\lambda > 0$,
\begin{equation}
\xi^\top a(t,x)\,\xi \geq \lambda\,|\xi|^2 \qquad \text{for all } (t,x) \in [0,T] \times \R^d, \ \xi \in \R^d.
\end{equation}

\item \textbf{Smoothness}: For any positive integer $\chi$, the functions $\mu$ and $\sigma$ belong to $C_b^\chi([0,T] \times \R^d)$ in the spatial variable, uniformly in $t$. More precisely, for each multi-index $\beta$ with $|\beta| \leq \chi$, the derivatives $\partial_x^\beta \mu(t,x)$ and $\partial_x^\beta \sigma(t,x)$ exist, are continuous in $(t,x)$, and are bounded uniformly over $[0,T] \times \R^d$.
\end{enumerate}
\end{assumption}

\begin{assumption}[BSDE driver]\label{ass:driver}
The driver $f: [0,T] \times \R^d \times \R \times \R^d \to \R$ satisfies:
\begin{enumerate}[label=(\alph*), leftmargin=2.5em]
\item \textbf{Lipschitz continuity in $(y,z)$}: There exists a constant $L_f > 0$ such that for all $t \in [0,T]$, $x \in \R^d$, and $(y_1, z_1), (y_2, z_2) \in \R \times \R^d$,
\begin{equation}
|f(t,x,y_1,z_1) - f(t,x,y_2,z_2)| \leq L_f\,(|y_1 - y_2| + |z_1 - z_2|).
\end{equation}

\item \textbf{Lipschitz continuity in $x$}: For all $(t,y,z)$, the map $x \mapsto f(t,x,y,z)$ is Lipschitz continuous with a constant that may depend on $(y,z)$ but grows at most linearly.

\item \textbf{Smoothness in $x$}: For fixed $(y,z)$, the map $x \mapsto f(t,x,y,z)$ belongs to $C_b^\chi(\R^d)$ uniformly in $(t,y,z)$.

\item \textbf{Integrability}: $\E\!\left[\int_0^T |f(t, X_t, 0, 0)|^2\,\diff t\right] < \infty$.
\end{enumerate}
\end{assumption}

The uniform ellipticity (Assumption~\ref{ass:forward}(c)) can be relaxed to a hypoellipticity or weak H\"ormander condition at the expense of more delicate heat kernel estimates, and the $C_b^\chi$ smoothness of the coefficients is inherited from \citep{Zhang2017}; we retain the stronger conditions for simplicity. Under Assumptions~\ref{ass:forward} and \ref{ass:driver}, the classical BSDE theory yields existence and uniqueness.

\begin{theorem}[Existence and uniqueness; {\citep{PardouxPeng1990, ElKarouiPengQuenez1997}}]
Let Assumptions~\ref{ass:forward} and \ref{ass:driver} hold, and suppose that $\psi: \R^d \to \R$ is a Borel measurable function satisfying $\E[|\psi(X_T)|^2] < \infty$. Then the FBSDE \eqref{eq:FBSDE} admits a unique solution $(X, Y, Z)$ with $X \in \mathcal{S}^2(0,T;\R^d)$ and $(Y, Z) \in \mathcal{B}^2(0,T)$. Moreover, $Y$ admits the representation $Y_t = u(t, X_t)$ for a deterministic function $u: [0,T] \times \R^d \to \R$.
\end{theorem}

\begin{theorem}[Nonlinear Feynman--Kac; {\citep{PardouxPeng1992, MaYong1999}}]
Under Assumptions~\ref{ass:forward} and \ref{ass:driver}, suppose additionally that $\psi \in C_b^2(\R^d)$ and that the coefficients $\mu$, $\sigma$, $f$ are sufficiently smooth. Then the function $u(t,x) := Y_t^{t,x}$, where $(X^{t,x}, Y^{t,x}, Z^{t,x})$ solves \eqref{eq:FBSDE} from $X_t = x$, belongs to $C^{1,2}([0,T) \times \R^d) \cap C([0,T] \times \R^d)$ and is the unique classical solution of the PDE
\begin{equation}
\begin{cases}
\partial_t u + \mathcal{A} u + f(t, x, u, \nabla_x u \cdot \sigma) = 0, & (t,x) \in [0,T) \times \R^d,\\
u(T, x) = \psi(x), & x \in \R^d,
\end{cases}
\end{equation}
where $\mathcal{A}$ is given by \eqref{eq:generator}. Furthermore, $Z_t = \nabla_x u(t, X_t)\,\sigma(t, X_t)$ for all $t \in [0, T)$.
\end{theorem}

\subsection{Picard Iteration Scheme}

The Picard iteration linearizes the nonlinear BSDE by solving a sequence of linear BSDEs, each using the previous solution as input to the driver.

\begin{definition}[Picard iteration]\label{def:Picard}
Given the FBSDE \eqref{eq:FBSDE} satisfying Assumptions~\ref{ass:forward} and \ref{ass:driver} with $\psi \in L^2(\Omega, \mathcal{F}_T, \Prob)$, the Picard iteration sequence $\{(Y^{(k)}, Z^{(k)})\}_{k \geq 0}$ is defined as follows.

\noindent\textbf{Initialization ($k = 0$).} Set
\begin{equation}
Y_t^{(0)} = \E[\psi(X_T) \mid \mathcal{F}_t], \qquad Z_t^{(0)} = D_t\,Y_T^{(0)},
\end{equation}
i.e.\ $(Y^{(0)}, Z^{(0)})$ solves the linear BSDE
\begin{equation}
\diff Y_t^{(0)} = Z_t^{(0)}\,\diff W_t, \qquad Y_T^{(0)} = \psi(X_T).
\end{equation}

\noindent\textbf{Recursion ($k \geq 1$).} Given $(Y^{(k-1)}, Z^{(k-1)})$, define $(Y^{(k)}, Z^{(k)})$ as the unique solution of the linear BSDE
\begin{equation}
\diff Y_t^{(k)} = -f(t, X_t, Y_t^{(k-1)}, Z_t^{(k-1)})\,\diff t + Z_t^{(k)}\,\diff W_t, \qquad Y_T^{(k)} = \psi(X_T).
\end{equation}
\end{definition}

By the Feynman--Kac formula, each iterate satisfies $Y_t^{(k)} = u^{(k)}(t, X_t)$, where $u^{(k)}$ solves a linear parabolic PDE with source term evaluated at the previous iterate (Appendix~\ref{app:picard-pde}). The Picard iterates converge geometrically \citep{PardouxPeng1990, Zhang2017}:

\begin{proposition}[Geometric convergence of Picard iteration; {\citep{PardouxPeng1990, Zhang2017}}]\label{prop:Picard-convergence}
Let $\delta = L_f\,(T - t)$, where $L_f$ is the Lipschitz constant of $f$ in $(y,z)$; here $\delta$ is the contraction parameter on the \emph{full} interval $[t,T]$, whereas in the time-discretized scheme of Sections~\ref{sec:prelim-time}--\ref{sec:prelim-error} the same symbol denotes $\delta = L_f\,\delta_t$ with $\delta_t$ the sub-interval length (the context will always make clear which is meant). If $2\delta < 1$ (i.e.\ $T - t < 1/(2L_f)$), then the Picard iterates satisfy
\begin{equation}
\|Y - Y^{(k)}\|_{\mathcal{S}^2}^2 + \|Z - Z^{(k)}\|_{\mathcal{H}^2}^2 \leq K\,\frac{(2\delta)^k}{1 - 2\delta},
\end{equation}
where $K = K(\|\psi(X_T)\|_{L^2}, \|f(\cdot, \cdot, 0, 0)\|, L_f, T)$ is a constant independent of $k$. In particular, the convergence rate is geometric with ratio $2\delta < 1$.
\end{proposition}

For longer time horizons, the time discretization (Section~\ref{sec:prelim-time}) ensures that each sub-interval satisfies $2\delta < 1$. In practice, $k = 2$ or $k = 3$ Picard iterations suffice for high accuracy.

\subsection{PDE Coefficient Expansion}\label{sec:prelim-coeff}

The second step expands the coefficients of $\mathcal{A}$ in a Taylor series about a fixed point $\bar{x} \in \R^d$, yielding the decomposition $\mathcal{A} = \sum_{i=0}^{\infty} \mathcal{A}_i^{\bar{x}}$, where each $\mathcal{A}_i^{\bar{x}}$ has coefficients that are homogeneous polynomials of degree $i$ in $(x - \bar{x})$, and the leading-order operator $\mathcal{A}_0^{\bar{x}}$ freezes the coefficients at $\bar{x}$. Under the uniform ellipticity of Assumption~\ref{ass:forward}(c), $\mathcal{A}_0^{\bar{x}}$ generates a Gaussian semigroup $\{\mathcal{P}_0^{\bar{x}}(t,T)\}$ with explicit mean $m_{\bar{x}}(t,x,T) = x + \mu(t, \bar{x})\,(T - t)$ and covariance $\Sigma_{\bar{x}}(t,T) = a(t, \bar{x})\,(T - t)$. Expanding the solution of the $k$-th Picard PDE as $u^{(k)} = \sum_{l=0}^{\infty} u_l^{(k),\bar{x}}$ then yields a hierarchy of constant-coefficient PDEs in which the $l$-th correction is driven by the perturbation operators $\mathcal{A}_i^{\bar{x}}$, $i \le l$, applied to lower-order terms; each member admits an explicit Duhamel (semigroup) representation. The explicit operators $\mathcal{A}_i^{\bar{x}}$, the Gaussian density, and the semigroup representations of the hierarchy are collected in Appendix~\ref{app:coeff-hierarchy}.

\subsection{Taylor Expansion of Terminal Conditions}\label{sec:prelim-Taylor}

The third step replaces $\psi$ and the source terms by Taylor polynomials about $\bar{x}$, enabling explicit evaluation of the semigroup representations.

\begin{definition}[Taylor expansion operator]\label{def:Taylor}
For $\bar{x} \in \R^d$, a non-negative integer $m$, and a function $g \in C^m(\R^d)$, the Taylor expansion operator $\mathcal{T}_m^{\bar{x}}$ is defined by
\begin{equation}
\mathcal{T}_m^{\bar{x}} g(x) = \sum_{|\beta| \leq m} \frac{1}{\beta!}\,\partial_x^\beta g(\bar{x})\,(x - \bar{x})^\beta.
\end{equation}
The Taylor remainder is $\mathcal{R}_m^{\bar{x}} g(x) = g(x) - \mathcal{T}_m^{\bar{x}} g(x)$, which satisfies the standard integral representation provided that $g \in C^{m+1}(\R^d)$.
\end{definition}

Applying $\mathcal{T}_m^{\bar{x}}$ to the terminal condition and the source terms of the expansion hierarchy yields polynomial data against a Gaussian semigroup; since the Gaussian semigroup maps polynomials to polynomials through explicit moment identities, all quantities reduce to closed-form expressions (Appendix~\ref{app:taylor-hierarchy}). Setting $\bar{x} = x$ (Step~4) yields the approximate solutions $u_{l,m}^{(k)}(t,x) := u_{l,m}^{(k),\bar{x}=x}(t,x)$, and the combined $(l,m)$-th order approximation is $u_m^{(k)}(t,x) = \sum_{j=0}^{l} u_{j,m}^{(k)}(t,x)$.

\subsection{Time Discretization}\label{sec:prelim-time}

To maintain accuracy over long time horizons, the method partitions $[t, T]$ into sub-intervals on which the local expansion is accurate.

\begin{definition}[Time discretization]\label{def:time-disc}
Fix a positive integer $n$ (the number of time steps) and define the uniform partition
\begin{equation}
t = t_0 < t_1 < \cdots < t_n = T, \qquad t_j = t + j\,\delta_t, \qquad \delta_t = \frac{T - t}{n}.
\end{equation}
The $(k, l, m, n)$-th order approximation $u^{(k,l,m,n)}(t, x)$ is constructed by the following backward recursion:
\begin{enumerate}[label=\textbf{(TD-\arabic*)}, leftmargin=4em]
\item On the last sub-interval $[t_{n-1}, T]$: solve the PDE expansion with terminal condition $\psi$ to obtain $u^{(k,l,m)}(s, x)$ for $s \in [t_{n-1}, T]$, using expansion center $\bar{x} = x$ and expansion orders $(l, m)$.

\item For $j = n-2, n-3, \ldots, 0$: on the sub-interval $[t_j, t_{j+1}]$, solve the PDE expansion with terminal condition $g_j(x) := u^{(k,l,m)}(t_{j+1}, x)$ (the approximate solution from the next sub-interval evaluated at time $t_{j+1}$) to obtain $u^{(k,l,m)}(s, x)$ for $s \in [t_j, t_{j+1}]$.
\end{enumerate}
The final approximation is $u^{(k,l,m,n)}(t, x) := u^{(k,l,m)}(t_0, x)$.
\end{definition}

The local expansion errors on each sub-interval are controlled by powers of $\delta_t = (T-t)/n$; the backward propagation remains stable thanks to the semigroup property. The complete pseudocode of the resulting algorithm is given in Appendix~\ref{app:full-algorithm}.

\subsection{Original Error Bounds}\label{sec:prelim-error}

We now state the convergence theorem of Zhang (\citeyear{Zhang2017}), which provides the theoretical foundation for the analytical expansion method. The following definitions and assumptions specify the required regularity.

\begin{definition}[Order of a function; {\citep[Definition 2.4.3]{Zhang2017}}]\label{def:order}
Let $g \in C_b^\chi(\R^d)$ with $\chi > m \geq 0$. The \emph{order} of $g$ with respect to $(m, \chi)$ is
\begin{equation}
\varrho = \varrho_g^{(m,\chi)} := \max_{m+1 \leq |\beta| \leq \chi} \frac{\|\partial_x^\beta g\|_\infty^{1/(|\beta| - m)}}{\|g\|_\infty^{1/(|\beta|-m)} + 1}.
\end{equation}
This measures how rapidly the high-order derivatives of $g$ grow relative to $g$ itself; for a mollified function $g_\varepsilon$, $\varrho$ grows as a negative power of $\varepsilon$ as $\varepsilon \to 0$. (We use $\varrho$ rather than $\rho$ to avoid confusion with the mollifier kernel $\rho_\varepsilon$.)
\end{definition}

\begin{assumption}[Smoothness of terminal condition; {\citep[Assumption 2.4.6]{Zhang2017}}]\label{ass:terminal-smooth}
The terminal condition $\psi$ belongs to $C_b^\chi(\R^d)$ for a sufficiently large positive integer $\chi \geq m + d + 2$.
\end{assumption}

\begin{assumption}[Coefficient regularity; {\citep[Assumption 2.4.5]{Zhang2017}}]\label{ass:coeff-regularity}
The coefficients $\mu$ and $\sigma$ belong to $C_b^\chi([0,T] \times \R^d)$ in the spatial variable (uniformly in $t$) for the same $\chi$ as in Assumption~\ref{ass:terminal-smooth}. The driver $f$ satisfies the same regularity in all its arguments and is globally Lipschitz in $(y,z)$ with constant $L_f$.
\end{assumption}

\begin{assumption}[Compatibility condition]\label{ass:compatibility}
The expansion orders satisfy $m > 2l$ and $2\delta = 2L_f\,\delta_t < 1$.
\end{assumption}

\begin{remark}[Relaxation of the original condition $m > 4l+2$]\label{rem:compatibility-relaxation}
The compatibility condition in the original $(k,l,m,n)$ convergence theorem of Zhang (\citeyear{Zhang2017}, Assumption~2.4.7) is the stronger requirement $m > 4l+2$. That requirement is intrinsic to the squared $(\mathcal{S}^2, \mathcal{H}^2)$-norm error bound \eqref{eq:main-error-bound} restated in Theorem~\ref{thm:convergence} below, in which the Taylor expansion error contributes the term $n^{2l+2}\,\delta_t^{m-2l}$. The error analysis developed in this paper instead uses the supremum-norm formulation proved in Appendix~\ref{app:exp-error}, in which the Taylor expansion error contributes $O(\delta_t^{(m-2l)/2})$ (multiplied by the appropriate $\varepsilon$-dependent constant in the mollified setting); for that formulation convergence requires only $m > 2l$. Assumption~\ref{ass:compatibility} therefore records the weaker condition, which is the one in force throughout Sections~\ref{sec:mollification}--\ref{sec:extension}; in particular, the headline choices $(l,m) = (0,2)$ and $(l,m) = (1,4)$ satisfy it, whereas they would violate the original condition $m > 4l+2$.  The supremum-norm formulation is verified in Appendix~\ref{app:exp-error} in full except for a single step---the local-to-global gain with $\varepsilon$-dependent constants---whose status is heuristic (Remark~\ref{rem:telescoping-status}); all rates derived from Theorem~\ref{thm:E2} inherit this caveat.
\end{remark}

We are now ready to state the main convergence theorem.

\begin{theorem}[Convergence of the analytical expansion; {\citep[Theorem 2.4.8]{Zhang2017}}]\label{thm:convergence}
Let Assumptions~\ref{ass:forward}, \ref{ass:driver}, \ref{ass:terminal-smooth}, and \ref{ass:coeff-regularity} hold, and assume the strengthened compatibility condition $m > 4l+2$ together with $2\delta < 1$ (cf.\ Remark~\ref{rem:compatibility-relaxation}). Then there exist constants $K$ and $C$, depending on $T$, $d$, the coefficient bounds, $L_f$, $\lambda$, and $\|\partial_x^\beta \psi\|_\infty$ for $|\beta| \leq \chi$, such that
\begin{equation}\label{eq:main-error-bound}
\|Y - Y^{(k,l,m,n)}\|_{\mathcal{S}^2}^2 + \|Z - Z^{(k,l,m,n)}\|_{\mathcal{H}^2}^2
\leq K\,\frac{(2\delta)^k}{1 - 2\delta} + C\,\delta_t^{l+2} + C\,n^{2l+2}\,\delta_t^{m - 2l},
\end{equation}
where $\delta = L_f\,\delta_t$ and $\delta_t = (T-t)/n$.
\end{theorem}

The three terms represent the \textbf{Picard error} (geometric in $k$), the \textbf{PDE coefficient expansion error} ($O(\delta_t^{l+2})$), and the \textbf{Taylor expansion error} ($O(n^{2l+2}\delta_t^{m-2l})$). Since $n^{2l+2}\delta_t^{m-2l} = O(n^{-(m-4l-2)})$, the squared-error bound \eqref{eq:main-error-bound} decays at the overall rate $O\bigl(n^{-\min(l+2,\; m-4l-2)}\bigr)$; in particular, the rate $O(n^{-(l+2)})$ in this squared-norm sense is attained only under the stronger requirement $m \geq 5l+4$, not merely $m > 4l+2$. In the supremum-norm formulation of Appendix~\ref{app:exp-error}, which is the one used in the mollification error analysis of Sections~\ref{sec:mollification}--\ref{sec:extension}, the Taylor expansion error contributes $O(\delta_t^{(m-2l)/2})$ instead, and convergence then requires only the compatibility condition of Assumption~\ref{ass:compatibility}.

\begin{remark}[Dependence of constants on $\psi$]\label{rem:constant-dependence}
The constant $C$ depends polynomially on the derivatives of $\psi$: $C = \widetilde{C}\cdot\bigl(1 + \max_{|\beta| \leq \chi} \|\partial_x^\beta \psi\|_\infty\bigr)^{p}$ for some $p > 0$ and a constant $\widetilde{C}$ depending only on the coefficients, Lipschitz constants, ellipticity, and $T$; when $\psi$ is non-smooth, the bound becomes vacuous. We are candid that this is a \emph{generic} polynomial bound: it predicts a far more pessimistic blow-up for mollified data than the precise exponent $\varepsilon^{-(m-2l-\alpha)}$ that the expansion error constant actually carries in the bookkeeping of Appendix~\ref{app:exp-error}, and it is retained here only as a qualitative explanation of why non-smooth $\psi$ destroys the bound. Moreover, the requirement $\chi \geq m + d + 2$ makes the constants dimension-dependent, and no dimension-independence claim is made anywhere in this paper (cf.\ Remark~\ref{rem:multi-d}).
\end{remark}

\subsection{The Non-Smooth Terminal Condition Problem}

We now analyze the mechanisms by which non-smooth terminal conditions cause the expansion method to fail, motivating the mollification approach of the subsequent sections.

\paragraph{Failure of the Taylor operator and Gibbs-type oscillations.}
For the European put payoff $\psi(x) = (K - x)^+$, the first distributional derivative is a Heaviside function and the second a Dirac mass at $K$, so $\mathcal{T}_m^{\bar{x}} \psi$ is classically undefined at $\bar{x} = K$ for $m \geq 1$; for the digital payoff $\mathds{1}_{\{x \geq K\}}$ it is meaningless for all $m \geq 1$. Near the non-smooth locus, polynomial approximation moreover exhibits Gibbs-type oscillations whose amplitude grows with $m$. We emphasize that the fundamental failure is \emph{structural}---the error bounds of Theorem~\ref{thm:convergence} are vacuous because the constants involve divergent derivative norms---and is not always visible as a large numerical error: in special linear models with closed-form transition densities, a direct implementation evaluated \emph{away} from the kink can still return machine-precision values (see the direct-evaluation comparison in the companion paper, Part~II \citep{Zhang2026II}). Such apparent accuracy is an artifact of the linear case; it provides no guarantee for nonlinear dynamics, for evaluation near the kink, or for the $Z$-component.

\paragraph{Blow-up of the error constants.}
The principal obstruction is the blow-up of the constant $C$ in \eqref{eq:main-error-bound} (Remark~\ref{rem:constant-dependence}). For smooth approximations $\psi_\varepsilon \to \psi$, the derivative norms diverge as
\begin{equation}\label{eq:derivative-blowup}
\|\partial_x^\beta \psi_\varepsilon\|_\infty \leq C_\beta\,\varepsilon^{-(|\beta| - \alpha)}, \qquad |\beta| \geq 1,
\end{equation}
where $\alpha$ is the H\"older exponent of $\psi$; the order function of Definition~\ref{def:order} inherits the same blow-up, so the sharp $\varepsilon$-exponent of the expansion error must be obtained by tracking \eqref{eq:derivative-blowup} through the expansion hierarchy itself. That bookkeeping is carried out in Appendix~\ref{app:exp-error} (Theorem~\ref{thm:E2}), and the resulting total error bound---regularization bias plus expansion error---is the starting point of the balanced couplings derived in Sections~\ref{sec:error} and \ref{sec:asymptotic}. A catalog of the non-smoothness types and H\"older exponents of common derivative payoffs is given in Appendix~\ref{app:payoff-catalog}.

\paragraph{Parabolic regularization.}
The parabolic PDE \eqref{eq:PDE} itself regularizes the terminal condition: even for non-smooth $\psi$, the solution $u(t,x)$ is smooth in $x$ for $t < T$, with derivative bounds that blow up like $(T-t)^{-|\beta|/2}$ as $t \to T$. Since the expansion method Taylor-expands $\psi$ itself (Step~3), this smoothing does not automatically resolve the difficulty; one may either treat the last time step specially (semigroup pre-smoothing) or introduce artificial regularization (mollification). This paper pursues the latter, incorporating elements of the former through the asymptotic correction terms of Section~\ref{sec:asymptotic}.

\begin{remark}[Connection to viscosity solutions]
When $\psi$ is merely continuous or discontinuous, the PDE \eqref{eq:PDE} admits a unique viscosity solution (in the sense of Crandall, Ishii, and Lions (\citeyear{CrandallIshiiLions1992})) and $Y_t = u(t, X_t)$ remains valid in this sense. Our mollification approach bridges the gap between the classical PDE framework and the viscosity setting by approximating the viscosity solution through a family of classical solutions with smooth terminal data $\psi_\varepsilon$.
\end{remark}

\begin{remark}[Multi-dimensional non-smoothness]\label{rem:multi-d}
In dimension $d \geq 2$, the non-smooth locus is typically a $(d-1)$-dimensional manifold. The mollification step itself is dimension-agnostic in one specific sense: the convolution $\psi_\varepsilon = \psi * \rho_\varepsilon$ is smooth everywhere, and the regularization error $E_1(\varepsilon) = O(\varepsilon^\alpha)$ depends only on the global H\"older regularity of $\psi$, not on the local geometry of the locus. The expansion error constants, however, depend on $d$ through the derivative order $\chi \geq m + d + 2$ and combinatorial multi-index factors (Remark~\ref{rem:constant-dependence}); accordingly, the practical scope of the method is moderate dimensions, as reflected in the numerical experiments reported in Part~II \citep{Zhang2026II}.
\end{remark}

\section{The Mollification Framework}\label{sec:mollification}

In this section we develop the mathematical foundations of the
mollification approach for regularizing non-smooth terminal conditions.
We introduce standard mollifiers, define mollified terminal conditions,
and establish the derivative estimates
(Theorem~\ref{thm:deriv-est}) on which the error analysis of
Section~\ref{sec:error} rests.

\subsection{Standard Mollifiers}

\begin{definition}[Standard Mollifier]\label{def:mollifier}
A function $\rho\in C_c^\infty(\R^d)$ is called a \emph{standard
mollifier} if it satisfies:
\begin{enumerate}[label=(\roman*)]
  \item $\rho(x)\ge 0$ for all $x\in\R^d$;
  \item $\supp(\rho)\subset B(0,1):=\{x\in\R^d:|x|<1\}$;
  \item $\int_{\R^d}\rho(x)\,\diff x=1$;
  \item $\rho$ is radially symmetric: $\rho(x)=\rho(|x|)$.
\end{enumerate}
\end{definition}

\begin{example}[Canonical Mollifier]
Define
\begin{equation}
  \eta(x):=
  \begin{cases}
    \exp\!\Bigl(\dfrac{-1}{1-|x|^2}\Bigr), & |x|<1,\\[6pt]
    0, & |x|\ge 1,
  \end{cases}
\end{equation}
and set $\rho(x):=c_d\,\eta(x)$ with
$c_d:=\bigl(\int_{\R^d}\eta(x)\,\diff x\bigr)^{-1}$.
In dimension $d=1$ one computes $c_1\approx 2.2523$.
\end{example}

\begin{remark}[Gaussian and exponentially decaying kernels]
The compact-support requirement in Definition~\ref{def:mollifier}(ii)
is used only to guarantee that all moments of $\rho$ and all $L^1$
norms of its derivatives are finite.  Every result of this section and
of Section~\ref{sec:error} therefore remains valid for symmetric
kernels with super-polynomial decay, notably the Gaussian kernel
$\rho(z)=(2\pi)^{-d/2}e^{-|z|^2/2}$ used in the numerical
implementation (Part~II \citep{Zhang2026II}): the compact support is
replaced by the bounds $\int|z|^k\rho(z)\,\diff z<\infty$ and
$\|\partial^\beta\rho\|_{L^1}<\infty$, which such kernels satisfy.
Whenever the term ``Gaussian mollifier'' appears below, it is
understood in this extended sense.
\end{remark}

\begin{definition}[Scaled Mollifier]
For $\varepsilon>0$,
\begin{equation}
  \rho_\varepsilon(x):=\varepsilon^{-d}\,\rho(x/\varepsilon),
  \qquad x\in\R^d.
\end{equation}
\end{definition}

\begin{lemma}[Properties of Scaled Mollifiers]\label{lem:scaled-props}
$\rho_\varepsilon$ satisfies:
\begin{enumerate}[label=(\roman*)]
  \item $\rho_\varepsilon\in C_c^\infty(\R^d)$, $\rho_\varepsilon\ge 0$;
  \item $\supp(\rho_\varepsilon)\subset B(0,\varepsilon)$;
  \item $\int_{\R^d}\rho_\varepsilon(x)\,\diff x=1$;
  \item $\partial_x^\beta\rho_\varepsilon(x)
         =\varepsilon^{-d-|\beta|}(\partial^\beta\rho)(x/\varepsilon)$;
  \item $\|\rho_\varepsilon\|_{L^\infty}
         =\varepsilon^{-d}\|\rho\|_{L^\infty}$;
  \item $\int_{\R^d}|y|^k\rho_\varepsilon(y)\,\diff y
         =\varepsilon^k\int_{\R^d}|z|^k\rho(z)\,\diff z$
         for every integer $k\ge 0$.
\end{enumerate}
\end{lemma}

\begin{proof}
(i)--(iii) follow immediately from the definition and the change of
variable $y=x/\varepsilon$.  For~(iv) the chain rule gives
\[
  \partial_x^\beta\rho_\varepsilon(x)
  =\varepsilon^{-d}\cdot\varepsilon^{-|\beta|}
   (\partial^\beta\rho)(x/\varepsilon).
\]
Property~(v) is a direct consequence.  For~(vi), setting $y=\varepsilon z$:
\[
  \int_{\R^d}|y|^k\rho_\varepsilon(y)\,\diff y
  =\int_{\R^d}|\varepsilon z|^k\varepsilon^{-d}\rho(z)\varepsilon^d\,\diff z
  =\varepsilon^k\int_{\R^d}|z|^k\rho(z)\,\diff z.
\]
\end{proof}

\begin{lemma}[Scaling of Mollifier Derivatives]\label{lem:deriv-scaling}
For any multi-index $\beta$,
\begin{equation}
  \|\partial_x^\beta\rho_\varepsilon\|_{L^1}
  =\varepsilon^{-|\beta|}\|\partial^\beta\rho\|_{L^1}.
\end{equation}
\end{lemma}

\begin{proof}
From Lemma~\ref{lem:scaled-props}(iv) and $y=x/\varepsilon$:
\[
  \int_{\R^d}|\partial_x^\beta\rho_\varepsilon(x)|\,\diff x
  =\varepsilon^{-d-|\beta|}\int_{\R^d}
    |(\partial^\beta\rho)(x/\varepsilon)|\,\diff x
  =\varepsilon^{-|\beta|}\int_{\R^d}
    |(\partial^\beta\rho)(y)|\,\diff y.
\]
\end{proof}

\subsection{Mollified Terminal Conditions}

\begin{definition}[Mollified Terminal Condition]
For $\varepsilon>0$,
\begin{equation}
  \psi_\varepsilon(x)
  :=(\psi*\rho_\varepsilon)(x)
  =\int_{\R^d}\psi(x-y)\,\rho_\varepsilon(y)\,\diff y.
\end{equation}
\end{definition}

\begin{lemma}[Smoothness]
If $\psi\in L^\infty(\R^d)$, then $\psi_\varepsilon\in C^\infty(\R^d)$ and
\begin{equation}
  \partial_x^\beta\psi_\varepsilon
  =\psi*\partial^\beta\rho_\varepsilon.
\end{equation}
\end{lemma}

\begin{proof}
Fix a multi-index $\beta$ and consider the $i$-th partial difference
quotient.  By the mean value theorem the integrand is dominated by
$\|\psi\|_\infty\|\partial_{x_i}\rho_\varepsilon\|_\infty
 \mathds{1}_{B(0,2\varepsilon)}$,
which is integrable.  The dominated convergence theorem justifies
differentiation under the integral sign.  Iteration gives the result
for general~$\beta$.
\end{proof}

\begin{lemma}[Uniform Boundedness]\label{lem:bounded}
$\|\psi_\varepsilon\|_\infty\le\|\psi\|_\infty$.
\end{lemma}

\begin{proof}
$|\psi_\varepsilon(x)|
 \le\int|\psi(x-y)|\rho_\varepsilon(y)\,\diff y
 \le\|\psi\|_\infty$.
\end{proof}

\begin{lemma}[Uniform Convergence]
If $\psi$ is uniformly continuous, then
$\|\psi_\varepsilon-\psi\|_\infty\to 0$ as $\varepsilon\to 0$.
\end{lemma}

\begin{proof}
Let $\omega_\psi$ be the modulus of continuity.  Then
$|\psi_\varepsilon(x)-\psi(x)|
 \le\int\omega_\psi(|y|)\rho_\varepsilon(y)\,\diff y
 \le\omega_\psi(\varepsilon)\to 0$.
\end{proof}

\begin{lemma}[Lipschitz Rate]\label{lem:lip-rate}
If $\psi$ is Lipschitz with constant $L_\psi$, then
$\|\psi_\varepsilon-\psi\|_\infty\le C_\rho L_\psi\varepsilon$,
where $C_\rho=\int|z|\rho(z)\,\diff z$.
\end{lemma}

\begin{proof}
$|\psi_\varepsilon(x)-\psi(x)|
 \le L_\psi\int|y|\rho_\varepsilon(y)\,\diff y
 =L_\psi\varepsilon\int|z|\rho(z)\,\diff z$.
\end{proof}

\begin{lemma}[H\"older Rate]\label{lem:holder-rate}
If $\psi\in C^{0,\alpha}(\R^d)$, $0<\alpha\le 1$, then
$\|\psi_\varepsilon-\psi\|_\infty
 \le C_{\rho,\alpha}[\psi]_{C^{0,\alpha}}\varepsilon^\alpha$.
\end{lemma}

\begin{proof}
$|\psi_\varepsilon(x)-\psi(x)|
 \le[\psi]_{C^{0,\alpha}}\int|y|^\alpha\rho_\varepsilon(y)\,\diff y
 =[\psi]_{C^{0,\alpha}}\varepsilon^\alpha
   \int|z|^\alpha\rho(z)\,\diff z$.
\end{proof}

\begin{lemma}[BV Rate]
In $d=1$, if $\psi$ has bounded variation,
$\|\psi_\varepsilon-\psi\|_{L^1}\le\varepsilon\,\mathrm{TV}(\psi)
 \int|z|\rho(z)\,\diff z$.
\end{lemma}

\begin{proof}
By Minkowski,
\begin{align*}
\|\psi_\varepsilon-\psi\|_{L^1}
 &\le\int\rho_\varepsilon(y)\|\tau_y\psi-\psi\|_{L^1}\,\diff y
  \le\mathrm{TV}(\psi)\int|y|\rho_\varepsilon(y)\,\diff y\\
 &=\varepsilon\,\mathrm{TV}(\psi)\int|z|\rho(z)\,\diff z.
\end{align*}
\end{proof}

\subsection{Derivative Estimates}

\begin{theorem}[Derivative Estimates]\label{thm:deriv-est}
Let $\psi\in C^{0,\alpha}(\R^d)$, $0<\alpha\le 1$.
For any multi-index $\beta$ with $|\beta|\ge 1$,
\begin{equation}
  \|\partial_x^\beta\psi_\varepsilon\|_\infty
  \le C_{|\beta|,\alpha}\,\varepsilon^{-(|\beta|-\alpha)}
      [\psi]_{C^{0,\alpha}},
\end{equation}
where
$C_{|\beta|,\alpha}
 :=\int_{\R^d}|z|^\alpha|(\partial^\beta\rho)(z)|\,\diff z<\infty$.
\end{theorem}

\begin{proof}
Since $\int\partial_x^\beta\rho_\varepsilon(y)\,\diff y=0$ for
$|\beta|\ge 1$, we may write
\[
  \partial_x^\beta\psi_\varepsilon(x)
  =\int[\psi(x-y)-\psi(x)]\partial_x^\beta\rho_\varepsilon(y)\,\diff y.
\]
Then
\begin{align*}
  |\partial_x^\beta\psi_\varepsilon(x)|
  &\le[\psi]_{C^{0,\alpha}}\int|y|^\alpha
       |\partial_x^\beta\rho_\varepsilon(y)|\,\diff y\\
  &=[\psi]_{C^{0,\alpha}}\int|y|^\alpha
    \varepsilon^{-d-|\beta|}|(\partial^\beta\rho)(y/\varepsilon)|\,\diff y.
\end{align*}
Setting $z=y/\varepsilon$:
\[
  |\partial_x^\beta\psi_\varepsilon(x)|
  \le[\psi]_{C^{0,\alpha}}\varepsilon^{\alpha-|\beta|}
      \int|z|^\alpha|(\partial^\beta\rho)(z)|\,\diff z.
\]
\end{proof}

\begin{corollary}[European Put]\label{cor:put-deriv}
For $\psi(x)=(K-x)^+$ ($\alpha=1$, $d=1$):
$\|\partial_x^k\psi_\varepsilon\|_\infty\le C_k\varepsilon^{-(k-1)}$, $k\ge 1$.
In particular $\|\psi_\varepsilon'\|_\infty\le C_1$,
$\|\psi_\varepsilon''\|_\infty\le C_2\varepsilon^{-1}$,
$\|\psi_\varepsilon'''\|_\infty\le C_3\varepsilon^{-2}$.
\end{corollary}

\begin{proof}
Theorem~\ref{thm:deriv-est} with $\alpha=1$,
$[\psi]_{C^{0,1}}=1$.
\end{proof}

\begin{corollary}[Digital Option]\label{cor:digital-deriv}
For $\psi(x)=\mathds{1}_{\{x\ge K\}}$ ($d=1$):
$\|\partial_x^k\psi_\varepsilon\|_\infty\le C_k\varepsilon^{-k}$, $k\ge 1$.
\end{corollary}

\begin{proof}
$|\partial_x^k\psi_\varepsilon(x)|
 \le\|\psi\|_\infty\|\partial^k\rho_\varepsilon\|_{L^1}
 =\varepsilon^{-k}\|\partial^k\rho\|_{L^1}$.
\end{proof}

\begin{remark}[Distributional interpretation]
In the distributional limit $\varepsilon \to 0$, the mollified digital option satisfies $\partial_x \psi_\varepsilon \to \delta_K$ (the Dirac mass at the strike) in the sense of distributions, and higher derivatives converge to derivatives of $\delta_K$. The blow-up rate $\varepsilon^{-k}$ for the $k$-th derivative is therefore sharp and reflects the singular nature of these distributional limits. This should be contrasted with the Lipschitz case (Corollary~\ref{cor:put-deriv}), where the first derivative remains bounded.
\end{remark}

\begin{remark}[The two derivative bounds for non-smooth $\psi$]
Theorem~\ref{thm:deriv-est} yields $\|\partial_x^k \psi_\varepsilon\|_\infty \le C\,\varepsilon^{-(k-\alpha)}\,[\psi]_{C^{0,\alpha}}$ for $\psi \in C^{0,\alpha}$, while Corollary~\ref{cor:digital-deriv} gives $\|\partial_x^k \psi_\varepsilon\|_\infty \le C\,\varepsilon^{-k}\,\|\psi\|_\infty$ for $\psi \in L^\infty$.  These bounds are not contradictory but complementary: the former exploits H\"older regularity $\alpha>0$, whereas the latter requires only boundedness and a direct $L^1$ estimate of $\partial^k\rho_\varepsilon$.  For a discontinuous payoff such as the digital option $\psi = \mathds{1}_{\{x \ge K\}}$, Theorem~\ref{thm:deriv-est} does not apply---a step function belongs to no H\"older class---and the correct derivative bound is the $\varepsilon^{-k}$ estimate of Corollary~\ref{cor:digital-deriv}.  Accordingly, the regularization error for discontinuous terminal conditions is quantified exclusively through the $L^2$ route of Corollary~\ref{cor:E1-disc} (effective exponent $\alpha_{\mathrm{eff}}=1/2$), not through any sup-norm derivative bound.
\end{remark}

\subsection{Explicit Mollification Examples}

\begin{example}[European Put]\label{ex:put}
For $\psi(x)=(K-x)^+$ in $d=1$, substituting $z=y/\varepsilon$:
\[
  \psi_\varepsilon(x)
  =\int_{-1}^{1}(K-x+\varepsilon z)^+\rho(z)\,\diff z.
\]
If $x\le K-\varepsilon$: $\psi_\varepsilon(x)=K-x$.
If $x\ge K+\varepsilon$: $\psi_\varepsilon(x)=0$.
If $|x-K|<\varepsilon$: smooth interpolation.
(These outer identities are exact for compactly supported kernels; for the Gaussian kernel used in the numerics they hold up to an exponentially small tail and are thus approximate.)
\end{example}

\begin{example}[Digital Option]\label{ex:digital}
$\psi(x)=\mathds{1}_{\{x\ge K\}}$ gives
\[
\psi_\varepsilon(x)=\int_{-\infty}^{x-K}\rho_\varepsilon(y)\,\diff y
 =:\Phi_\varepsilon(x-K),
\]
a smooth step function; here
$\Phi_\varepsilon(s):=\int_{-\infty}^{s}\rho_\varepsilon(y)\,\diff y$
denotes the cumulative distribution function of the
kernel~$\rho_\varepsilon$.
\end{example}

\begin{example}[Bull Spread]
$\psi(x)=(x-K_1)^+-(x-K_2)^+$ with $K_1<K_2$.  By linearity,
$\psi_\varepsilon=((x-K_1)^+)_\varepsilon-((x-K_2)^+)_\varepsilon$,
with two smooth transition regions near $K_1$ and $K_2$.
\end{example}

\begin{figure}[htbp]
\centering
\includegraphics[width=0.85\textwidth]{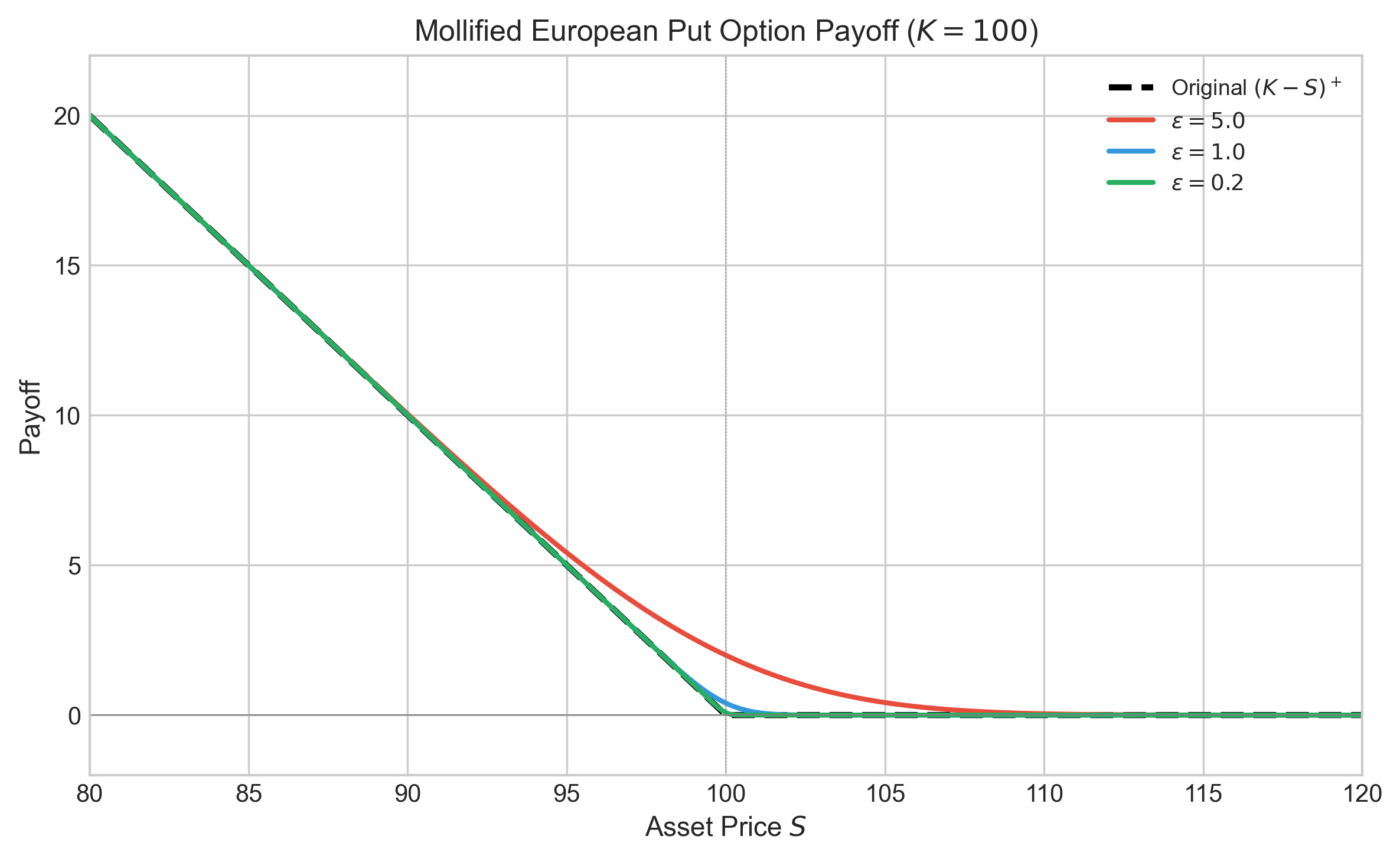}
\caption{Mollified European put option payoff for various~$\varepsilon$.}
\label{fig:mollified-put}
\end{figure}

\begin{figure}[htbp]
\centering
\includegraphics[width=0.85\textwidth]{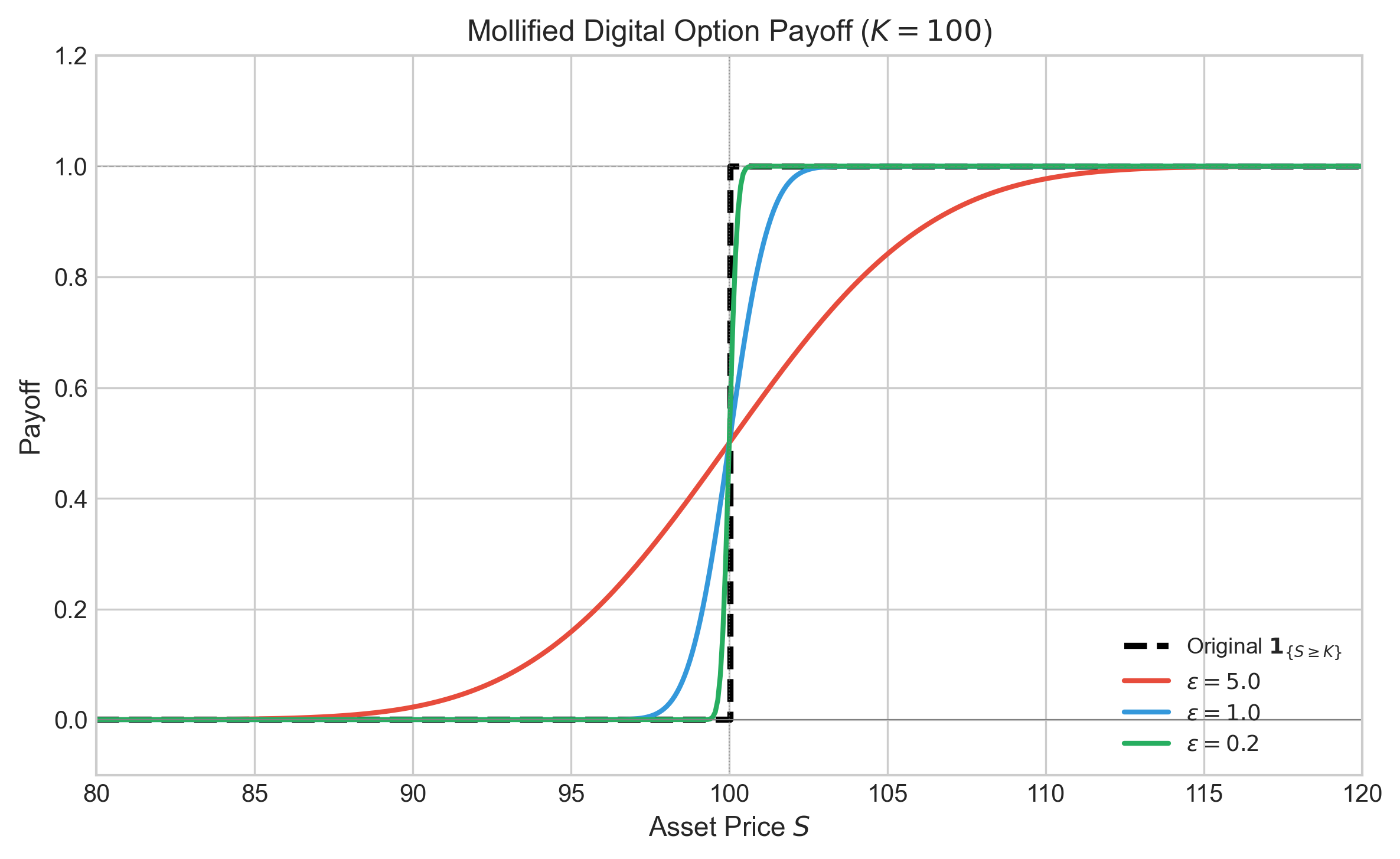}
\caption{Mollified digital option payoff for various~$\varepsilon$.}
\label{fig:mollified-digital}
\end{figure}

\begin{figure}[htbp]
\centering
\includegraphics[width=0.85\textwidth]{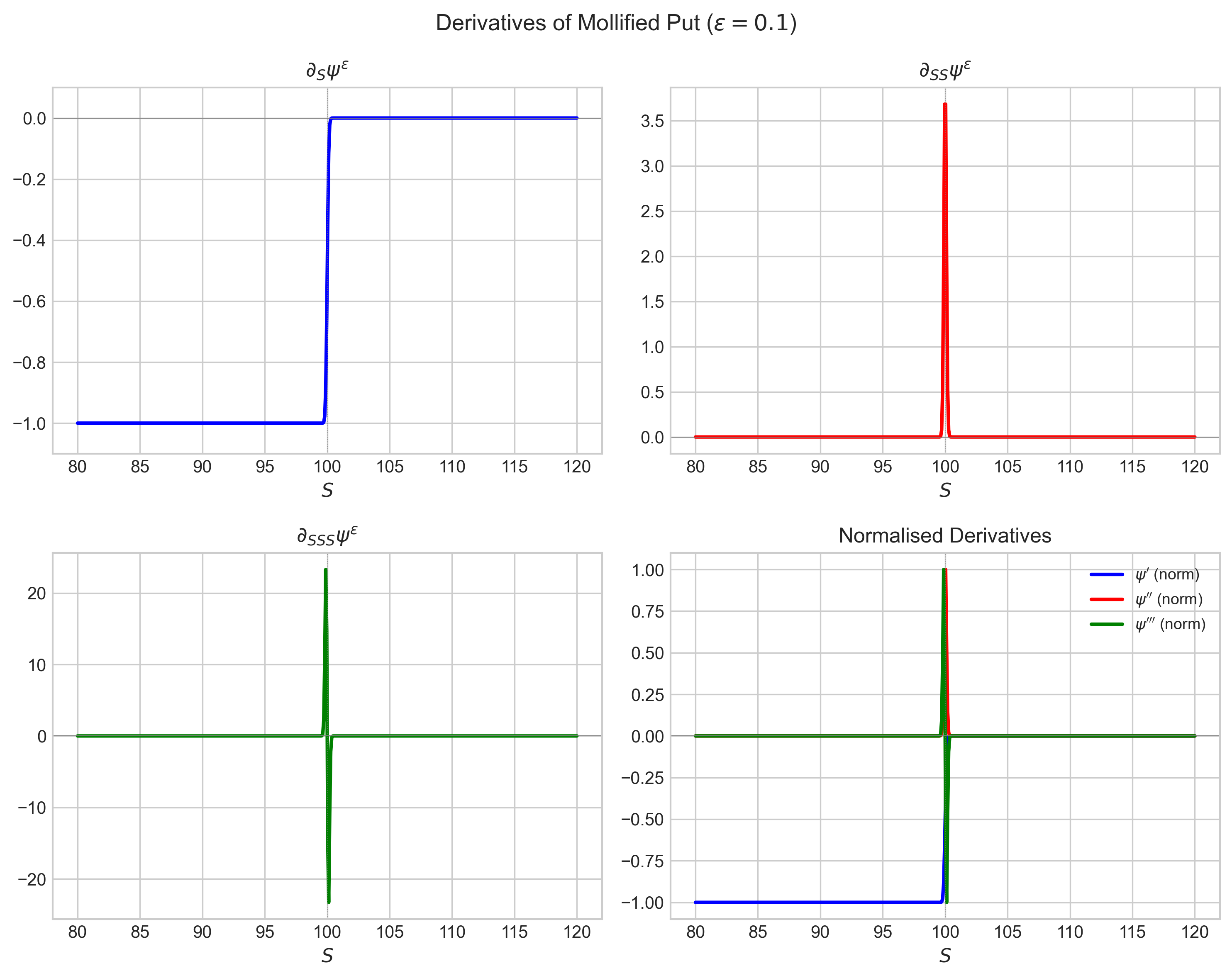}
\caption{First three derivatives of the mollified European put payoff.}
\label{fig:mollified-derivatives}
\end{figure}

Figures~\ref{fig:mollified-put} and~\ref{fig:mollified-digital}
display the mollified put and digital payoffs of
Examples~\ref{ex:put} and~\ref{ex:digital} for several values
of~$\varepsilon$, illustrating the smooth transition layer of
width~$O(\varepsilon)$ around the kink/strike.
Figure~\ref{fig:mollified-derivatives} shows the first three
derivatives of the mollified put payoff, whose sup-norm growth
$\|\partial^\beta\psi_\varepsilon\|_\infty\le C\varepsilon^{-(|\beta|-1)}$
as $\varepsilon\to0$ is the content of Theorem~\ref{thm:deriv-est}.

\section{Error Analysis}\label{sec:error}

We now carry out the central error analysis.  The key idea is to
decompose the total error into a \emph{regularization component}
(from replacing $\psi$ by~$\psi_\varepsilon$) and an
\emph{expansion component} (from the $(l,m,n)$-truncation of the
PDE expansion applied to the smoothed problem).

\subsection{Error Decomposition}\label{subsec:error-decomp}

Let $u$ denote the classical solution of
\[
  (\partial_t+\mathcal{A})u+f=0,\quad u(T,\cdot)=\psi(\cdot),
\]
and let $u_\varepsilon$ denote the solution with terminal condition
$\psi_\varepsilon$.  Let $\bar u_{\varepsilon,l,m,n}$ be the
$(l,m,n)$-order expansion approximation computed with $\psi_\varepsilon$.

\begin{proposition}[Error Decomposition]
\begin{equation}
  \|u-\bar u_{\varepsilon,l,m,n}\|
  \le\underbrace{\|u-u_\varepsilon\|}_{E_1(\varepsilon)}
   +\underbrace{\|u_\varepsilon-\bar u_{\varepsilon,l,m,n}\|}_{E_2(\varepsilon,l,m,n)}.
\end{equation}
\end{proposition}

\begin{proof}
Triangle inequality.
\end{proof}

Throughout Sections~\ref{sec:error}--\ref{sec:asymptotic}, unsubscripted norms $\|\cdot\|$ in error bounds denote the $(\mathcal{S}^2,\mathcal{H}^2)$ product norm for process-level estimates and the $L^\infty$ or $L^2(\R^d)$ norm for function-level estimates, as clear from context. In the numerical experiments reported in Part~II \citep{Zhang2026II}, we report pointwise errors $|Y_0 - \hat{Y}_0|$ and $|Z_0 - \hat{Z}_0|$ evaluated at $t=0$, $x=x_0$.

\subsection{Regularization Error \texorpdfstring{$E_1(\varepsilon)$}{E1}}

\begin{theorem}[BSDE Stability]\label{thm:bsde-stability}
Let $(Y,Z)$ and $(Y_\varepsilon,Z_\varepsilon)$ denote the
solutions of the BSDE \eqref{eq:FBSDE} with terminal conditions
$\psi$ and $\psi_\varepsilon$ respectively.  Under
Assumptions~\ref{ass:forward}--\ref{ass:driver}, suppose
additionally that both $\psi$ and $\psi_\varepsilon$ are bounded
(i.e.\ $\psi,\psi_\varepsilon\in L^\infty(\R^d)$).  Then
\begin{equation}\label{eq:stability}
  \E\!\left[\sup_{t\in[0,T]}|Y_t-Y_{\varepsilon,t}|^2\right]
  +\E\!\left[\int_0^T|Z_t-Z_{\varepsilon,t}|^2\,\diff t\right]
  \le C\|\psi-\psi_\varepsilon\|_\infty^2,
\end{equation}
where $C$ depends only on $T$, the Lipschitz constant of~$f$, and the
bounds on~$\mu,\sigma$.
\end{theorem}

\begin{remark}[On the boundedness assumption]\label{rem:bsde-stability-L2}
The $L^\infty$ stability estimate \eqref{eq:stability} requires boundedness of both $\psi$ and $\psi_\varepsilon$, which is satisfied in all our applications (since mollification preserves boundedness: $\|\psi_\varepsilon\|_\infty \leq \|\psi\|_\infty$ by Lemma~\ref{lem:bounded}).  For unbounded terminal conditions satisfying only $\E[|\psi(X_T)|^2] < \infty$, the stability estimate holds in the weaker $L^2$ form:
\[
  \E\!\left[\sup_{t\in[0,T]}|Y_t-Y_{\varepsilon,t}|^2\right]
  +\E\!\left[\int_0^T|Z_t-Z_{\varepsilon,t}|^2\,\diff t\right]
  \le C\,\E\!\left[|\psi(X_T)-\psi_\varepsilon(X_T)|^2\right],
\]
where $C$ is the same constant.  The $L^\infty$ version is used throughout this paper because it yields cleaner error bounds; the $L^2$ version can be substituted at the cost of replacing the supremum norms by $L^2(\Prob)$ norms throughout the analysis.
\end{remark}

\begin{proof}
Define $\delta Y_t:=Y_t-Y_{\varepsilon,t}$ and
$\delta Z_t:=Z_t-Z_{\varepsilon,t}$.  Then
\[
  \diff(\delta Y_t)
  =-\bigl[f(t,X_t,Y_t,Z_t)-f(t,X_t,Y_{\varepsilon,t},Z_{\varepsilon,t})\bigr]\diff t
   +\delta Z_t\,\diff W_t,
\]
with $\delta Y_T=\psi(X_T)-\psi_\varepsilon(X_T)$.
By the Lipschitz condition on~$f$,
\[
  |f(t,X_t,Y_t,Z_t)-f(t,X_t,Y_{\varepsilon,t},Z_{\varepsilon,t})|
  \le L_f(|\delta Y_t|+|\delta Z_t|).
\]
Applying It\^o's formula to $e^{\beta t}|\delta Y_t|^2$ with
$\beta=2L_f+2L_f^2+1$:
\begin{align*}
  e^{\beta t}|\delta Y_t|^2
  &+\int_t^T e^{\beta s}|\delta Z_s|^2\,\diff s\\
  &=e^{\beta T}|\delta Y_T|^2
    +\int_t^T e^{\beta s}\bigl[-\beta|\delta Y_s|^2
      \\
  &\qquad
      +2\delta Y_s\,(f_s-f_{\varepsilon,s})\bigr]\diff s
    -2\int_t^T e^{\beta s}\delta Y_s\delta Z_s\,\diff W_s.
\end{align*}
Using Young's inequality
$2|\delta Y_s||f_s-f_{\varepsilon,s}|
 \le 2L_f|\delta Y_s|^2+2L_f|\delta Y_s||\delta Z_s|
 \le(2L_f+L_f^2)|\delta Y_s|^2+|\delta Z_s|^2$,
the choice of $\beta$ ensures the drift term is non-positive.
Taking expectations and using
$|\delta Y_T|\le\|\psi-\psi_\varepsilon\|_\infty$ a.s.:
\[
  e^{\beta t}\E[|\delta Y_t|^2]
  +\E\!\left[\int_t^T e^{\beta s}|\delta Z_s|^2\,\diff s\right]
  \le e^{\beta T}\|\psi-\psi_\varepsilon\|_\infty^2.
\]
For the supremum estimate, the BDG inequality applied to the
stochastic integral yields
\[
  \E\!\left[\sup_{s\in[t,T]}|\delta Y_s|^2\right]
  \le C\|\psi-\psi_\varepsilon\|_\infty^2,
\]
completing the proof.
\end{proof}

\begin{corollary}[Lipschitz Terminal]\label{cor:E1-lip}
If $\psi$ is Lipschitz with constant $L_\psi$, then
$E_1(\varepsilon)\le C L_\psi\varepsilon$.
\end{corollary}

\begin{proof}
Combine Theorem~\ref{thm:bsde-stability} with Lemma~\ref{lem:lip-rate}.
The pointwise (sup-norm) version
$\|u-u_\varepsilon\|_{L^\infty([0,T]\times\R^d)}\le CL_\psi\varepsilon$
is proved in Appendix~\ref{app:reg-error}.
\end{proof}

\begin{corollary}[H\"older Terminal]\label{cor:E1-holder}
If $\psi\in C^{0,\alpha}(\R^d)$ (H\"older continuous with exponent
$\alpha$ and semi-norm $[\psi]_{C^{0,\alpha}}$), then
$E_1(\varepsilon)\le C[\psi]_{C^{0,\alpha}}\varepsilon^\alpha$.
\end{corollary}

\begin{proof}
Combine Theorem~\ref{thm:bsde-stability} with Lemma~\ref{lem:holder-rate}.
The pointwise (sup-norm) version
$\|u-u_\varepsilon\|_{L^\infty([0,T]\times\R^d)}
 \le C[\psi]_{C^{0,\alpha}}\varepsilon^\alpha$
is proved in Appendix~\ref{app:reg-error}.
\end{proof}

\begin{corollary}[Discontinuous Terminal]\label{cor:E1-disc}
For a bounded measurable $\psi$ that is piecewise Lipschitz with
finitely many jump discontinuities (e.g.\ digital option),
$E_1(\varepsilon)\le C\varepsilon^{1/2}$.
More precisely, in the notation of the H\"older framework, this
corresponds to an \emph{effective regularization exponent}
$\alpha_{\mathrm{eff}} = 1/2$, which accounts for the parabolic
smoothing effect of the forward diffusion.  The constant $C$ depends
on $T-t$ through the Gaussian strip bound $O(\varepsilon/\sqrt{T-t})$
in the proof below and therefore degenerates like $(T-t)^{-1/2}$ as
$t\uparrow T$: the estimate is applied at fixed $t<T$ (equivalently,
with a constant depending on the horizon $T-t$ of the interval on
which the scheme is run), not uniformly up to the terminal layer.
\end{corollary}

\begin{proof}
In the PDE framework, the maximum principle gives
$\|u(\cdot,t)-u_\varepsilon(\cdot,t)\|_\infty
 \le\|\psi-\psi_\varepsilon\|_\infty$.
For $\psi=\mathds{1}_{\{x\ge K\}}$,
$\|\psi-\psi_\varepsilon\|_\infty\le 1$, but this $L^\infty$ bound
is too crude.  Instead, we exploit the parabolic smoothing of the
forward diffusion: for $t < T$, the transition density
$\Gamma(t,x;T,y)$ is a smooth Gaussian kernel, and by a direct
$L^2(\Prob)$ computation,
\[
  \E[|\psi(X_T)-\psi_\varepsilon(X_T)|^2]
  \;\le\;
  C\,\varepsilon\,\sum_{j}|\Delta\psi(x_j)|^2,
\]
where $\{x_j\}$ are the jump locations and $|\Delta\psi(x_j)|$ the
jump magnitudes.  This follows from the fact that the probability
mass in a strip of width $\varepsilon$ around each jump point $x_j$
is $O(\varepsilon/\sqrt{T-t})$ under the Gaussian density, and the
$L^2$ error vanishes away from these strips.  Combining with
Theorem~\ref{thm:bsde-stability} (in its $L^2$ form, cf.\
Remark~\ref{rem:bsde-stability-L2}) yields
$E_1(\varepsilon) \le C\varepsilon^{1/2}$.
The constant~$C$ depends on the number of jump points, the jump
amplitudes, and the diffusion parameters through the Gaussian density
bounds.  In higher dimensions the same estimate holds when the jump
set of $\psi$ is a finite union of smooth hypersurfaces: the strip
argument is applied to the law of $X_T$ in the direction normal to
each jump hypersurface, whose one-dimensional marginal density is
bounded whenever the diffusion is non-degenerate in that direction,
and the $O(\varepsilon)$ probability-mass bound then follows exactly
as above.  We omit the routine geometric details.
\end{proof}

\begin{remark}[Notation and status of the effective regularization exponent]
Throughout this paper, $\alpha$ denotes the H\"older exponent of $\psi$
in the classical sense: $\alpha=1$ for Lipschitz functions,
$0<\alpha<1$ for H\"older-$\alpha$ functions.
For discontinuous (BV) terminal conditions, $\psi$ does not belong to
any H\"older class, but the parabolic smoothing yields an effective
regularization rate $E_1(\varepsilon) = O(\varepsilon^{\alpha_{\mathrm{eff}}})$
with $\alpha_{\mathrm{eff}} = 1/2$.
Two distinct uses of the value $1/2$ must be separated.
(i)~In the \emph{regularization} error $E_1$ the exponent $1/2$ is
proved ($L^2(\Prob)$ sense, Corollary~\ref{cor:E1-disc}).
(ii)~Substituting $\alpha=1/2$ into the $\varepsilon$-exponent of the
\emph{expansion} bound \eqref{eq:E2-bound} is not justified by any
theorem of this paper: for BV data Theorem~\ref{thm:E2} applies only
with $\alpha=0$.  The rigorously proved BV couplings and rates of
Section~\ref{subsec:optimal} therefore use $\alpha=0$
(Theorem~\ref{thm:opt-bv} and the BV case of Theorem~\ref{thm:opt-weak})---here and
throughout, ``rigorous'' is relative to Theorem~\ref{thm:E2}, whose
complete verification contains one disclosed heuristic step
(Remark~\ref{rem:telescoping-status}); couplings
obtained via the formal substitution (ii) are reported solely as
heuristics in Remark~\ref{rem:bv-heuristic-status}.
\end{remark}

\subsection{Expansion Error
  \texorpdfstring{$E_2(\varepsilon,l,m,n)$}{E2}}

Since $\psi_\varepsilon\in C^\infty(\R^d)$, the analytical expansion of~\citep{Zhang2017} applies to the mollified problem, but its constants depend on $\varepsilon$.  The terminal-layer cancellation used to sum the Taylor defects with one additional power of $\delta_t$ is isolated as an assumption rather than treated as a proved step.

\begin{assumption}[Uniform local-to-global Taylor gain]\label{ass:telescoping-gain}
For the parameters under consideration, the Taylor-truncation defects of the $(l,m,n)$ expansion aggregate to $C\varepsilon^{-(m-2l-\alpha)}\delta_t^{(m-2l)/2}$, with a constant uniform along the time grid, including the terminal steps.  The precise terminal-layer bookkeeping required by this assumption is described in Remark~\ref{rem:telescoping-status}.
\end{assumption}

Assumption~\ref{ass:telescoping-gain} is standard in form for smooth data with $\varepsilon$-independent constants, but is not proved here with $\varepsilon$-dependent constants.  In particular it remains open in the flagship case $m-2l=2$.  To display all grid-dependent factors, define
\begin{equation}
\Lambda_{l,\alpha}(\varepsilon,n):=
\begin{cases}
\varepsilon^{-(1-\alpha)}, & l\ \text{even and }\alpha<1,\\
1+\log n, & l\ \text{even and }\alpha=1,\\
1, & l\ \text{odd}.
\end{cases}
\end{equation}
The coefficient-expansion layer is propagated additively and does not use Assumption~\ref{ass:telescoping-gain}.

\begin{theorem}[Conditional expansion error with $\varepsilon$-dependence]\label{thm:E2}
Let Assumptions~\ref{ass:forward}--\ref{ass:driver} and~\ref{ass:telescoping-gain} hold, let $\psi\in C^{0,\alpha}(\R^d)$, $0<\alpha\le1$, and assume $m>2l$, $0\le l\le1$, and $\varepsilon^2\ge\delta_t$.  For $l\ge1$, assume the coefficient bounds of Assumption~\ref{ass:forward} for $|\beta|\le2m$.  Then
\begin{equation}\label{eq:E2-bound}
\begin{aligned}
\|u_\varepsilon-\bar u_{\varepsilon,l,m,n}\|_{L^\infty([0,T]\times\R^d)}
&\le C\varepsilon^{-(m-2l-\alpha)}\delta_t^{(m-2l)/2}\\
&\quad+C\delta_t^{\lceil(l+1)/2\rceil}\Lambda_{l,\alpha}(\varepsilon,n)
 +C\delta_t^{l+1}\varepsilon^{-(2l+2-\alpha)} .
\end{aligned}
\end{equation}
Here $C$ is independent of $\varepsilon$ and $n$; every remaining grid dependence is explicit in $\Lambda_{l,\alpha}$.  The first term uses Assumption~\ref{ass:telescoping-gain}; the other two are obtained by additive propagation.  The terminal-step term is displayed in the form proved for $l\le1$.  For bounded measurable $\psi$ the same statement holds with $\alpha=0$ and is the only expansion bound used here for BV/digital data.  No unconditional convergence assertion for $m-2l=2$ follows from this theorem.
\end{theorem}

\begin{proof}
Appendix~\ref{app:exp-error} derives the one-step estimates and the coefficient-expansion terms.  The Taylor defect at level $p$ is $O(\varepsilon^{-(m-2p-\alpha)}\delta_t^{(m-2p)/2})$; in the regime $\varepsilon^2\ge\delta_t$ the deepest level $p=l$ dominates.  Assumption~\ref{ass:telescoping-gain} is precisely the assertion that these defects aggregate to the first term of~\eqref{eq:E2-bound}.  The coefficient layer accumulates additively to the second term, including the factor recorded in~\eqref{eq:app-K1-global}, and the terminal step gives the third term for $l\le1$.  Remark~\ref{rem:telescoping-status} separates these proved additive steps from the unproved terminal-layer cancellation.
\end{proof}

\begin{remark}[Compatibility condition]
The supremum-norm bound \eqref{eq:E2-bound} requires only $m>2l$,
which guarantees that all terms vanish as $\delta_t\to 0$.  This is
weaker than the condition $m>4l+2$ of the original
$(\mathcal{S}^2,\mathcal{H}^2)$ formulation of
\citep[Theorem~2.4.8]{Zhang2017}; see
Remark~\ref{rem:compatibility-relaxation} for the comparison of the
two formulations.
\end{remark}

\begin{proposition}[Unconditional additive propagation bound]\label{prop:additive-E2}
Assume the hypotheses of Theorem~\ref{thm:E2}, except Assumption~\ref{ass:telescoping-gain}, and write $q=m-2l$.  Then
\begin{equation}
\begin{aligned}
\|u_\varepsilon-\bar u_{\varepsilon,l,m,n}\|_\infty
&\le C(T-t)\left(\delta_t^{\lceil(l+1)/2\rceil}\Lambda_{l,\alpha}(\varepsilon,n)
+\varepsilon^{-(q-\alpha)}\delta_t^{q/2-1}\right)\\
&\quad+C\delta_t^{l+1}\varepsilon^{-(2l+2-\alpha)} .
\end{aligned}
\end{equation}
Consequently an unconditional positive polynomial rate requires $q>2$.  Balancing $E_1=O(\varepsilon^\alpha)$ against the Taylor term gives
\[
\varepsilon^*\sim\delta_t^{a},\qquad a=\frac12-\frac1q .
\]
After substitution into all terms, the polynomial exponent is
\[
\begin{gathered}
R=\min\left\{\alpha a,\ A-a(1-\alpha)\mathds{1}_{\{l\ \mathrm{even},\alpha<1\}},\
l+1-(2l+2-\alpha)a\right\},\\
A=\left\lceil\frac{l+1}{2}\right\rceil .
\end{gathered}
\]
Thus the total error is $O(\delta_t^R)$ when $\alpha<1$, and is $O(\delta_t^{R-\eta})$ for arbitrarily small $\eta>0$ when $\alpha=1$, the loss absorbing $1+\log n$.  For $q=2$, $a=0$ and this route yields no decay.
\end{proposition}

\begin{proof}
The bound is the additive propagation of Appendix~\ref{app:exp-error}, Steps~4--5, and does not invoke Assumption~\ref{ass:telescoping-gain}.  The balance gives $\varepsilon^{q}\sim(T-t)\delta_t^{q/2-1}$, hence $a=1/2-1/q$ at fixed $T-t$.  The exponent $R$ is obtained by direct substitution; positivity for $q>2$ follows from $a>0$ and $l+1-(2l+2-\alpha)a>0$.
\end{proof}

\begin{remark}[The mollification--expansion trade-off]

\end{remark}

\subsection{Balanced Parameter Coupling}\label{subsec:optimal}
We now minimize the total error
$E(\varepsilon,n):=E_1(\varepsilon)+E_2(\varepsilon,l,m,n)$ over
$\varepsilon>0$ for fixed $(l,m,n)$ with $m>2l$.  We present two
couplings: a \emph{crude coupling}, obtained by balancing the
sup-norm/$L^2$ regularization bounds of
Corollaries~\ref{cor:E1-lip}--\ref{cor:E1-disc} against
Theorem~\ref{thm:E2}; and an \emph{enhanced coupling} (henceforth
also the \emph{weak coupling}) for symmetric
mollifiers, obtained by balancing the weak (pointwise, $t<T$)
regularization error $E_1^{\mathrm{w}}=O(\varepsilon^2)$ of
Proposition~\ref{prop:weak-E1} against the same expansion bound.

\begin{theorem}[Conditional Crude Balanced Coupling --- Lipschitz Case]\label{thm:opt-lip}
Under Assumption~\ref{ass:telescoping-gain}, let $\psi$ be Lipschitz ($\alpha=1$) and let $0\le l\le1$, $m>2l$.
Then the choice
\begin{equation}
  \varepsilon^*(n)=C_0\left(\frac{T-t}{n}\right)^{1/2}
\end{equation}
yields the total error bound
\begin{equation}
  \|u-\bar u_{\varepsilon^*,l,m,n}\|_{L^\infty([0,T]\times\R^d)}
  =O\!\left(\left(\frac{T-t}{n}\right)^{1/2}\right).
\end{equation}
\end{theorem}

\begin{proof}
By Corollary~\ref{cor:E1-lip}, $E_1(\varepsilon)=O(\varepsilon)$, and
by Theorem~\ref{thm:E2},
$E_2=O\bigl(\varepsilon^{-(m-2l-1)}\delta_t^{(m-2l)/2}
  +\delta_t^{\lceil(l+1)/2\rceil}\bigr)$.
Balancing the $\varepsilon$-dependent terms,
$\varepsilon\sim\varepsilon^{-(m-2l-1)}\delta_t^{(m-2l)/2}$, gives
$\varepsilon^{m-2l}\sim\delta_t^{(m-2l)/2}$, i.e.\
$\varepsilon^*\sim\delta_t^{1/2}$, independently of $(l,m)$.
Substituting back,
$(\varepsilon^*)^{-(m-2l-1)}\delta_t^{(m-2l)/2}
 =\delta_t^{-(m-2l-1)/2}\delta_t^{(m-2l)/2}=\delta_t^{1/2}$,
so both $\varepsilon$-dependent terms are $O(\delta_t^{1/2})$.  The
terminal-step term of \eqref{eq:E2-bound} is
$\delta_t^{l+1}(\varepsilon^*)^{-(2l+1)}=\delta_t^{1/2}$, of exactly
the same order, and the coefficient-expansion term
$\delta_t^{\lceil(l+1)/2\rceil}=O(\delta_t)$ is of higher order.
The coupling $\varepsilon^*\sim\delta_t^{1/2}$ satisfies the regime
condition $\varepsilon^2\ge\delta_t$ of Theorem~\ref{thm:E2} with
equality.
\end{proof}

\begin{theorem}[Conditional Crude Balanced Coupling --- BV Case]\label{thm:opt-bv}
Let $\psi$ be bounded and piecewise Lipschitz with finitely many jump
discontinuities, and let $0\le l\le1$, $m>2l$; write $q:=m-2l$.  Assumption~\ref{ass:telescoping-gain} is in force.  The
regularization error is $E_1(\varepsilon)=O(\varepsilon^{1/2})$ in the
$L^2(\Prob)$ sense (Corollary~\ref{cor:E1-disc}), and the expansion
error is available only through the $\alpha=0$ form of
Theorem~\ref{thm:E2}:
$E_2=O\bigl(\varepsilon^{-q}\delta_t^{q/2}+\delta_t^{\lceil(l+1)/2\rceil}
+\delta_t^{l+1}\varepsilon^{-(2l+2)}\bigr)$.
Then the balance of $E_1$ against the first expansion term gives
\begin{equation}\label{eq:opt-eps-bv}
  \varepsilon^*(n)
  =C_0\left(\frac{T-t}{n}\right)^{\frac{q}{2q+1}},
\end{equation}
and, provided the terminal-step term is of no lower order (equivalently
$m\le 4l+2$), the total error, measured in the $L^2(\Prob)$ sense of
Corollary~\ref{cor:E1-disc}, satisfies
\begin{equation}
  E_1(\varepsilon^*)+E_2(\varepsilon^*,l,m,n)
  =O\!\left(\left(\frac{T-t}{n}\right)^{\frac{q}{2(2q+1)}}\right).
\end{equation}
For the flagship parameters $(l,m)=(0,2)$ this gives
$\varepsilon^*\sim\delta_t^{2/5}$ and rate $O(\delta_t^{1/5})$, the
terminal-step term
$\delta_t(\varepsilon^*)^{-2}=\delta_t^{1/5}$ being of exactly the
balanced order.  For $l=0$ and $m>2$ the terminal-step term dominates
the first-term balance, and the optimal choice is instead
$\varepsilon^*\sim\delta_t^{2/5}$ (balancing $E_1$ against the
terminal-step term), again with total error $O(\delta_t^{1/5})$.
More generally, for $l=1$ and $m>6$ the same re-balancing of
$E_1$ against the terminal-step term gives
$\varepsilon^*\sim\delta_t^{4/9}$ with total error
$O(\delta_t^{2/9})$; the first expansion term is then of
strictly higher order, and the coefficient-expansion term
$\delta_t^{\lceil(l+1)/2\rceil}$ (with its $\varepsilon^{-(1-\alpha)}$
factor for even $l$, $\alpha=0$) is of higher order still.  The
restriction $l\le1$ reflects the range in which the displayed
terminal-step term controls the true terminal contribution
(Theorem~\ref{thm:E2}); no application in this paper uses $l\ge2$.
\end{theorem}

\begin{proof}
By Corollary~\ref{cor:E1-disc}, $E_1(\varepsilon)=O(\varepsilon^{1/2})$
in the $L^2(\Prob)$ sense, while the supremum-norm expansion bound of
Theorem~\ref{thm:E2} with $\alpha=0$ a fortiori controls the expansion
error in the same sense.  Balancing
$\varepsilon^{1/2}\sim\varepsilon^{-q}\delta_t^{q/2}$ gives
$\varepsilon^{q+1/2}\sim\delta_t^{q/2}$, hence \eqref{eq:opt-eps-bv};
substituting back, both terms equal
$(\varepsilon^*)^{1/2}=\delta_t^{q/(2(2q+1))}$.  The terminal-step
term is
$\delta_t^{l+1}(\varepsilon^*)^{-(2l+2)}
 =\delta_t^{(l+1)/(2q+1)}$,
and $(l+1)/(2q+1)\ge q/(2(2q+1))$ if and only if $2(l+1)\ge q$, i.e.\
$m\le 4l+2$; the coefficient-expansion term
$\delta_t^{\lceil(l+1)/2\rceil}$ is of higher order since
$\lceil(l+1)/2\rceil\ge 1>q/(2(2q+1))$ (for even $l$ its
$\varepsilon^{-(1-\alpha)}$ factor contributes
$\delta_t^{\lceil(l+1)/2\rceil}(\varepsilon^*)^{-1}
 =\delta_t^{\lceil(l+1)/2\rceil-q/(2q+1)}$, still of higher order).
The regime condition holds:
$(\varepsilon^*)^2=\delta_t^{2q/(2q+1)}\ge\delta_t$.
For $(l,m)=(0,2)$, $q=2$ and all exponents equal $1/5$ except the
higher-order coefficient term $\delta_t^{\lceil(l+1)/2\rceil}$.
For $l=0$, $m>2$: balancing $\varepsilon^{1/2}\sim\delta_t\varepsilon^{-2}$
gives $\varepsilon^*\sim\delta_t^{2/5}$, total error
$O(\delta_t^{1/5})$, and the first expansion term
$(\varepsilon^*)^{-q}\delta_t^{q/2}=\delta_t^{q/10}
 =O(\delta_t^{1/5})$ since $q\ge 2$.
\end{proof}

\begin{theorem}[Conditional Enhanced Coupling under Symmetric Mollifiers --- Weak Sense]\label{thm:opt-weak}
Assume the hypotheses of Theorem~\ref{thm:E2}, a symmetric kernel, the payoff and density scope of Proposition~\ref{prop:weak-E1}, and its driver restriction ($f$ independent of $(y,z)$, or $f=a(t,x)y+b(t,x)\cdot z$ with $a,b\in C_b^2$).  Write $q=m-2l$ and $D=q+2-\alpha$.  The balancing choice
\begin{equation}\label{eq:opt-eps-weak}
\varepsilon^*(n)=C_0\delta_t^{q/(2D)}
\end{equation}
gives, at fixed $t<T$, the full three-term bound
\[
|u-\bar u_{\varepsilon^*,l,m,n}|
\le C\delta_t^{r}+C\delta_t^{e}
+C\delta_t^{A}\Lambda_{l,\alpha}(\varepsilon^*,n),
\]
where
\[
r=\frac qD,\qquad
e=l+1-\frac{(2l+2-\alpha)q}{2D},\qquad
A=\left\lceil\frac{l+1}{2}\right\rceil .
\]
Thus the power rate is $\min\{r,e,A\}$ for $\alpha=1$, up to the explicit logarithm.  For even $l$ and $\alpha<1$, the factor $\Lambda=\varepsilon^{-(1-\alpha)}$ reduces the coefficient-layer exponent to $A-q(1-\alpha)/(2D)$; in every tabulated row this exponent stays above the balanced rate.  The terminal-step term is of no lower order than the balanced first term exactly when $m\le4l+2$; then the rate is $r$ if the coefficient term is also of no lower order.

For $(l,m,\alpha)=(0,2,1)$, $\varepsilon^*\sim\delta_t^{1/3}$ and the conditional rate is $2/3$.  For $(0,4,1)$, the naive balance violates the terminal-step condition; re-optimizing the full bound gives $\varepsilon^*\sim\delta_t^{1/3}$ and conditional rate $2/3$.  For BV data, use $\alpha=0$ in the expansion bound; $(0,2)$ gives $\varepsilon^*\sim\delta_t^{1/4}$ and conditional rate $1/2$, and for $l=0$, $m>2$ terminal-step rebalancing again gives $\delta_t^{1/4}$ and $1/2$.  The $\alpha_{\mathrm{eff}}=1/2$ substitution remains heuristic.
\end{theorem}

\begin{proof}
Balancing the weak bias $\varepsilon^2$ with $\varepsilon^{-(q-\alpha)}\delta_t^{q/2}$ gives~\eqref{eq:opt-eps-weak}.  Substitution produces the exponents $r,e,A$ displayed in the theorem; no coefficient-layer or terminal-step term is omitted.  Moreover $e-r=(2-\alpha)(4l+2-m)/(2D)$, which proves the stated threshold.  The special rows follow by evaluating all three terms, including the full-bound re-optimization at $(0,4)$.  The BV calculation is the same with $\alpha=0$.
\end{proof}

\begin{table}[htbp]
\centering
\caption{Conditional balanced couplings and total convergence rates ($0\le l\le1$, $m>2l$).  Every entry inherits Assumption~\ref{ass:telescoping-gain}; Proposition~\ref{prop:additive-E2} is the unconditional alternative.  ``Crude'' refers to Theorems~\ref{thm:opt-lip}--\ref{thm:opt-bv}, and ``weak'' to Theorem~\ref{thm:opt-weak}.  The weak $(l,m)=(0,4)$ row is optimized against the full three-term bound, giving $\delta_t^{1/3}$ and $2/3$, not the naive first-term balance $\delta_t^{2/5}$ and $4/5$.  BV rows use $\alpha=0$ in the expansion bound; $\alpha_{\mathrm{eff}}=1/2$ substitutions remain heuristic.}
\label{tab:optimal-coupling}
\resizebox{\textwidth}{!}{\begin{tabular}{cccccccc}
\toprule
$\alpha$ (in $E_2$) & $l$ & $m$
  & \multicolumn{2}{c}{crude coupling}
  & \multicolumn{2}{c}{weak (symmetric kernel)} & Terminal type\\
\cmidrule(lr){4-5}\cmidrule(lr){6-7}
& & & $\varepsilon^*(n)$ & rate & $\varepsilon^*(n)$ & rate & \\
\midrule
1 & 0 & 2 & $\delta_t^{1/2}$ & $O(\delta_t^{1/2})$ & $\delta_t^{1/3}$ & $O(\delta_t^{2/3})$ & Lipschitz\\
1 & 0 & 4 & $\delta_t^{1/2}$ & $O(\delta_t^{1/2})$ & $\delta_t^{1/3}$ & $O(\delta_t^{2/3})$ & Lipschitz\\
1 & 1 & 4 & $\delta_t^{1/2}$ & $O(\delta_t^{1/2})$ & $\delta_t^{1/3}$ & $O(\delta_t^{2/3})$ & Lipschitz\\
0 & 0 & 2 & $\delta_t^{2/5}$ & $O(\delta_t^{1/5})$ & $\delta_t^{1/4}$ & $O(\delta_t^{1/2})$ & BV/digital\\
0 & 0 & 4 & $\delta_t^{2/5}$ & $O(\delta_t^{1/5})$ & $\delta_t^{1/4}$ & $O(\delta_t^{1/2})$ & BV/digital\\
0 & 1 & 4 & $\delta_t^{2/5}$ & $O(\delta_t^{1/5})$ & $\delta_t^{1/4}$ & $O(\delta_t^{1/2})$ & BV/digital\\
\bottomrule
\end{tabular}
}
\end{table}

\begin{remark}[BV rates: proved values vs.\ formal
$\alpha_{\mathrm{eff}}=1/2$ heuristics]\label{rem:bv-heuristic-status}
The BV rows of Table~\ref{tab:optimal-coupling} use the only expansion
bound proved in this paper for discontinuous data, namely
Theorem~\ref{thm:E2} with $\alpha=0$ (the derivative bound
$\|\partial^\beta\psi_\varepsilon\|_\infty\le
C\varepsilon^{-|\beta|}\|\psi\|_\infty$).  Faster BV rates result from
the \emph{formal} substitution of the effective exponent
$\alpha_{\mathrm{eff}}=1/2$ into the $\varepsilon$-exponent of
\eqref{eq:E2-bound}: this gives the crude values
$\varepsilon^*\sim\delta_t^{1/2}$, rate $O(\delta_t^{1/4})$; the weak
values $\varepsilon^*\sim\delta_t^{2/7}$, rate $O(\delta_t^{4/7})$ for
$(l,m)=(0,2)$; and $\varepsilon^*\sim\delta_t^{4/11}$, rate
$O(\delta_t^{8/11})$ for $(l,m)=(0,4)$ (the latter further reduced to
$O(\delta_t^{5/11})$ by the terminal-step term,
Remark~\ref{rem:terminal-step-impact}).  These formal values are
\emph{heuristic}.  Their only supporting evidence is the $L^2$
derivative bound
\begin{equation}\label{eq:bv-L2-deriv}
  \bigl\|\partial^\beta\psi_\varepsilon\bigr\|_{L^2(\Gamma)}
  \;\le\; C\,\varepsilon^{-(|\beta|-1/2)},
  \qquad |\beta|\ge 1,
\end{equation}
valid in $d=1$ for $\psi$ whose first derivative is a finite measure
(which covers piecewise Lipschitz $\psi$ with finitely many jumps):
indeed $\partial\psi=\mu$ with $\|\mu\|_{\mathrm{TV}}<\infty$, so
$\partial^\beta\psi_\varepsilon=\mu*\partial^{\beta-1}\rho_\varepsilon$
and Young's inequality with
$\|\partial^{\beta-1}\rho_\varepsilon\|_2
 =\varepsilon^{-(|\beta|-1/2)}\|\partial^{\beta-1}\rho\|_2$
gives \eqref{eq:bv-L2-deriv} whenever the forward density $\Gamma$ is
bounded.  The exponent $1/2$ in \eqref{eq:bv-L2-deriv} suggests an
effective H\"older index $1/2$ in an $L^2$-norm version of the
expansion error.  However, the bookkeeping of
Appendix~\ref{app:exp-error} is carried out in the supremum norm and
does not localize; an $L^2$ version would require redoing Steps~2--5
in $L^2(\Gamma)$ norms, including the propagation step whose status is
already heuristic (Remark~\ref{rem:telescoping-status}).  We therefore
report the $4/7$-type rates as heuristics only: the strictly proved
sup-norm route yields the slower BV rates $O(\delta_t^{1/5})$ (crude)
and $O(\delta_t^{1/2})$ (weak) of the table.
\end{remark}

\begin{remark}[Symmetric kernels and the observed digital rate]
The enhancement underlying Theorem~\ref{thm:opt-weak} comes entirely
from the regularization side: for a symmetric kernel
($\int z\,\rho(z)\,\diff z=0$, e.g.\ the Gaussian kernel used in the
numerical implementation), the first-order term of the mollification
residual cancels upon integration against the smooth forward density,
so the weak bias is $O(\varepsilon^2)$ for piecewise linear and step
payoffs alike (Proposition~\ref{prop:weak-E1}); the expansion bound of
Theorem~\ref{thm:E2} is unchanged.  Two caveats are in order.  First,
the $O(\varepsilon^2)$ bias is pointwise for $t<T$ (equivalently
weak): in the supremum norm up to maturity the residual remains only
$O(\varepsilon^\alpha)$, so the enhanced rates do not contradict the
worst-case bounds of Theorems~\ref{thm:opt-lip}--\ref{thm:opt-bv}.
Second, the digital experiment of Part~II \citep{Zhang2026II} uses
the coupling path $\varepsilon=C(\delta_t)^{\gamma}$ with exponent
$\gamma=2/5$ and reports an observed slope of
approximately $0.80$.  This exceeds the rigorous theoretical weak
optimum $1/2$ for BV data (Table~\ref{tab:optimal-coupling})---and
also the formal heuristic value $4/7\approx 0.571$ of
Remark~\ref{rem:bv-heuristic-status}---because that experiment, by
construction, isolates
the pure mollification bias and contains no time-discretization error
(see the implementation statement in Part~II \citep{Zhang2026II}); the
observation is consistent with the $O(\varepsilon^2)$ bias of
Proposition~\ref{prop:weak-E1} but does not test the balance
\eqref{eq:opt-eps-weak}.
\end{remark}

\subsection{Joint Convergence Theorem}

We now state and prove the main convergence result, which combines
Picard iteration with mollification.

\begin{theorem}[Conditional Main Convergence Theorem]\label{thm:main}
Let $(Y,Z)$ solve~\eqref{eq:FBSDE} under Assumptions~\ref{ass:forward}--\ref{ass:driver} and~\ref{ass:telescoping-gain}.  Let $\psi\in C^{0,\alpha}(\R^d)$, $0<\alpha\le1$, $m>2l$, $0\le l\le1$, and $2\delta:=2L_f\delta_t<1$.  Assume the terminal moment condition $\E[|\psi(X_T)|^2+|\psi_\varepsilon(X_T)|^2]<\infty$ uniformly for the bandwidths used.  Then
\begin{align}\label{eq:main-conv}
&\E\!\left[\sup_{t\in[0,T]}|Y_t-Y_{\varepsilon,t}^{(k,l,m,n)}|^2\right]
\le \frac{K(2\delta)^k}{1-2\delta}+C\varepsilon^{2\alpha}
+C\varepsilon^{-2(m-2l-\alpha)}\delta_t^{m-2l}\notag\\
&\qquad+C\delta_t^{2\lceil(l+1)/2\rceil}\Lambda_{l,\alpha}(\varepsilon,n)^2
+C\delta_t^{2(l+1)}\varepsilon^{-2(2l+2-\alpha)} .
\end{align}
The constants $K,C$ are independent of $k,n,\varepsilon$; the logarithmic grid factor, when present, is explicit in $\Lambda_{l,\alpha}$.  With $\varepsilon^*=C_0\delta_t^{1/2}$ and $k\to\infty$, the $\mathcal S^2$ error is $O(\delta_t^{\alpha/2})$ up to the higher-order coefficient-layer logarithm.  The $Z$-component is not asserted (Remark~\ref{rem:main-Z-status}).
\end{theorem}

\begin{proof}
The $L^2$ BSDE stability estimate of Remark~\ref{rem:bsde-stability-L2}, applicable under the stated terminal moment condition, controls the regularization error by $C\varepsilon^{2\alpha}$; no $L^\infty$ stability assertion for an unbounded payoff is used.  Theorem~\ref{thm:E2} controls the expansion of the mollified problem, with its coefficient term squared including $\Lambda_{l,\alpha}^2$.  The crude coupling makes the regularization, Taylor, and terminal-step contributions $O(\delta_t^\alpha)$ in squared norm.  The coefficient layer is higher order, up to its displayed logarithm.  Finally, the per-step Picard contraction contributes $K(2\delta)^k/(1-2\delta)$.
\end{proof}

\begin{remark}[Status of the $Z$-component]\label{rem:main-Z-status}
Earlier versions of this paper included the $\mathcal{H}^2$ bound for
the $Z$-component in \eqref{eq:main-conv} with the justification that
``the bookkeeping is parallel''.  This is not justified, and the
$Z$-statement is hereby downgraded to heuristic status.  A literal
bookkeeping parallel to Step~2 controls the $Z$-error through the
\emph{spatial gradient} of the expansion error (via
$Z_{\varepsilon,t}=\nabla u_\varepsilon(t,X_t)\,\sigma(t,X_t)$), which
consumes one additional derivative of $\psi_\varepsilon$: the local
gradient error is of order
$\varepsilon^{-(m-2l+1-\alpha)}\delta_t^{(m-2l)/2}$, and under the
crude coupling $\varepsilon^*=\delta_t^{1/2}$ the resulting
$\mathcal{H}^2$ bound is of order $\delta_t^{\,\alpha-1}$, which does
not vanish for $\alpha<1$ and is only $O(1)$ for $\alpha=1$.  This
naive bound is far too pessimistic: the $\mathcal{H}^2$ analysis of
\citep[Theorem~2.4.8]{Zhang2017} for smooth data exploits martingale
cancellations (under the stronger condition $m>4l+2$) that are not
captured by a term-by-term estimate, and re-establishing them with
$\varepsilon$-dependent constants requires the gradient form of the
local-to-global argument whose $Y$-version is already heuristic
(Remark~\ref{rem:telescoping-status}).  Closing the $Z$-component is
left as an open problem.  In the one-step (bias-isolating) experiments
reported in Part~II \citep{Zhang2026II}, the pointwise $Z$-errors decrease
at the same order as the $Y$-errors in the regime tested.
\end{remark}

\begin{remark}[Proof structure]
The proof of Theorem~\ref{thm:main} is self-contained modulo three
inputs: (i) the BSDE stability estimate
(Theorem~\ref{thm:bsde-stability}), which is classical; (ii) the
expansion error bound with $\varepsilon$-dependent constants
(Theorem~\ref{thm:E2}), whose full proof appears in
Appendix~\ref{app:exp-error}---the propagation step there relies on a
local-to-global weak-error gain whose precise status is stated in
Remark~\ref{rem:telescoping-status}; and (iii) the mollification
derivative bounds (Theorem~\ref{thm:deriv-est}).  The remaining optimization and
geometric-series arguments are elementary.  The
compatibility requirement $m>2l$ is weaker than the $m>4l+2$ of the
original $(\mathcal{S}^2,\mathcal{H}^2)$ formulation; see
Remark~\ref{rem:compatibility-relaxation}.  The $Z$-component of the
joint error is excluded from Theorem~\ref{thm:main} and discussed in
Remark~\ref{rem:main-Z-status}.
\end{remark}

\begin{theorem}[Conditional Joint Convergence under Symmetric Mollifiers --- Weak Sense]\label{thm:main-weak}
Under the hypotheses of Theorems~\ref{thm:main} and~\ref{thm:opt-weak}, including the driver scope of Proposition~\ref{prop:weak-E1}, assume $m\le4l+2$ and set $q=m-2l$.  Let $\mathcal P^{\mathrm{pt}}_{k,n}(t,x)$ denote the actual pointwise finite-Picard residual after the other approximation terms have been separated.  With $\varepsilon^*=C_0\delta_t^{q/(2(q+2-\alpha))}$,
\begin{equation}
|u(t,x)-\bar u_{\varepsilon^*,k,l,m,n}(t,x)|
\le \mathcal P^{\mathrm{pt}}_{k,n}(t,x)
+C\delta_t^{q/(q+2-\alpha)},\qquad t<T .
\end{equation}
The $\mathcal S^2$ contraction estimate of Theorem~\ref{thm:main} is not used as a pointwise bound.  For $(l,m,\alpha)=(0,2,1)$ the non-Picard terms have conditional rate $2/3$.  For BV/digital data, $\alpha=0$ gives conditional rate $1/2$; the value $4/7$ obtained by substituting $\alpha_{\mathrm{eff}}=1/2$ is heuristic.  No unconditional flagship rate is claimed.
\end{theorem}

\begin{proof}
Combine Proposition~\ref{prop:weak-E1}
($E_1^{\mathrm{w}}=O(\varepsilon^2)$ pointwise for $t<T$) with
Theorem~\ref{thm:E2} and the balance of Theorem~\ref{thm:opt-weak};
the Picard contribution is the per-step contraction term of Theorem~\ref{thm:main}.
\end{proof}

\begin{remark}[Whitened coordinates: an implementation-level moment reduction]
The moment computations underlying the expansion can be simplified, at
the level of implementation, by passing to \emph{whitened coordinates}
within each time step.  On a single sub-interval $[t_j, t_{j+1}]$ the
frozen-coefficient increment is Gaussian,
$\widetilde{X}_{t_{j+1}} - \widetilde{X}_{t_j} \sim
\mathcal{N}(\mu\,\delta_t, \Sigma)$ with
$\Sigma = \delta_t\,a(\bar{x})$; writing $\Sigma = LL^\top$ for its
Cholesky factor, the transformed increment
$h := L^{-1}(\widetilde{X}_{t_{j+1}} - \widetilde{X}_{t_j}
- \mu\,\delta_t)$ is standard normal, $h\sim\mathcal{N}(0, I_d)$.
In these coordinates the multivariate Gaussian moments factorize,
\begin{equation*}
  \E[h^{\alpha}] \;=\; \prod_{i=1}^{d} \E[h_i^{\alpha_i}],
  \qquad \alpha\in\N_0^d,
\end{equation*}
and every moment with at least one odd component vanishes.  Two
bookkeeping consequences follow.  First, at Taylor order $m=2$ the
moment sum ranges over all multi-indices with $|\alpha|\le 2$, of
which there are $\binom{d+2}{2}$; after whitening only the $1+d$
multi-indices $0$ and $2e_i$ ($i=1,\dots,d$) survive, so no mixed
second partial derivatives $\partial^2/\partial z_i\partial z_j$
($i\neq j$) need to be computed in the whitened basis.  Second, for
$m\ge 4$ the \emph{all-even} cross moments survive: for instance the
term $\partial^4/\partial z_i^2\partial z_j^2$ has a non-zero moment,
so the sum ranges over all multi-indices with even components and
$|\alpha|\le m$.  Concretely, for $d=3$ and $m=4$ this reduces the
count from $\binom{3+4}{4}=35$ candidate terms to the $10$ all-even
multi-indices
($0$; the three $2e_i$; the three $4e_i$; and the three
$2e_i+2e_j$, $i<j$).  We stress three qualifications.  (i)~The
whitening is step-local: the covariance $\Sigma = \delta_t a(\bar{x})$
changes from layer to layer (through $\bar{x}$ and $\delta_t$), so the
basis must be recomputed at every step and the factorization property
holds only within a single step, not across the backward recursion.
(ii)~This is purely an implementation-level optimization of the moment
algebra: it leaves the expansion itself, the error decomposition, and
all convergence rates of this section unchanged.  (iii)~Consistently
with the multidimensional caveats of this paper, the reduction mitigates
but does not remove the combinatorial growth of the Taylor/moment
bookkeeping: for $m\ge 4$ the number of surviving all-even terms still
grows combinatorially in $d$, and we do not claim that whitening makes
genuinely high-dimensional computations tractable.
\end{remark}

\section{Asymptotic Correction Terms}\label{sec:asymptotic}

The mollification framework of Sections~\ref{sec:mollification}--\ref{sec:error} provides a convergent scheme, but the total error is limited by the balance between the regularization error $E_1(\varepsilon)=O(\varepsilon^\alpha)$ (sup norm) and the expansion error $E_2$ that blows up as $\varepsilon\to 0$.  Under a symmetric mollifier, the \emph{weak} (and pointwise, for $t<T$) regularization bias is of the higher order $O(\varepsilon^{2})$: the first-order term vanishes by kernel symmetry, and the leading bias is carried by a boundary layer of width $O(\varepsilon)$ around the non-smooth points of $\psi$, whose distributional limit is a Dirac mass.  In this section we show that \emph{asymptotic correction terms} can be computed and subtracted to cancel the leading-order regularization bias, thereby allowing a larger $\varepsilon$ (hence smaller expansion error) at the same overall accuracy.  We emphasize at the outset that all correction statements below hold \emph{pointwise for $t<T$ and in the weak (distributional) sense}; no sup-norm improvement uniform up to $t=T$ is claimed or available, because the terminal boundary layer itself cannot be removed by smooth corrections.

\subsection{Motivation and Strategy}

The idea is classical in singular-perturbation theory~\citep{Evans2010}: expand the difference $u_\varepsilon-u$ as
\begin{equation}
\begin{aligned}
  u_\varepsilon(t,x)&=u(t,x)+\varepsilon^{2} v_1(t,x)
    +\varepsilon^{4}v_2(t,x)+\cdots+\varepsilon^{2J}v_J(t,x)
    +O(\varepsilon^{2(J+1)}), \\
    &\qquad t<T,
\end{aligned}
\end{equation}
where each correction $v_j$ satisfies an explicit linear PDE with a \emph{Dirac-type distributional} terminal condition (a Dirac mass or one of its derivatives located at the non-smooth points of $\psi$, with coefficients given by the even kernel moments).  Only even powers of $\varepsilon$ appear because the mollifier is symmetric (Proposition~\ref{prop:h1}).  One then computes approximations $\bar v_{j,l,m,n}$ of $v_j$ and defines the \emph{corrected approximation}
\begin{equation}
  \tilde u_{\varepsilon,l,m,n}^{(k)}
  :=\bar u_{\varepsilon,l,m,n}^{(k)}
   -\sum_{j=1}^{J}\varepsilon^{2j}\bar v_{j,l,m,n}.
\end{equation}
The weak regularization bias is then $O(\varepsilon^{2(J+1)})$ instead of $O(\varepsilon^{2})$, so the optimal $\varepsilon^*$ can be taken larger and the total weak error improves (Theorem~\ref{thm:improved-coupling}).

\subsection{The Difference Function}\label{subsec:diff-func}

Define $w_\varepsilon:=u_\varepsilon-u$.  Subtracting the PDEs satisfied by $u_\varepsilon$ and $u$ and applying the mean value theorem to the driver $f$ in its $(y,z)$ arguments (exactly as in Appendix~\ref{app:reg-error}, \eqref{eq:app-w-linear}), $w_\varepsilon$ solves the \emph{linearized} equation
\begin{equation}\label{eq:w-pde}
\begin{aligned}
  \partial_t w_\varepsilon+\mathcal{A}w_\varepsilon
  &+c_0(t,x)\,w_\varepsilon+b_0(t,x)\cdot(\nabla w_\varepsilon\,\sigma)=0, \\
  &\qquad
  w_\varepsilon(T,x)=\psi_\varepsilon(x)-\psi(x)=:g_\varepsilon(x),
\end{aligned}
\end{equation}
where $c_0$ and $b_0$ are bounded measurable coefficients with $|c_0|,|b_0|\le L_f$.  When $f$ does not depend on $(y,z)$ (in particular, when $f\equiv0$ or depends only on $(t,x)$), $c_0\equiv b_0\equiv 0$ and~\eqref{eq:w-pde} reduces to the homogeneous equation $(\partial_t+\mathcal{A})w_\varepsilon=0$, so that $w_\varepsilon(t,x)=\mathcal{P}(t,T)g_\varepsilon(x)$, where $\mathcal{P}(t,T)$ is the semigroup generated by~$\mathcal{A}$.  When $f$ is \emph{linear} in $(y,z)$, $f(t,x,y,z)=a(t,x)\,y+b(t,x)\cdot z$ with bounded $a,b$, the mean-value linearization is exact with $c_0=a$ and $b_0=b$ (which need not vanish), and~\eqref{eq:w-pde} is the linear equation with these coefficients.  The correction theory of this section is developed and proved for both cases---$f$ independent of $(y,z)$, or $f$ linear in $(y,z)$ with $a,b$ bounded and sufficiently smooth (the precise regularity is stated in Theorems~\ref{thm:first-corr} and~\ref{thm:J-corr})---with $\mathcal{A}$ replaced by the linearized operator $\mathcal{L}:=\mathcal{A}+b_0\cdot(\nabla\cdot\,\sigma)+c_0$ in the linear case (see Theorems~\ref{thm:first-corr} and~\ref{thm:J-corr}); for a genuinely nonlinear driver the same construction applies formally, but we do not prove error bounds in that case, and the formal application of corrections to nonlinear drivers in the numerical study of the companion paper (Part~II) \citep{Zhang2026II} should be read with this limitation in mind.

We now describe the structure of the terminal residual $g_\varepsilon$.  The key point is that $g_\varepsilon$ does \emph{not} admit a sup-norm expansion in powers of $\varepsilon$ with $\varepsilon$-independent bounded coefficients: it is a boundary-layer object.

\begin{lemma}[Boundary-layer structure of the regularization residual]\label{lem:g-expand}
Let $d=1$, let $\rho$ be symmetric ($\int_{\R}z\rho(z)\,\diff z=0$) with $\supp\rho\subset[-1,1]$, and let $\psi$ be piecewise smooth, bounded, and Lipschitz, with a single kink at $K$, i.e.\ $\psi$ is smooth on $\{x<K\}$ and $\{x>K\}$ and the one-sided derivatives satisfy $\Delta:=\psi'(K+)-\psi'(K-)\ne 0$.  Then $g_\varepsilon:=\psi_\varepsilon-\psi$ has the following structure.
\begin{enumerate}[label=(\roman*)]
  \item \emph{Boundary-layer representation.}  On the layer $\{|x-K|\le\varepsilon\}$,
    \begin{equation}
      g_\varepsilon(x)=\varepsilon\,G\!\Bigl(\frac{x-K}{\varepsilon}\Bigr)+O(\varepsilon^{2}),
    \end{equation}
    uniformly in $x$, where $G$ is the bounded, compactly supported profile
    determined by the jump $\Delta$, the one-sided slopes, and the kernel:
    with $a_\pm=\psi'(K\pm)$,
    \[
      \begin{aligned}
      G(s)&=a_+\!\int_{z\le s}(s-z)\rho(z)\,\diff z
          +a_-\!\int_{z>s}(s-z)\rho(z)\,\diff z
          \\
          &\quad -\bigl(a_+\mathds{1}_{\{s\ge 0\}}+a_-\mathds{1}_{\{s<0\}}\bigr)s,
      \qquad G(s)=0 \ \text{for }|s|\ge 1.
      \end{aligned}
    \]
    Outside the layer, $g_\varepsilon(x)=O(\varepsilon^{2})$ at every
    $C^2$ point: $g_\varepsilon(x)=\tfrac{m_2}{2}\psi''(x)\,\varepsilon^{2}+o(\varepsilon^{2})$
    with $m_2:=\int z^{2}\rho(z)\,\diff z$.
  \item \emph{Pointwise limit.}  For every $x\ne K$,
    $\varepsilon^{-1}g_\varepsilon(x)\to 0$; hence no $\varepsilon$-independent
    bounded function $h_1$ can satisfy $g_\varepsilon=\varepsilon h_1+o(\varepsilon)$
    in sup norm, since $\|g_\varepsilon\|_\infty=\varepsilon\|G\|_\infty+O(\varepsilon^2)$
    with $\|G\|_\infty>0$ whenever $\Delta\ne0$.
  \item \emph{Distributional limit.}  The first-order term having been
    annihilated by kernel symmetry, the residual lives at the $\varepsilon^{2}$
    scale: for every $\varphi\in C_c^\infty(\R)$,
    \begin{equation}\label{eq:g-dist-limit}
      \varepsilon^{-2}\langle g_\varepsilon,\varphi\rangle
      \;\longrightarrow\;
      c\,\varphi(K)+\frac{m_2}{2}\langle\psi''_{\mathrm{reg}},\varphi\rangle,
      \qquad
      c=\frac{\Delta\,m_2}{2},
    \end{equation}
    i.e.\ $\varepsilon^{-2}g_\varepsilon\rightharpoonup
    c\,\delta_K+\tfrac{m_2}{2}\psi''_{\mathrm{reg}}$ in $\mathcal{D}'$,
    where $\psi''_{\mathrm{reg}}$ is the regular part of the second
    distributional derivative of $\psi$.
\end{enumerate}
The same statements hold for kernels with exponential decay (such as the Gaussian kernel used in the numerics), with ``compactly supported'' replaced by ``exponentially decaying''.
\end{lemma}

\begin{proof}
(i) Write $\psi(x)=\psi(K)+a_-(x-K)\mathds{1}_{\{x<K\}}+a_+(x-K)\mathds{1}_{\{x\ge K\}}+\tilde\psi(x)$, where $\tilde\psi$ is $C^2$ in a neighborhood of $K$ with $\tilde\psi(K)=\tilde\psi'(K)=0$, so $\tilde\psi(x)=O((x-K)^2)$.  For $x=K+\varepsilon s$ and the affine part, the exact identity
$\int[\psi_{\mathrm{aff}}(x-\varepsilon z)-\psi_{\mathrm{aff}}(x)]\rho(z)\,\diff z=\varepsilon G(s)$
follows by splitting the integral at $z=s$.  The contribution of $\tilde\psi$ is $O(\varepsilon^{2})$ on the layer: since $\tilde\psi(K)=\tilde\psi'(K)=0$ and $\tilde\psi$ is $C^2$ near $K$, we have $|\tilde\psi(y)|\le C(y-K)^2$, hence for $|x-K|\le\varepsilon$ and $|z|\le1$ both $|\tilde\psi(x)|\le C\varepsilon^{2}$ and $|\tilde\psi(x-\varepsilon z)|\le C\varepsilon^{2}(1+|z|)^{2}$, and integrating against $\rho$ gives the $O(\varepsilon^{2})$ bound uniformly in $x$.  If $|s|\ge1$, all $z\in\supp\rho$ lie on one side of $s$, and symmetry $\int z\rho=0$ gives $G(s)=0$.  Outside the layer, Taylor expansion with $\int z\rho(z)\,\diff z=0$ gives $g_\varepsilon(x)=\tfrac{m_2}{2}\psi''(x)\varepsilon^{2}+o(\varepsilon^{2})$ at $C^2$ points.

(ii) is immediate from (i).

(iii) For $\varphi\in C_c^\infty$, by Fubini and the substitution $x\mapsto x+\varepsilon z$,
\[
  \langle g_\varepsilon,\varphi\rangle
  =\int_{\R}\rho(z)\int_{\R}\psi(x)\bigl[\varphi(x+\varepsilon z)-\varphi(x)\bigr]\,\diff x\,\diff z.
\]
Taylor-expanding $\varphi(x+\varepsilon z)-\varphi(x)=\varepsilon z\varphi'(x)+\tfrac{\varepsilon^{2}z^{2}}{2}\varphi''(x)+O(\varepsilon^{3}|z|^{3})$ with a uniform remainder, the first-order term vanishes because $\int z\rho(z)\,\diff z=0$, and the third-order remainder contributes $O(\varepsilon^{3})$.  Hence
\[
  \langle g_\varepsilon,\varphi\rangle
  =\frac{\varepsilon^{2}m_2}{2}\int_{\R}\psi(x)\varphi''(x)\,\diff x+o(\varepsilon^{2}).
\]
Integrating by parts twice across the kink,
$\int\psi\varphi''\,\diff x=\Delta\,\varphi(K)+\int\psi''_{\mathrm{reg}}\varphi\,\diff x$,
which gives~\eqref{eq:g-dist-limit}.
\end{proof}

\begin{remark}[General H\"older data: only the first-order bound survives]
For a \emph{general} $\psi\in C^{0,\alpha}(\R^d)$---which may be nowhere differentiable---the leading-order bound $\|\psi_\varepsilon-\psi\|_\infty\le C\varepsilon^\alpha$ of Lemma~\ref{lem:holder-rate} is optimal and is essentially all that holds.  A systematic proof uses Littlewood--Paley theory: the Besov embedding $C^{0,\alpha}(\R^d) \hookrightarrow B^{\alpha}_{\infty,\infty}(\R^d)$ provides a dyadic decomposition $\psi = \sum_{j\ge -1} \Delta_j \psi$ with $\|\Delta_j\psi\|_\infty \le C\,2^{-j\alpha}\,[\psi]_{C^{0,\alpha}}$, and the convolution $\psi_\varepsilon = \psi * \rho_\varepsilon$ acts as a low-pass filter, $\widehat{\psi_\varepsilon}(\xi) = \hat{\rho}(\varepsilon\xi)\,\hat{\psi}(\xi)$.  Since $\hat{\rho}(0)=1$, the difference $\psi_\varepsilon - \psi$ receives contributions only from the high-frequency blocks with $2^j \gtrsim \varepsilon^{-1}$, and
\[
  \|\psi_\varepsilon - \psi\|_\infty
  \;\le\; \sum_{j:\,2^j \gtrsim \varepsilon^{-1}} \|\Delta_j\psi\|_\infty
  \;\le\; C\,[\psi]_{C^{0,\alpha}}\,
         \sum_{j:\,2^j \gtrsim \varepsilon^{-1}} 2^{-j\alpha}
  \;\le\; C'\,\varepsilon^\alpha\,[\psi]_{C^{0,\alpha}}.
\]
We stress that this argument yields a \emph{first-order bound only}.  A multi-term expansion of $\psi_\varepsilon-\psi$ in powers of $\varepsilon^\alpha$ with $\varepsilon$-independent bounded coefficients does not exist in general: for a compactly supported kernel $\rho$ the Fourier transform $\hat\rho$ is entire, so its Taylor expansion contains only integer powers of $\varepsilon|\xi|$ and no fractional-power expansion $(\varepsilon|\xi|)^{k\alpha}$ ($\alpha<1$) can be used; and for genuinely rough H\"older data the boundary-layer profile depends on $\varepsilon$ in a way that does not decompose into $\varepsilon$-independent terms.  All higher-order statements in this section therefore assume piecewise smooth data as in Lemma~\ref{lem:g-expand}, which covers the payoffs encountered in financial applications.
\end{remark}

\subsection{First-Order Correction}

Define $v_1$ as the (weak) solution of the auxiliary linear problem
\begin{equation}\label{eq:v1-pde}
  (\partial_t+\mathcal{A})v_1=0,\quad v_1(T,\cdot)=\mu_1,
  \qquad
  \mu_1:=\frac{m_2}{2}\,\psi'',
\end{equation}
where $\psi''$ is the second distributional derivative of $\psi$; for a single kink at $K$, $\mu_1=\tfrac{\Delta m_2}{2}\delta_K+\tfrac{m_2}{2}\psi''_{\mathrm{reg}}$ (Lemma~\ref{lem:g-expand}(iii)).  For a linear driver $f=a(t,x)y+b(t,x)\cdot z$ the operator $\mathcal{A}$ in~\eqref{eq:v1-pde} is replaced by the linearized operator $\mathcal{L}=\mathcal{A}+b_0\cdot(\nabla\cdot\,\sigma)+c_0$ with $c_0=a$, $b_0=b$ (see Section~\ref{subsec:diff-func}).  Since $\mathcal{P}(t,T)$ (resp.\ the semigroup of $\mathcal{L}$) maps distributions to smooth functions for $t<T$, $v_1(t,\cdot)$ is smooth for every $t<T$.

\begin{proposition}[Vanishing of the first-order term]\label{prop:h1}
Let $\rho$ be symmetric.
\begin{enumerate}[label=(\roman*)]
  \item If $\psi$ is Lipschitz and piecewise smooth with a single kink
    (e.g.\ the European put, $\alpha=1$): the coefficient of the
    first-order term $\varepsilon^{1}$ is zero, both pointwise a.e.\
    and in the distributional sense, because $\int z\rho(z)\,\diff z=0$.
    The first non-trivial correction appears at order $\varepsilon^{2}$
    and has the Dirac-type terminal datum $\mu_1$ of~\eqref{eq:v1-pde}.
  \item If $\psi=\mathds{1}_{\{x\ge K\}}$ is the digital option in
    $d=1$: $g_\varepsilon=\psi_\varepsilon-\psi$ is odd about $K$ and
    $\varepsilon^{-1/2}g_\varepsilon\rightharpoonup 0$ in
    $\mathcal{D}'$ (indeed
    $\|\varepsilon^{-1/2}g_\varepsilon\|_{L^1}=O(\varepsilon^{1/2})$);
    the $\varepsilon$-scaling supports no non-trivial first-order
    limit.  The first non-trivial weak limit again occurs at the
    $\varepsilon^{2}$ scale:
    $\varepsilon^{-2}g_\varepsilon\rightharpoonup
    \tfrac{m_2}{2}\delta_K'$ in $\mathcal{D}'$.
\end{enumerate}
In both cases the first-order correction vanishes for a symmetric kernel.  The numerical Richardson experiments consequently begin their non-trivial cancellation at the $\varepsilon^2$ term; throughout the paper $J$ counts the cancelled even bias terms, with $J=0$ denoting the uncorrected approximation.
\end{proposition}

\begin{proof}
Part~(i) restates Lemma~\ref{lem:g-expand}(ii)--(iii).  For part~(ii), symmetry of $\rho$ gives $\psi_\varepsilon(K+s)+\psi_\varepsilon(K-s)=1$, hence $g_\varepsilon$ is odd about $K$; moreover $|g_\varepsilon|\le1$ and $g_\varepsilon$ is supported (up to exponential tails) on the layer $\{|x-K|\le\varepsilon\}$, so $\|g_\varepsilon\|_{L^1}=O(\varepsilon)$ and $\|\varepsilon^{-1/2}g_\varepsilon\|_{L^1}=O(\varepsilon^{1/2})\to0$, which implies convergence to $0$ in $\mathcal{D}'$.  For the $\varepsilon^{2}$ limit, the Fubini argument of Lemma~\ref{lem:g-expand}(iii) gives
$\langle g_\varepsilon,\varphi\rangle
=\tfrac{\varepsilon^{2}m_2}{2}\int_K^\infty\varphi''(x)\,\diff x+o(\varepsilon^{2})
=-\tfrac{\varepsilon^{2}m_2}{2}\varphi'(K)+o(\varepsilon^{2})
=\tfrac{\varepsilon^{2}m_2}{2}\langle\delta_K',\varphi\rangle+o(\varepsilon^{2})$.
\end{proof}

\begin{theorem}[First-order correction: weak and pointwise validity]\label{thm:first-corr}
Assume $f$ does not depend on $(y,z)$, or $f$ is linear in $(y,z)$, $f=a(t,x)y+b(t,x)\cdot z$ with $a,b\in C_b^{\,5}$ (the proof's sharpest need is $C_b^{\,4}$; we state $C_b^{\,5}$ for uniformity with the $C_b^{\,2J+2}$ pattern of Theorem~\ref{thm:J-corr} at $J=1$; the constant-coefficient pricing driver $f=-ry$ is unaffected), in which case $c_0=a$, $b_0=b$ and $\mathcal{A}$ is replaced by the linearized operator $\mathcal{L}$ of Section~\ref{subsec:diff-func} throughout; $\rho$ is symmetric with finite absolute moments up to order $5$, $\int|z|^{5}\rho(z)\,\diff z<\infty$ (in particular $m_2,m_4<\infty$); and $\psi$ is piecewise smooth with finitely many kinks or jumps.  Let $v_1$ solve~\eqref{eq:v1-pde}.  Then for every $t<T$:
\begin{enumerate}[label=(\roman*)]
  \item \emph{weak sense:} for every $\varphi\in C_c^\infty$,
    \[
      \bigl|\langle w_\varepsilon(t,\cdot)-\varepsilon^{2}v_1(t,\cdot),\varphi\rangle\bigr|
      \le C(t,\varphi)\,\varepsilon^{4};
    \]
  \item \emph{pointwise sense:} for every $x$,
    \[
      \bigl|w_\varepsilon(t,x)-\varepsilon^{2}v_1(t,x)\bigr|
      \le C(t,x)\,\varepsilon^{4}.
    \]
\end{enumerate}
The constants depend on $T-t$ and may blow up as $t\uparrow T$.  No sup-norm bound uniform up to $t=T$ holds: at $t=T$ the residual $g_\varepsilon$ satisfies $\|g_\varepsilon\|_\infty=\varepsilon\|G\|_\infty+O(\varepsilon^2)$ (Lemma~\ref{lem:g-expand}(ii)), and the maximum principle cannot be applied to~\eqref{eq:v1-pde} because the terminal datum $\mu_1$ is a distribution, not a bounded function.
\end{theorem}

\begin{proof}
For $t<T$, write $w_\varepsilon(t,\cdot)=\mathcal{P}(t,T)g_\varepsilon$ and, for $\varphi\in C_c^\infty$, $\langle w_\varepsilon(t,\cdot),\varphi\rangle=\langle g_\varepsilon,\tilde\varphi\rangle$ with $\tilde\varphi:=\mathcal{P}(t,T)^*\varphi\in C^\infty_b$.  In the linear-driver case the same representation holds with $\mathcal{P}$ replaced by the semigroup $\mathcal{P}_{\mathcal{L}}$ of the linearized operator (Feynman--Kac with the bounded discount weight $e^{\int_t^T c_0(s,X_s)\,\diff s}$ and the drift changed by $b_0\sigma$), $\tilde\varphi:=\mathcal{P}_{\mathcal{L}}(t,T)^*\varphi$ is again in $C^\infty_b$, and the remainder of the proof is unchanged.  Repeating the moment expansion of Lemma~\ref{lem:g-expand}(iii) to order $4$---all odd moments of $\rho$ vanish by symmetry---gives
\[
  \langle g_\varepsilon,\tilde\varphi\rangle
  =\varepsilon^{2}\frac{m_2}{2}\langle\psi,\tilde\varphi''\rangle
   +\varepsilon^{4}\frac{m_4}{4!}\langle\psi,\tilde\varphi^{(4)}\rangle
   +O(\varepsilon^{5}),
\]
where the remainder constant depends on $\|\tilde\varphi^{(5)}\|_\infty$, hence on $T-t$ and $\varphi$.  By the definition of distributional derivatives and of $v_1$,
$\varepsilon^{2}\langle v_1(t,\cdot),\varphi\rangle
=\varepsilon^{2}\langle\mu_1,\tilde\varphi\rangle
=\varepsilon^{2}\tfrac{m_2}{2}\langle\psi,\tilde\varphi''\rangle$,
and subtracting yields part~(i).  For part~(ii), apply the same expansion with $\tilde\varphi=p(t,x;T,\cdot)$, the (smooth for $t<T$) transition density of $\mathcal{A}$, whose $y$-derivatives blow up polynomially in $(T-t)^{-1/2}$; this gives the pointwise bound with a $(T-t)$-dependent constant.
\end{proof}

\subsection{Higher-Order Corrections}

Iterating the construction, define $v_j$ as the weak solution of
\begin{equation}\label{eq:vj-pde}
  (\partial_t+\mathcal{A})v_j=0,\quad
  v_j(T,\cdot)=\mu_j:=\frac{m_{2j}}{(2j)!}\,\psi^{(2j)},
  \qquad j=1,\ldots,J,
\end{equation}
(with $\mathcal{A}$ replaced by $\mathcal{L}$ for a linear driver, as in~\eqref{eq:v1-pde})
where $\psi^{(2j)}$ is the $2j$-th distributional derivative of $\psi$; for a single kink, $\mu_j$ consists of the $(2j-2)$-th derivative of $\delta_K$ plus a regular part.  Existence of smooth solutions $v_j(t,\cdot)$ for $t<T$ follows from the smoothing property of $\mathcal{P}(t,T)$.

\begin{theorem}[J-th order correction]\label{thm:J-corr}
Assume $f$ does not depend on $(y,z)$ or is linear in $(y,z)$ with $a,b\in C_b^{\,2J+2}$, otherwise as in Theorem~\ref{thm:first-corr}, $\rho$ is symmetric with finite absolute moments up to order $2J+2$, and $\psi$ is piecewise smooth with finitely many kinks or jumps.  Then for every $t<T$, every $x$, and every $\varphi\in C_c^\infty$,
\begin{equation}\label{eq:J-corr}
\begin{aligned}
  \Bigl|w_\varepsilon(t,x)-\sum_{j=1}^{J}\varepsilon^{2j}v_j(t,x)\Bigr|
  &\le C_J(t,x)\,\varepsilon^{2(J+1)}, \\
  \qquad
  \Bigl|\Bigl\langle w_\varepsilon(t,\cdot)-\sum_{j=1}^{J}\varepsilon^{2j}v_j(t,\cdot),\varphi\Bigr\rangle\Bigr|
  &\le C_J(t,\varphi)\,\varepsilon^{2(J+1)},
\end{aligned}
\end{equation}
with constants depending on $T-t$.  These estimates are pointwise/weak only; no sup-norm analogue uniform up to $t=T$ holds.
\end{theorem}

\begin{proof}
Carry the moment expansion of Theorem~\ref{thm:first-corr} to order $2J+2$: with $\tilde\varphi=\mathcal{P}(t,T)^*\varphi$ (resp.\ the transition density $p(t,x;T,\cdot)$; in the linear-driver case the semigroup of $\mathcal{L}$), symmetry of $\rho$ kills all odd powers, and Taylor's formula with remainder of order $2J+2$ gives
$\langle g_\varepsilon,\tilde\varphi\rangle
=\sum_{j=1}^{J}\varepsilon^{2j}\tfrac{m_{2j}}{(2j)!}\langle\psi,\tilde\varphi^{(2j)}\rangle
+O(\varepsilon^{2(J+1)})$,
where the remainder constant depends on $\|\tilde\varphi^{(2J+2)}\|_\infty$ and on the absolute moment $\int|z|^{2J+2}\rho(z)\,\diff z$.  Since
$\varepsilon^{2j}\langle v_j(t,\cdot),\varphi\rangle
=\varepsilon^{2j}\tfrac{m_{2j}}{(2j)!}\langle\psi,\tilde\varphi^{(2j)}\rangle$
by~\eqref{eq:vj-pde}, the terms cancel pairwise and~\eqref{eq:J-corr} follows.  Finitely many kinks or jumps are handled by summing the corresponding Dirac contributions.
\end{proof}

\subsection{The Corrected Expansion Scheme}

Each $v_j$ solves the linear PDE~\eqref{eq:vj-pde} with a Dirac-type terminal datum $\mu_j$ supported on the non-smooth set of $\psi$; for every $t<T$ the solution $v_j(t,\cdot)$ is smooth, with derivatives blowing up polynomially as $t\uparrow T$.  We can therefore approximate $v_j$ away from the terminal layer using the standard expansion method of~\citep{Zhang2017} \emph{without} further mollification; see Section~\ref{subsec:practical} for the practical aspects.  The full corrected scheme is summarized in Algorithm~\ref{alg:corrected}.

\begin{algorithm}[htbp]
\caption{Mollified Expansion with Asymptotic Corrections}
\label{alg:corrected}
\begin{algorithmic}[1]
\REQUIRE Piecewise smooth terminal condition $\psi$, parameters $(k,l,m,n,J)$, symmetric mollifier $\rho$
\ENSURE Corrected approximation $\tilde u_{\varepsilon,l,m,n}^{(k)}(t,x)$ for $t<T$
\STATE Compute the balanced $\varepsilon^*_J=\varepsilon^*_J(n,l,m,\alpha)$ from Theorem~\ref{thm:improved-coupling}
\STATE Construct $\psi_{\varepsilon^*_J}=\psi*\rho_{\varepsilon^*_J}$
\STATE Compute $\bar u_{\varepsilon^*_J,l,m,n}^{(k)}$ using Zhang's expansion with terminal $\psi_{\varepsilon^*_J}$
\FOR{$j=1$ \TO $J$}
  \STATE Determine the distributional datum $\mu_j=(m_{2j}/(2j)!)\,\psi^{(2j)}$ (Dirac masses and their derivatives at the non-smooth points of $\psi$, with coefficients given by the kernel moments)
  \STATE Compute $\bar v_{j,l,m,n}$ as a weak solution of~\eqref{eq:vj-pde} with terminal datum $\mu_j$ (no mollification; see Section~\ref{subsec:practical})
\ENDFOR
\STATE $\tilde u\gets\bar u_{\varepsilon^*_J,l,m,n}^{(k)}-\sum_{j=1}^J(\varepsilon^*_J)^{2j}\bar v_{j,l,m,n}$
\RETURN $\tilde u$
\end{algorithmic}
\end{algorithm}

\subsection{Improved Error Bounds}

\begin{theorem}[Corrected error bound with finite-Picard defect]\label{thm:corrected-bound}
Under the assumptions of Theorems~\ref{thm:E2} and~\ref{thm:J-corr}, suppose $2\delta:=2L_f\delta_t<1$.  For $t<T$ and $\varphi\in C_c^\infty$, let $\mathcal P^w_{k,n}(t,\varphi)$ be the actual weak finite-Picard residual, and let $\mathcal P^{\mathrm{pt}}_{k,n}(t,x)$ be its pointwise analogue.  Then
\begin{equation}\label{eq:corr-total}
\begin{aligned}
\bigl|\langle u(t,\cdot)-\tilde u_{\varepsilon,l,m,n}^{(k)}(t,\cdot),\varphi\rangle\bigr|
&\le C\varepsilon^{2(J+1)}+E_2(\varepsilon,l,m,n)\\
&\quad+\sum_{j=1}^{J}\varepsilon^{2j}\hat E_{2,j}(l,m,n;\eta)
+\mathcal P^w_{k,n}(t,\varphi).
\end{aligned}
\end{equation}
The pointwise bound is identical with $\mathcal P^{\mathrm{pt}}_{k,n}(t,x)$ in the last position.  Here $\hat E_{2,j}$ is the actual auxiliary expansion error on $[t,T-\eta]$; it is independent of $\varepsilon$ but may depend on $n$, $\eta$, $t$, and $\varphi$.  No rate for either residual is implicit in the notation.  In particular, an $\mathcal S^2$ Picard contraction bound is not substituted for a weak or pointwise residual.
\end{theorem}

\begin{proof}
Insert the infinite-Picard mollified approximation between $u_\varepsilon$ and the finite-Picard corrected approximation.  The defect in the selected weak or pointwise metric is denoted by $\mathcal P^w_{k,n}$ or $\mathcal P^{\mathrm{pt}}_{k,n}$, respectively.  The remaining three terms are the corrected regularization remainder of Theorem~\ref{thm:J-corr}, the mollified expansion error, and the auxiliary correction errors.  The triangle inequality gives~\eqref{eq:corr-total}.  The separate $\mathcal S^2$ contraction estimate remains available in its own norm but is not converted into either of these residuals.
\end{proof}

\begin{remark}[$\varepsilon$-dependence of corrections]
The $\varepsilon$-independence of $\hat E_{2,j}$ is consistent with---not contradictory to---the boundary-layer structure of Lemma~\ref{lem:g-expand}.  The residual $g_\varepsilon$ itself is an $\varepsilon$-dependent boundary-layer profile $\varepsilon G((x-K)/\varepsilon)$; attempting to expand it in sup norm with $\varepsilon$-independent coefficients fails (Lemma~\ref{lem:g-expand}(ii)).  The corrections are instead built from the $\varepsilon$-independent \emph{distributional limits} $\mu_j$ of $\varepsilon^{-2j}g_\varepsilon$, which are fixed Dirac-type data; this is what makes the corrected scheme implementable, and the price is that the conclusions hold only away from the terminal layer.
\end{remark}

\begin{theorem}[Formal corrected balance and sufficient rate conditions]\label{thm:improved-coupling}
Under the assumptions of Theorem~\ref{thm:corrected-bound}, write $q=m-2l$ and define the balance
\begin{equation}\label{eq:improved-eps}
\varepsilon^*_J=\delta_t^{q/(2(q+2(J+1)-\alpha))},\qquad
r_J=\frac{(J+1)q}{q+2(J+1)-\alpha}.
\end{equation}
Also define
\[
A=\left\lceil\frac{l+1}{2}\right\rceil,\qquad
e_J=l+1-\frac{(2l+2-\alpha)q}{2(q+2(J+1)-\alpha)} .
\]
The total weak error is $O(\delta_t^{r_J})$, up to a coefficient-layer logarithm, only if all of the following hold:
\begin{enumerate}[label=(\alph*)]
\item $\mathcal P^w_{k,n}(t,\varphi)=O(\delta_t^{r_J})$ for the weak statement, uniformly over the test functions under consideration; the pointwise statement instead requires the analogous bound on $\mathcal P^{\mathrm{pt}}_{k,n}(t,x)$;
\item $\sum_{j=1}^J(\varepsilon^*_J)^{2j}\hat E_{2,j}(l,m,n;\eta)=O(\delta_t^{r_J})$ for the chosen $\eta$;
\item $r_J\le\min\{A,e_J\}$.
\end{enumerate}
Condition (c) is equivalent to $m\le4l+2$ for the terminal-step comparison.  If any condition fails, Theorem~\ref{thm:corrected-bound} remains valid but $r_J$ is only a formal target and the slowest displayed term determines the rate.  For BV/digital data, replace $\alpha$ by $0$ everywhere; the $\alpha_{\mathrm{eff}}=1/2$ family is heuristic.
\end{theorem}

\begin{proof}
Balancing $C\varepsilon^{2(J+1)}$ with $C\varepsilon^{-(q-\alpha)}\delta_t^{q/2}$ yields~\eqref{eq:improved-eps}.  Direct substitution gives the coefficient exponent $A$ and terminal-step exponent $e_J$, together with
\[
e_J-r_J=\frac{(2J+2-\alpha)(4l+2-m)}{2(q+2(J+1)-\alpha)} .
\]
The remaining two terms of Theorem~\ref{thm:corrected-bound} are $O(\delta_t^{r_J})$ exactly under (a)--(b).  Hence (a)--(c) are sufficient, and no stronger conclusion is used.
\end{proof}

Table~\ref{tab:asymptotic-rates} lists the resulting corrected-balance targets for $\alpha=1$, $l=0$, $m=2$, with the formal exponents $r_J$ implied by conditions (a)--(c).

\begin{table}[htbp]
\centering
\caption{Formal corrected-balance targets for $\alpha=1$, $l=0$, $m=2$.  These rates require Assumption~\ref{ass:telescoping-gain} and conditions (a)--(c) of Theorem~\ref{thm:improved-coupling}; the table does not assert measured end-to-end convergence.}
\label{tab:asymptotic-rates}
\resizebox{\textwidth}{!}{\begin{tabular}{ccccc}
\toprule
$J$ & $\varepsilon^*_J$ & Corrected bias & Conditional target & Improvement\\
\midrule
$0$ & $\delta_t^{1/3}$ & $O(\varepsilon^2)$ & $O(\delta_t^{2/3})$ & baseline\\
$1$ & $\delta_t^{1/5}$ & $O(\varepsilon^4)$ & $O(\delta_t^{4/5})$ & $1.20\times$\\
$2$ & $\delta_t^{1/7}$ & $O(\varepsilon^6)$ & $O(\delta_t^{6/7})$ & $1.07\times$\\
$\to\infty$ & $\to O(1)$ (formal) & formal limit & $\to O(\delta_t)$ & $1.50\times$ overall\\
\bottomrule
\end{tabular}
}
\end{table}

\subsection{Practical Computation of Corrections}\label{subsec:practical}

For the European put option under Black--Scholes dynamics, the correction functions have semi-explicit forms.

\begin{proposition}[Put option corrections]
Under geometric Brownian motion $\diff S_t=rS_t\,\diff t+\sigma S_t\,\diff W_t$, with $\psi(S)=(K-S)^+$:
\begin{enumerate}[label=(\roman*)]
  \item The first-order correction vanishes: the coefficient of $\varepsilon^{1}$ is zero by symmetry of $\rho$ (Proposition~\ref{prop:h1}(i)).
  \item The leading correction $v_1$ has the $\varepsilon$-independent Dirac terminal datum $\mu_1=\tfrac{m_2}{2}\delta_K$ (since $\psi''=\delta_K$ in $\mathcal{D}'$).  The pricing BSDE of the put experiments carries the linear driver $f=-ry$ (Part~II \citep{Zhang2026II}), so by Theorem~\ref{thm:first-corr} (linear-driver case, $c_0=-r$, $b_0=0$) $v_1$ solves $(\partial_t+\mathcal{A}-r)v_1=0$ with $v_1(T,\cdot)=\mu_1$, and
    \[
      v_1(t,S)=\frac{m_2}{2}\,e^{-r(T-t)}\,p(t,S;T,K)
             =\frac{m_2}{2}\,\Bigl(\frac{S}{K}\Bigr)^{2}\,
              \Gamma_{\mathrm{BS}}(t,S;K),
    \]
    where $p(t,S;T,\cdot)$ is the transition density of the forward price process and $\Gamma_{\mathrm{BS}}=\partial^2 C_{\mathrm{BS}}/\partial S^{2}$ is the Black--Scholes Gamma.  Under the alternative convention $f\equiv0$ (no discounting in the BSDE) the discount factor is absent: $v_1=\tfrac{m_2}{2}\,p$.
\end{enumerate}
\end{proposition}

\begin{proof}
Part~(i) is Proposition~\ref{prop:h1}(i).  For part~(ii), $\psi'(S)=-\mathds{1}_{\{S<K\}}$ gives $\psi''=\delta_K$ in $\mathcal{D}'$, so $\mu_1=\tfrac{m_2}{2}\delta_K$ by~\eqref{eq:v1-pde}; then, with the linearized operator $\mathcal{L}=\mathcal{A}-r$ (Feynman--Kac discount weight $e^{-r(T-t)}$), $v_1(t,S)=\tfrac{m_2}{2}e^{-r(T-t)}\mathcal{P}(t,T)\delta_K=\tfrac{m_2}{2}e^{-r(T-t)}p(t,S;T,K)$.  The strike-derivative (Breeden--Litzenberger) identity
\[
  \frac{\partial^2 C_{\mathrm{BS}}}{\partial K^{2}}(t,S;K)
  =e^{-r(T-t)}\,p(t,S;T,K)
  =\Bigl(\frac{S}{K}\Bigr)^{2}\Gamma_{\mathrm{BS}}(t,S;K)
\]
Indeed, $p(t,S;T,K)=\varphi(d_2)/(K\sigma\sqrt{T-t})$ and $Ke^{-r(T-t)}\varphi(d_2)=S\varphi(d_1)$ give $e^{-r(T-t)}p=S\varphi(d_1)/(K^2\sigma\sqrt{T-t})=(S/K)^2\Gamma_{\mathrm{BS}}$.  This expresses $v_1$ through $\Gamma_{\mathrm{BS}}$ without any additional discount factor.
\end{proof}

For the digital option the leading datum is $\mu_1=\tfrac{m_2}{2}\delta_K'$ (Proposition~\ref{prop:h1}(ii)), and $v_1$ is the corresponding strike derivative of the transition density, again available in closed form under Black--Scholes.  For BV/digital payoffs only the $\alpha=0$ expansion bookkeeping is used.  The corrected balances $\varepsilon^*_J=\delta_t^{1/(2J+4)}$ and exponents $(J+1)/(J+2)$ are conditional on Theorem~\ref{thm:improved-coupling}(a)--(c); the $\alpha_{\mathrm{eff}}=1/2$ values remain heuristic.

Two computational routes are available for the auxiliary problems~\eqref{eq:vj-pde}: (a)~under Black--Scholes dynamics, $v_j$ is a derivative of the transition density and is evaluated by the closed-form Gaussian integrals of Appendix~\ref{app:gaussian}; (b)~in general, $v_j(T-\eta,\cdot)$ is smooth for any $\eta>0$ (parabolic smoothing of the measure datum) and can be computed by one direct integration against the Gaussian kernel, after which the standard expansion scheme of~\citep{Zhang2017} is applied on $[0,T-\eta]$ with the smooth terminal profile $v_j(T-\eta,\cdot)$.  In both cases the corrections are computed and used only in the weak/pointwise sense for $t<T$, consistent with Theorems~\ref{thm:first-corr}--\ref{thm:corrected-bound}.

\begin{figure}[htbp]
\centering
\includegraphics[width=0.85\textwidth]{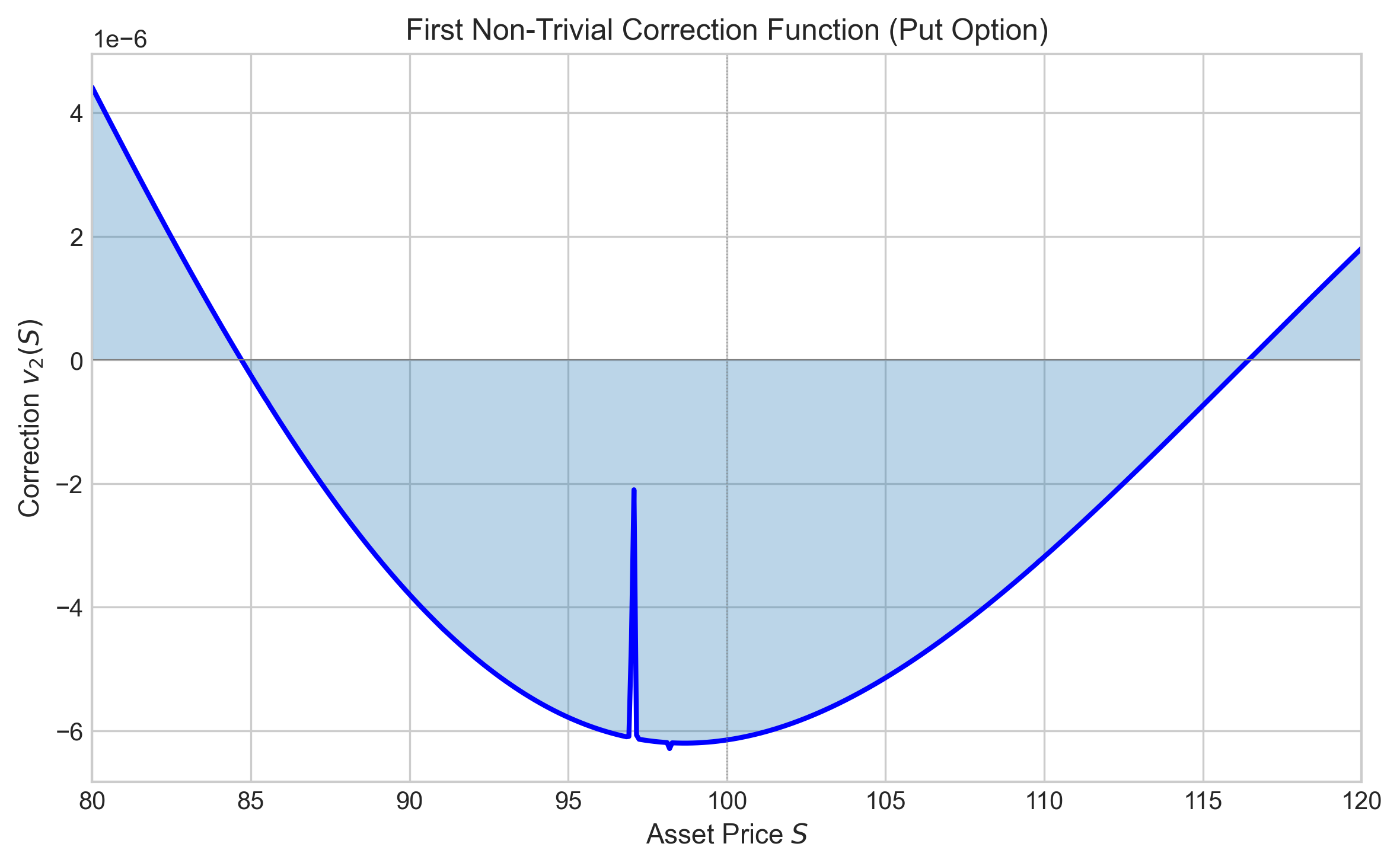}
\caption{Analytic-formula illustration of the leading correction function $v_1$ for the European put; no auxiliary-PDE solver is validated by this plot.}
\label{fig:correction-v1}
\end{figure}

The auxiliary problems above define a theoretical computational route.  Each auxiliary solve adds an expansion problem whose cost must be counted separately; it is not an $O(1)$ runtime guarantee.  The numerical tables instead use Richardson extrapolation in $\varepsilon$ and therefore do not validate an implementation of $v_j$ or the corrected total-error rates.  Figure~\ref{fig:correction-v1} is an analytic-formula illustration, not a numerical validation of an auxiliary-PDE solver.

\section{Extension to the Second Expansion Scheme}\label{sec:extension}

The first expansion scheme, described in Section~\ref{sec:prelim} and originating from Chapter~2 of~\citep{Zhang2017}, performs all Taylor expansions at a \emph{variable} expansion center~$\bar{x}$; after the expansion is computed, the center is set equal to the spatial variable via $\bar{x}=x$ (Step~4 of the algorithm in Section~\ref{sec:intro-expansion}).  The resulting Taylor coefficients are therefore \emph{functions} of the variable center, and the diagonal evaluation step adds both algebraic complexity and computational cost.

The \emph{second expansion scheme}, introduced in Chapter~3 of~\citep{Zhang2017}, eliminates this overhead by replacing the variable center~$\bar{x}$ with the \emph{fixed initial condition}~$x_0$, which is simply a constant.  This single modification has far-reaching consequences: all Taylor coefficients need only be evaluated once at the fixed point~$x_0$; the diagonal evaluation step (Step~4) is no longer required; and the resulting algorithm is substantially simpler, faster, and easier to implement.  The mollification construction itself is independent of the expansion center.  Its terminal-step error concentration, however, is proved here only in the driver-free case $f\equiv0$: a general driver adds a source integral and a Picard defect to every one-step map.  Under that restriction the fixed center produces an $n$-independent terminal-representation term (Theorem~\ref{thm:E2-second}), so the result is confined to moderate horizons and moderate $|x-x_0|$; digital-data convergence remains open (Remark~\ref{rem:E2-second-digital}).

\subsection{The Constant-Center Taylor Expansion}

In the first expansion scheme, the Taylor expansion operator $\mathcal{T}_m^{\bar{x}}$ acts on the terminal condition and source terms about the variable center~$\bar{x}$:
\begin{equation}
\mathcal{T}_m^{\bar{x}} g(x) = \sum_{|\beta| \leq m} \frac{1}{\beta!}\,\partial_x^\beta g(\bar{x})\,(x - \bar{x})^\beta.
\end{equation}
After expansion and solution, the center is set to~$\bar{x}=x$ (Step~4), producing the final approximation.

In the second expansion scheme, the Taylor expansion operator uses the \emph{fixed} initial condition $x_0 \in \R^d$ as the expansion center:
\begin{equation}
\mathcal{T}_m^{x_0} g(x) = \sum_{|\beta| \leq m} \frac{1}{\beta!}\,\partial_x^\beta g(x_0)\,(x - x_0)^\beta.
\end{equation}
The critical distinction is that the derivatives $\partial_x^\beta g(x_0)$ are now \emph{constants} (evaluated at the fixed point~$x_0$) rather than functions of a variable center.  This simplification has a cascade of beneficial effects throughout the algorithm.

\begin{remark}[Relation to the thesis formulation]\label{rem:thesis-formulation}
In the notation of \citep{Zhang2017} (Chapter~3, equation~(3.3.5)), the approximate solution on each time sub-interval is given by
\[
  u_{k,m,n}^{\bar{x}}(t,x)
  = \int_{\R^d} \Gamma_0^{\bar{x}}(t,x;\,t_{i+1},y)\;
    \mathcal{T}_m^{x_0}\bigl[u_{k,m,n}^{y}(t_{i+1},y)\bigr]\,\diff y,
\]
where $\Gamma_0^{\bar{x}}$ is the Gaussian kernel associated with the frozen-coefficient operator $\mathcal{A}_0^{\bar{x}}$.  The PDE coefficient expansion (Step~2) may still use a variable or frozen center, but the Taylor expansion of the terminal condition and source terms (Step~3) is performed at the \emph{constant}~$x_0$, which is the defining feature of the second scheme.  We caution the reader that the notation of \citep[eq.~(3.3.5)]{Zhang2017} has been transcribed into the conventions of this paper: the kernel center $\bar{x}$ and the Taylor center $x_0$ displayed above correspond to the frozen-coefficient center and the fixed initial point, respectively, of the thesis, and a direct symbol-by-symbol comparison with the thesis should be read with this notational correspondence in mind.
\end{remark}

\subsection{Algorithmic Simplifications}

The use of a constant expansion center~$x_0$ leads to three principal simplifications compared with the first scheme:

\begin{enumerate}[label=(\roman*), leftmargin=3em, itemsep=4pt]
\item \textbf{Elimination of Step~4.}
In the first scheme, the approximate solution $u_{l,m}^{(k),\bar{x}}(t,x)$ is computed as a function of both~$x$ and the expansion center~$\bar{x}$, and the final approximation is obtained by setting $\bar{x} = x$.  In the second scheme, the expansion center is always~$x_0$ and does not vary; hence no diagonal evaluation is needed.  This removes an entire layer of algebraic manipulation.

\item \textbf{Constant Taylor coefficients.}
Since $\partial_x^\beta \psi(x_0)$ and $\partial_x^\beta f^{(k)}(t, x_0, \cdot)$ are constants (for fixed~$x_0$, $t$, and iterate~$k$), the Gaussian semigroup applied to the polynomial $\mathcal{T}_m^{x_0}\psi$ produces a polynomial in~$x$ whose coefficients are explicit constants involving the Gaussian moments and the fixed derivatives.  No re-expansion or re-evaluation as a function of the center is required.

\item \textbf{Simpler polynomial moments.}
The $n$-th order polynomial moment of the Gaussian kernel $\Gamma_0^{\bar{x}}$ centered at a variable point~$\bar{x}$ is a polynomial in both $x$ and $\bar{x}$.  When $\bar{x}$ is replaced by the constant~$x_0$, this moment becomes a polynomial in~$x$ alone, and many cross-terms vanish.  The number of distinct terms that must be tracked is consequently reduced, leading to faster evaluation.
\end{enumerate}

These simplifications make the second scheme particularly attractive for practical computation: for the same expansion orders~$(l,m)$ and time-step number~$n$, the second scheme typically requires significantly less CPU time than the first, as verified in the numerical experiments of~\citep{Zhang2017} (Section~3.5).

\subsection{Mollification for the Second Scheme}

The mollification regularization operates on the terminal condition only and is independent of the choice of expansion center.  We define $\psi_\varepsilon = \psi * \rho_\varepsilon$ exactly as in Section~\ref{sec:mollification}.

\begin{proposition}[Compatibility with the constant-center expansion]
All properties established in Section~\ref{sec:mollification}---smoothness, boundedness, derivative estimates, and approximation order estimates---hold for the mollified terminal condition $\psi_\varepsilon$ in the second expansion scheme.  In particular, the derivative estimates
\begin{equation}
  \|\partial^\beta \psi_\varepsilon\|_\infty
  \leq C\,\varepsilon^{-(|\beta|-\alpha)},
  \qquad |\beta| \geq 1,
\end{equation}
remain valid, and the Taylor expansion $\mathcal{T}_m^{x_0}\psi_\varepsilon$ is well-defined for every $m \geq 0$ and every choice of initial condition~$x_0$.
\end{proposition}

\begin{proof}
The mollification acts on the spatial variable~$x$ at the fixed terminal time~$T$ and depends neither on the expansion center~$\bar{x}$ of the first scheme nor on the initial condition~$x_0$ of the second scheme.  All estimates in Section~\ref{sec:mollification} are stated in terms of global norms $\|\partial^\beta \psi_\varepsilon\|_\infty$ and therefore apply verbatim regardless of the expansion center.
\end{proof}

\subsection{Error Analysis for the Second Scheme}

\begin{theorem}[Regularization error---second scheme]
Under the same assumptions as Theorem~\ref{thm:bsde-stability}, the regularization error for the second expansion scheme satisfies the same bound:
\begin{equation}
  \E\!\left[\sup_{t}|Y_t-Y_{\varepsilon,t}|^2\right]
  +\E\!\left[\int_0^T|Z_t-Z_{\varepsilon,t}|^2\,\diff t\right]
  \le C\|\psi-\psi_\varepsilon\|_\infty^2.
\end{equation}
\end{theorem}

\begin{proof}
The regularization error depends only on $\|\psi - \psi_\varepsilon\|_\infty$ through BSDE stability and is independent of the expansion procedure.  The proof is identical to that of Theorem~\ref{thm:bsde-stability}.
\end{proof}

\begin{theorem}[Expansion error---second scheme, driver-free case]\label{thm:E2-second}
Let the assumptions of Theorem~\ref{thm:E2} hold (in particular $0\le l\le1$ and $m>2l$), and assume $f\equiv0$, with the coefficient expansion performed at the frozen per-step center as in the first scheme and the terminal datum expanded at the fixed center $x_0$ (cf.\ Remark~\ref{rem:thesis-formulation}).  Then the expansion error of the second scheme at $(t,x)$ satisfies
\begin{equation}\label{eq:E2-second}
\begin{aligned}
  E_2^{(2)}(t,x)
  &\;\le\;
  E_{\mathrm{cent}}(t,x)
  \;+\;C\,\delta_t^{\lceil(l+1)/2\rceil}
  \;+\;C\,\delta_t^{l+1}\,\varepsilon^{-(2l+2-\alpha)}, \\
  &\qquad \delta_t=(T-t)/n,
\end{aligned}
\end{equation}
where the \emph{centering error}
\begin{equation}\label{eq:Ecent}
  E_{\mathrm{cent}}(t,x)
  :=\bigl|\E\bigl[\bigl(\psi_\varepsilon-\mathcal{T}_m^{x_0}\psi_\varepsilon\bigr)(X_T)
     \,\big|\,X_t=x\bigr]\bigr|
\end{equation}
is the entire Taylor-truncation contribution, concentrated in the terminal step; in particular it does not depend on $n$ and does not vanish as $n\to\infty$.  The last two terms are the coefficient-expansion part, identical to the first scheme (Appendix~\ref{app:exp-error}, Steps~4--5, including the terminal-step bookkeeping~\eqref{eq:app-terminal-step}).  The centering error admits the following bounds:
\begin{enumerate}[label=(\roman*)]
\item \emph{general data:} with
  $Q_{m+1}:=\max_{|\beta|=m+1}\|\partial^\beta\psi_\varepsilon\|_\infty
   \le C\,\varepsilon^{-(m+1-\alpha)}$,
  \[
    E_{\mathrm{cent}}(t,x)
    \le C_m\,Q_{m+1}\,\bigl(|x-x_0|^{m+1}+(T-t)^{(m+1)/2}\bigr);
  \]
\item \emph{piecewise affine (kink) data:} if $\psi$ is piecewise
  affine with kinks $\{K_i\}$ and slope jumps $\{\Delta_i\}$, $m\ge1$,
  and $|x_0-K_i|>\varepsilon$ for all $i$, then
  \[
    \begin{aligned}
    E_{\mathrm{cent}}(t,x)
    &\le \sum_i|\Delta_i|\,\E\bigl[|X_T-K_i|\,\big|\,X_t=x\bigr] + C\varepsilon \\
    &\le \sum_i|\Delta_i|\,\bigl(|x-K_i|+C\sqrt{T-t}\,\bigr)+C\varepsilon,
    \end{aligned}
  \]
  which is $\varepsilon$-independent at leading order but does not
  vanish with $n$;
\item \emph{digital data:} for $\psi=\mathds{1}_{\{x\ge K\}}$ and
  $|x_0-K|>\varepsilon$,
  \[
    E_{\mathrm{cent}}(t,x)
    \le \Prob\bigl(X_T\ge K\,\big|\,X_t=x\bigr)
        +O\bigl(\varepsilon/\sqrt{T-t}\,\bigr)=O(1),
  \]
  while the bound of part~(i) carries $Q_{m+1}=O(\varepsilon^{-(m+1)})$
  and diverges under every coupling $\varepsilon^*(n)\to0$; hence no
  convergence guarantee is available for digital data
  (Remark~\ref{rem:E2-second-digital}).
\end{enumerate}
The $l=0$ case (which contains the flagship $(l,m)=(0,2)$) does not
invoke the local-to-global gain: the terminal defect is transported by
the exact semigroup and the coefficient-consistency terms accumulate
additively (Step~2 of the proof), so the $l=0$ bound is unconditional
apart from the constant bookkeeping.  The case $l\ge1$ inherits the
caveat of Remark~\ref{rem:telescoping-status} via Step~3 of the proof.
The centering floor \eqref{eq:Ecent} is independent of this
distinction.
\end{theorem}

\begin{proof}
\emph{Step 1: polynomial preservation.}
$\mathcal{T}_m^{x_0}$ is the identity on the space $\Pi_m$ of
polynomials of degree $\le m$ in $x$, and the frozen-coefficient
Gaussian semigroup maps $\Pi_m$ into $\Pi_m$: for
$p(y)=\sum_{|\beta|\le m}c_\beta\,(y-x_0)^\beta$,
\[
  \mathcal{P}_0^{\bar x}(t_j,t_{j+1})\,p\,(x)
  =\sum_{|\beta|\le m}c_\beta\,
   \E\Bigl[\bigl((x-x_0)+\mu_0\,\delta_t
   +\sigma_0\sqrt{\delta_t}\,Z\bigr)^\beta\Bigr],
\]
and the binomial expansion with Gaussian moments shows that this is a
polynomial of degree $\le|\beta|$ in $x$.  Hence, at the frozen level
($l=0$), every iterate of the scheme lies in $\Pi_m$ and the
truncation $\mathcal{T}_m^{x_0}$ acts trivially on it at every
interior step.

\emph{Step 2: error recursion at $l=0$; terminal-step concentration.}
Write $\bar u^{(2)}_j$ for the scheme's iterate at step $j$ and
$e_j:=u_\varepsilon(t_j,\cdot)-\bar u^{(2)}_j$.  With $P_j$ the exact
one-step semigroup and
$\widetilde P_j:=\mathcal{P}_0^{\bar x}(t_j,t_{j+1})$ the frozen one,
the scheme at $l=0$ reads
$\bar u^{(2)}_j=\widetilde P_j[\mathcal{T}_m^{x_0}\bar u^{(2)}_{j+1}]
 =\widetilde P_j[\bar u^{(2)}_{j+1}]$
(the truncation acts trivially by Step~1), while
$u_\varepsilon(t_j,\cdot)=P_j[u_\varepsilon(t_{j+1},\cdot)]$.  Hence
\[
  \begin{aligned}
  e_j
  &=P_j[e_{j+1}]+(P_j-\widetilde P_j)\bigl[\bar u^{(2)}_{j+1}\bigr],
  \qquad
  e_{n-1}
  =P_{n-1}\bigl[\psi_\varepsilon-\mathcal{T}_m^{x_0}\psi_\varepsilon\bigr] \\
  &\qquad
  +(P_{n-1}-\widetilde P_{n-1})\bigl[\mathcal{T}_m^{x_0}\psi_\varepsilon\bigr].
  \end{aligned}
\]
Iterating, the terminal defect (the first term) is transported to $(t,x)$ by the exact
semigroup and yields \eqref{eq:Ecent}: the entire Taylor-truncation
error of the scheme is concentrated in the terminal step, is
independent of $n$, and does not vanish as $n\to\infty$.  The
coefficient-consistency terms
$(P_j-\widetilde P_j)[\bar u^{(2)}_{j+1}]$ (including the second
terminal term above) are bounded by the
$K_1\delta_t$ bookkeeping of Appendix~\ref{app:exp-error}
(Steps~2 and~4), whose constants involve derivatives of the
polynomial iterates, bounded inductively by the interior-regularity
constants of $u_\varepsilon(t_{j+1},\cdot)$ ($\varepsilon$-independent
for $j\le n-2$) and by $\varepsilon^{-(2-\alpha)}$ on the terminal
step, giving \eqref{eq:app-terminal-step}.  Additive summation as in Step~5 of
Appendix~\ref{app:exp-error} produces the last two terms of
\eqref{eq:E2-second}; this propagation uses no local-to-global gain.
For $\alpha=0$ the additive sum acquires an additional $1+\log n$
factor that \eqref{eq:E2-second} suppresses; the convergence use of
the theorem is for $\alpha>0$, and no convergence guarantee for
digital data is claimed (Remark~\ref{rem:E2-second-digital}).

\emph{Step 3: the case $l\ge1$.}
The perturbation operators $\mathcal{A}_i^{\bar x}$ have polynomial
coefficients of degree $i$, so the iterates leave $\Pi_m$ within each
step and are re-truncated by $\mathcal{T}_m^{x_0}$; the truncation of
the degree-$>m$ components of the perturbation output produces
additional per-step terms.  Their coefficients carry the perturbation
scales $\delta_t^{i/2}$ and ($\varepsilon$-independent, by interior
regularity, at interior steps) derivatives of the iterates at $x_0$,
but summing them without loss of a power of $\delta_t$ requires the
same local-to-global cancellation already flagged as heuristic in
Remark~\ref{rem:telescoping-status}.  The $l\ge1$ statement therefore
inherits that heuristic status.

\emph{Step 4: bounds for $E_{\mathrm{cent}}$.}
(i) Taylor's formula with remainder gives
$|\psi_\varepsilon(y)-\mathcal{T}_m^{x_0}\psi_\varepsilon(y)|
 \le Q_{m+1}\,|y-x_0|^{m+1}/(m+1)!$,
and, since $\mu,\sigma$ are bounded,
$\E\bigl[|X_T-x_0|^{m+1}\,\big|\,X_t=x\bigr]
 \le C_m\bigl(|x-x_0|^{m+1}+(T-t)^{(m+1)/2}\bigr)$;
combining the two and using
$Q_{m+1}\le C\,\varepsilon^{-(m+1-\alpha)}$ from
\eqref{eq:app-sharp-deriv} (the crude bound for $\alpha=0$) gives the
claim.

(ii) Since $|x_0-K_i|>\varepsilon$ for all $i$, the derivatives of
$\psi_\varepsilon$ of order $\ge2$ vanish at $x_0$ (for a compactly
supported kernel; exponentially small otherwise), so for every
$m\ge1$,
$\mathcal{T}_m^{x_0}\psi_\varepsilon(y)
 =\psi_\varepsilon(x_0)+\psi_\varepsilon'(x_0)(y-x_0)$
is the tangent line $L$ of the affine piece containing $x_0$.  The
difference $\psi-L$ vanishes on that piece, is continuous, and its
slope changes by $\Delta_i$ at $K_i$; hence
$|\psi(y)-L(y)|\le\sum_i|\Delta_i|\,|y-K_i|$ for all $y$.
Moreover $\|\psi_\varepsilon-\psi\|_\infty\le C\varepsilon$ (the
Lipschitz mollification bound) and
$|\mathcal{T}_m^{x_0}\psi_\varepsilon-L|=O(\varepsilon)$, so the
remainder $r:=\psi_\varepsilon-\mathcal{T}_m^{x_0}\psi_\varepsilon$
satisfies $|r(y)|\le\sum_i|\Delta_i|\,|y-K_i|+C\varepsilon$.
Taking expectations and using
$\E[|X_T-K_i|\,|\,X_t=x]\le|x-K_i|+C\sqrt{T-t}$ gives the claim.

(iii) For $\psi=\mathds{1}_{\{x\ge K\}}$ and $|x_0-K|>\varepsilon$,
$\mathcal{T}_m^{x_0}\psi_\varepsilon$ is the constant
$\psi_\varepsilon(x_0)$ (up to exponentially small kernel tails);
the remainder satisfies
$|r(y)|\le\mathds{1}_{\{y\text{ on the far side of }K\}}
 +2\,\mathds{1}_{\{|y-K|\le\varepsilon\}}+O(\varepsilon)$;
taking expectations gives
$E_{\mathrm{cent}}(t,x)
 \le\Prob^{t,x}(X_T\text{ beyond }K)+O\bigl(\varepsilon/\sqrt{T-t}\bigr)
 =O(1)$,
while part~(i) with $Q_{m+1}=O(\varepsilon^{-(m+1)})$ diverges as
$\varepsilon\to0$.
\end{proof}

\begin{remark}[Digital data under the second scheme: downgraded to
heuristic]\label{rem:E2-second-digital}
For digital terminal data the only rigorous bound on the centering
error is the $O(1)$ bound of Theorem~\ref{thm:E2-second}(iii), while
the derivative-based bound of part~(i) carries $\varepsilon^{-(m+1)}$
and diverges under every coupling $\varepsilon^*(n)\to0$.  Whether
the second scheme converges for digital data---e.g.\ through
cancellations between the fixed-center polynomial and the transported
kernels across steps---is not established by the present analysis.
All convergence statements of this paper for the second scheme applied
to digital (non-Lipschitz) data are therefore downgraded to heuristic
status.
\end{remark}

\begin{corollary}[Convergence behavior of the second scheme]
Assume $f\equiv0$ as in Theorem~\ref{thm:E2-second}.  For piecewise affine (kink) data, the $n$-dependent part of
$E_2^{(2)}$ is
$O\bigl(\delta_t^{\lceil(l+1)/2\rceil}+\delta_t^{l+1}\varepsilon^{-(2l+2-\alpha)}\bigr)$,
to which the balanced couplings of Table~\ref{tab:optimal-coupling}
apply unchanged (subject to the same $m\le 4l+2$ proviso as in
Theorems~\ref{thm:opt-bv} and~\ref{thm:opt-weak}: for $m>4l+2$ the
terminal-step term dominates and the rebalanced couplings of those
theorems apply); but the total error additionally contains the
centering floor
$E_{\mathrm{cent}}
 =O\bigl(\sum_i|\Delta_i|\bigl(|x-K_i|+\sqrt{T-t}\bigr)+\varepsilon\bigr)$,
which does not vanish as $n\to\infty$.  The rates of
Table~\ref{tab:optimal-coupling} therefore describe the second scheme
only in the regime where this floor is negligible (short horizons, or
$x_0$ separated from the kinks by many diffusion standard
deviations); the earlier claim that the second scheme ``achieves the
same convergence order in $n$ as the first'' is superseded by
Theorem~\ref{thm:E2-second}.  For digital data no rate statement is
available (Remark~\ref{rem:E2-second-digital}).
\end{corollary}

\subsection{Asymptotic Corrections for the Second Scheme}

In the driver-free setting of Theorem~\ref{thm:E2-second}, the weak/pointwise expansion $u_\varepsilon = u + \sum_{j=1}^{J} \varepsilon^{2j} v_j + O(\varepsilon^{2(J+1)})$ ($t<T$) of Section~\ref{sec:asymptotic} carries over to the second scheme.  Each correction function $v_j$ satisfies the same auxiliary PDE as in Section~\ref{sec:asymptotic}:
\begin{equation}
  (\partial_t + \mathcal{A})\,v_j = 0, \qquad v_j(T,\cdot) = \mu_j,
\end{equation}
with the Dirac-type distributional datum $\mu_j$ of~\eqref{eq:vj-pde}, and can be solved within the second expansion framework by applying $\mathcal{T}_m^{x_0}$ to the smooth profile $v_j(T-\eta,\cdot)$, $\eta>0$ (cf.\ Section~\ref{subsec:practical}).  The correction identities of Section~\ref{sec:asymptotic} extend with the same weak/pointwise interpretation and the formal-rate qualifications of Theorem~\ref{thm:improved-coupling}, subject to the centering error of Theorem~\ref{thm:E2-second}: applying $\mathcal{T}_m^{x_0}$ to $v_j(T-\eta,\cdot)$ incurs the general-data bound of part~(i) with $Q_{m+1}=\max_{|\beta|=m+1}\|\partial^\beta v_j(T-\eta,\cdot)\|_\infty$, which grows as $\eta\downarrow0$, so the corrected second scheme is likewise confined to moderate horizons and neighborhoods of $x_0$, and to smooth or kink data.

\subsection{Computational Comparison}

Table~\ref{tab:scheme-comparison} contrasts the two schemes along the ingredients that enter the error analysis.

\begin{table}[htbp]
\centering
\caption{Structural comparison of the first and second expansion schemes.}
\label{tab:scheme-comparison}
\resizebox{\textwidth}{!}{\begin{tabular}{>{\raggedright\arraybackslash}p{3.2cm}>{\raggedright\arraybackslash}p{4.0cm}>{\raggedright\arraybackslash}p{7.6cm}}
\toprule
Feature & First scheme & Second scheme\\
\midrule
Expansion center & Variable $\bar{x}$ & Fixed $x_0$ (initial condition)\\
Taylor coefficients & Functions of $\bar{x}$ & Constants\\
Step~4 (diagonal evaluation) & Required & Not needed\\
Polynomial moments & Functions of $(x,\bar{x})$ & Functions of $x$ only\\
Mollification compatible & Yes & Yes\\
Asymptotic corrections & Yes & Yes\\
Balanced coupling $\varepsilon^*(n)$ & Same formula & Same formula for the $n$-dependent part (centering floor additional, Theorem~\ref{thm:E2-second})\\
\bottomrule
\end{tabular}}
\end{table}

The computational advantage of the second scheme is most pronounced in higher-dimensional problems ($d \geq 2$), where the number of multi-index terms in the Taylor expansion grows rapidly with~$m$ and the elimination of the variable center~$\bar{x}$ removes a significant fraction of the algebraic complexity.  We note that the choice of the initial condition~$x_0$ as the expansion center introduces a genuine dependence of the approximation quality on the distance $|x - x_0|$: since the center is not reset between time steps, the Taylor-truncation error is concentrated in the terminal step and is independent of~$n$ (Theorem~\ref{thm:E2-second}); it is of order $\varepsilon^{-(m+1-\alpha)}\bigl(|x-x_0|^{m+1}+(T-t)^{(m+1)/2}\bigr)$ for general data, an $\varepsilon$-independent centering floor $O\bigl(\sum_i|\Delta_i|(|x-K_i|+\sqrt{T-t})\bigr)$ for kink data, and uncontrolled ($O(1)$) for digital data.  The second scheme is therefore most accurate in a moderate neighborhood of~$x_0$, at moderate horizons~$T$, and for smooth or kink data.

To quantify the computational savings, a systematic wall-clock
comparison of the two schemes requires a shared, optimized
implementation and is left to future work.  At the level of operation
counts, however, the saving is structural: in the first scheme every
Taylor coefficient $\partial^\beta\psi_\varepsilon$ and every expanded
PDE coefficient must be re-evaluated at each variable center
$\bar{x}$ and at each time step, whereas in the second scheme all such
coefficients are computed \emph{once} at the fixed center~$x_0$.  The
expected speedup therefore grows with the number of multi-indices
$\binom{m+d}{m}$ and is most pronounced for larger~$m$ and~$d$;
in one dimension with moderate~$m$ the saving is a constant-factor
reduction of the symbolic workload, while in higher dimensions the
elimination of the variable-center (diagonal evaluation) step removes
a significant fraction of the algebraic complexity.  These are
structural estimates based on operation counts, not measured timings.

\begin{remark}[When to prefer each scheme]
The first scheme achieves optimal local accuracy by centering the expansion at the evaluation point~$x$ itself, and is therefore preferable when high pointwise accuracy over a wide spatial domain is required.  The second scheme sacrifices some spatial adaptivity in exchange for significantly reduced computation, and is preferable when speed is the primary concern---for example, in real-time pricing applications or when the solution is needed only in a moderate neighborhood of~$x_0$; it carries an $n$-independent centering error floor (Theorem~\ref{thm:E2-second}), so it should not be used at long horizons or for digital data without further analysis.  Since both schemes share the same mollification and correction infrastructure, the choice between them is otherwise purely an efficiency--accuracy trade-off.
\end{remark}

\section{Gaussian Moment Compression for the Constant-Center Scheme}\label{sec:moment-compression}

The error analysis of Sections~\ref{sec:error}--\ref{sec:extension} is dimension-free, but the \emph{computational} viability of the expansion schemes in several spatial dimensions hinges on a question that the preceding sections deliberately set aside: \emph{how expensive is the Gaussian conditional-expectation step, and can its cost be controlled a priori?}  This section answers that question for the constant-center (second) scheme of Section~\ref{sec:extension}, for which the answer is cleanest and the practical gains largest.  We isolate a \emph{moment layer} of the algorithm---the evaluation of Gaussian moments of polynomials---from the \emph{driver layer} (Picard iteration, source-term evaluation), and show that the moment layer admits a complete, certifiable compression theory: whitening eliminates the factorial pairing explosion of Wick--Isserlis moment evaluation; a Stein-recursion moment table is built in time linear in the number of retained moments and is immune to covariance degeneracy; and an a priori weight formula turns moment truncation into a knapsack problem with a \emph{proved} error budget.  The driver layer is entirely unaffected, so the compression applies equally to the drivers independent of $(y,z)$ treated in this paper and to the general drivers $f(t,x,y,z)$ treated by the second expansion scheme in \citep[Chapter~3]{Zhang2017}.  The numerical statements in this section are measured on a five-dimensional OU testbed with genuinely $(y,z)$-dependent drivers, and on ten- and twenty-dimensional extensions, all reported in the companion paper (Part~II) \citep{Zhang2026II}; the implementation is a self-contained script of moderate size that reproduces every reported table in under a minute on a single CPU core.

\emph{Section notation.}  Throughout this section, $\alpha, \beta \in \N_0^d$ and $\gamma \in \N_0^r$ are \emph{multi-indices} (state and noise respectively); they are unrelated to the H\"older exponent~$\alpha$ and the coupling exponent~$\gamma$ of Section~\ref{sec:error}.  As everywhere in the paper, $d$ is the state dimension and $r \le d$ the number of Brownian factors.  The conditional-expectation matrix is denoted $\mathsf{L}$ (the letter $\mathcal{L}$ is reserved for the linearized generator of Section~\ref{sec:asymptotic}); the retained moment set is denoted $\Lambda(\tau)$, not to be confused with the Euler matrix $A$ of one step~\eqref{eq:euler-step-mc}.

\subsection{The Two Cost Layers of Gaussian Conditional Expectation}

After the frozen-coefficient expansion of the forward generator (Step~2 of the algorithm in Section~\ref{sec:intro-expansion}), every backward step of the time-discretized scheme reduces to Gaussian conditional expectations of truncated polynomials.  Concretely, on a sub-interval of length $\delta_t$ with Euler step
\begin{equation}\label{eq:euler-step-mc}
X_{i+1} \;=\; A\,X_i \;+\; c\,\delta_t \;+\; B\,\xi_i,
\qquad \xi_i \sim \mathcal{N}(0, I_r),
\end{equation}
with $A = I - \Theta\,\delta_t \in \R^{d\times d}$, $B = \sigma\sqrt{\delta_t} \in \R^{d\times r}$, and $c = \Theta\mu$ for the Ornstein--Uhlenbeck testbed reported in Part~II \citep{Zhang2026II} (the letter $c$ is used to avoid confusion with the driver $f$), the scheme must evaluate, for a polynomial jet $P(\Delta) = \sum_{|\beta|\le m} c_\beta\,\Delta^\beta$ representing the value function at the next time slice (expanded about the conditional mean path),
\begin{equation}\label{eq:cond-exp-mc}
(\mathsf{L} P)(\Delta)
\;=\;
\E\!\left[\,P\!\left(A\Delta + B\xi\right)\,\right]
\;=\;
\sum_{|\beta|\le m} c_\beta \sum_{\alpha \le \beta}
\binom{\beta}{\alpha}\,(A\Delta)^{\alpha}\;\E\!\left[(B\xi)^{\beta-\alpha}\right],
\end{equation}
where $\binom{\beta}{\alpha} = \prod_{k=1}^d \binom{\beta_k}{\alpha_k}$.  The cost of~\eqref{eq:cond-exp-mc} decomposes into two conceptually distinct layers:

\begin{enumerate}[label=(\roman*), leftmargin=3em, itemsep=4pt]
\item \textbf{The monomial layer.}  The jet $P$ has $\binom{m+d}{d}$ coefficients; the output is again a polynomial of degree $\le m$ in $\Delta \in \R^d$.  This count is the dimension of the approximation space and is therefore an information-theoretic lower bound: no basis-preserving transformation can reduce it.  It can only be circumvented structurally---by lowering the effective dimension of the terminal data (ridge or low-ANOVA structure), by truncating anisotropically, or by reducing $m$ while accepting a larger truncation remainder (Section~\ref{subsec:equivalence}).
\item \textbf{The moment layer.}  Evaluating each output coefficient requires the Gaussian moments $\E[(B\xi)^{\gamma}]$, $|\gamma|\le m$.  A naive symbolic treatment evaluates these through the Wick--Isserlis pairing formula: $\E[\eta^{2k}] = (2k-1)!!$ for a unit variance scalar, and mixed moments $\E[\xi^\gamma]$ with a non-diagonal covariance sum over all $(|\gamma|-1)!!$ perfect matchings---a factorial explosion that historically dominates the bookkeeping of multidimensional expansion implementations.
\end{enumerate}

The content of Sections~\ref{subsec:whitening}--\ref{subsec:budget} is that the moment layer---the layer that actually explodes in practice---can be \emph{completely} tamed: its cost becomes linear in the number of retained moments, and the retained set itself can be pruned with a certified error budget.  The monomial layer cannot be compressed in general, and we make no such claim; the dimensions that occur in derivatives practice nevertheless keep the monomial layer small, as discussed in Part~II \citep{Zhang2026II}.

\subsection{Whitening and Moment Factorization}\label{subsec:whitening}

The first compression device is a linear change of the noise coordinates that diagonalizes the covariance.

\begin{lemma}[Whitening factorizes the moment table]\label{lem:whitening}
Let $\xi \sim \mathcal{N}(0, \Sigma)$ on $\R^d$ with $\Sigma = LL^{\top}$ of rank $r \le d$ (a factor $L \in \R^{d\times r}$ always exists, e.g.\ from the eigendecomposition, and no invertibility of $\Sigma$ is required).  Write $\xi \stackrel{\mathrm{law}}{=} L\eta$ with $\eta \sim \mathcal{N}(0, I_r)$.  Then for every multi-index $\gamma \in \N_0^r$,
\begin{equation}\label{eq:white-moments}
\E[\eta^{\gamma}] \;=\; \prod_{k=1}^{r} \mu(\gamma_k),
\qquad
\mu(g) = \begin{cases} (g-1)!!, & g \text{ even},\\ 0, & g \text{ odd},\end{cases}
\end{equation}
and the moments of the original vector are recovered by finite linear combination:
$\E[\xi^{\beta}] = \sum_{|\gamma|\le|\beta|} q_{\beta\gamma}\,\E[\eta^{\gamma}]$,
where $q_{\beta\gamma}$ is the coefficient of $\eta^\gamma$ in $\prod_{k=1}^d (L_{k\cdot}\eta)^{\beta_k}$.
\end{lemma}

\begin{proof}
For jointly Gaussian variables, vanishing correlations imply independence; the coordinates of $\eta$ are therefore independent scalar $\mathcal{N}(0,1)$ variables, and $\E[\eta^\gamma] = \prod_k \E[\eta_k^{\gamma_k}]$ factors.  The one-dimensional moments are $\E[\eta_k^{g}] = (g-1)!!$ for even $g$ and $0$ for odd $g$ (the odd case is the symmetry of the centered Gaussian law---the parity bookkeeping already visible in the expansion hierarchy).  The recovery formula is the multinomial expansion of $\xi = L\eta$.
\end{proof}

Two consequences are immediate.  First, in whitened coordinates \emph{at least half of the moment table vanishes}: all multi-indices with at least one odd coordinate have zero moment, which is $1 - 2^{-r}$ of the $2^r$ parity classes (three quarters for the $r=2$ testbed).  The surviving entries are pure double factorials---no pairing enumeration is ever performed.  Second, the number of distinct moments that can ever arise is $\binom{m+r}{r}$, a polynomial in $m$ for fixed $r$; the factorial $(m-1)!!$ growth of the pairing formula is a bookkeeping artifact, not an intrinsic cost.

\begin{remark}[Optimality of whitening among linear maps]
Whitening is optimal among \emph{linear} reparametrizations in the following sense: for jointly Gaussian $\xi$, a linear map $\xi \mapsto M\xi$ produces independent components if and only if $M\Sigma M^{\top}$ is diagonal; any non-diagonal choice leaves nonzero mixed moments $\E[(M\xi)^{\gamma}]$ with odd total structure, inflating the moment table that subsequent stages must store.  Nonlinear reparametrizations (e.g.\ transport to a product measure) can reduce the table further but destroy the polynomial structure that the expansion method relies on; we do not pursue them.
\end{remark}

\subsection{A Stein-Recursion Moment Table}\label{subsec:stein}

Whitening is what makes the \emph{storage} small; the following recursion is what makes the \emph{computation} linear, and it applies directly in the original (possibly correlated or degenerate) coordinates.

\begin{lemma}[Stein recursion for Gaussian moments]\label{lem:stein}
Let $\xi \sim \mathcal{N}(0,\Sigma)$ on $\R^d$, and write $M(\beta) = \E[\xi^\beta]$ for $\beta \in \N_0^d$.  Then $M(0) = 1$, $M(e_i) = 0$, and
\begin{equation}\label{eq:stein-recursion}
M(\beta + e_i) \;=\; \sum_{k=1}^{d} \Sigma_{ik}\,\beta_k\, M(\beta - e_k),
\qquad \beta \in \N_0^d,
\end{equation}
with the convention $M(\beta - e_k) = 0$ if $\beta_k = 0$.  In particular the full table $\{M(\beta): |\beta|\le m\}$ is generated by $O(1)$ work per entry, requires no factorization or inversion of $\Sigma$, and remains valid when $\Sigma$ is singular.
\end{lemma}

\begin{proof}
Stein's identity $\E[\xi_i\,g(\xi)] = \sum_k \Sigma_{ik}\,\E[\partial_k g(\xi)]$ for the centered Gaussian, applied with $g(\xi) = \xi^\beta$, gives~\eqref{eq:stein-recursion}.  The recursion closes within the table because the right-hand side involves only strictly smaller multi-indices and multiplies only by entries of $\Sigma$---never by entries of an inverse or a Cholesky factor---so rank deficiency is harmless.
\end{proof}

\begin{remark}[Two independent kernels agree to machine precision]
We implemented the moment layer twice: once via~\eqref{eq:stein-recursion} in the physical coordinates (with $\Sigma = BB^{\top}$ of rank $r=2$ inside a $d=5$ state space---a genuinely singular covariance), and once via whitening plus~\eqref{eq:white-moments} in the two-dimensional noise coordinates.  the two conditional-expectation matrices $\mathsf{L}$ of~\eqref{eq:cond-exp-mc} agree to $1.7\times10^{-18}$ in the maximum entrywise norm on the five-dimensional testbed reported in the companion paper (Part~II) \citep{Zhang2026II}, and the moment table contains $28$ entries at order $m=6$---to be compared with the $\binom{6+5}{5} = 462$ monomial coefficients of the jet, and with the $5!! = 15$ pairings that a naive Wick expansion would enumerate \emph{per} sixth-order moment (a count that would already reach $11!! = 10{,}395$ for a single mixed moment of total order twelve).
\end{remark}

\begin{remark}[Decoupling from the state dimension]
When the forward noise is driven by $r < d$ Brownian factors, parametrizing the step~\eqref{eq:euler-step-mc} directly by $\xi \in \R^r$ places the moment table in $\N_0^r$: its size $\binom{m+r}{r}$ is then \emph{independent of the state dimension $d$}.  In our implementation the moment table stays at $15$ entries ($m=4$, $r=2$) while the state monomial basis grows from $126$ ($d=5$) through $1{,}001$ ($d=10$) to $10{,}626$ ($d=20$); the matrix $\mathsf{L}$ at $d=20$, $m=4$ ($10{,}626\times10{,}626$ sparse-structured) is assembled in $1.6$ seconds on one CPU core.  This is the precise sense in which the moment layer is \emph{solved}: its cost scales with the number of noise factors and the expansion order, not with the state dimension.
\end{remark}

\subsection{Jet-Conditioned Error Budgets and Knapsack Truncation}\label{subsec:budget}

Whitening and the Stein recursion make the moment table cheap to \emph{compute}; the next device makes it cheap to \emph{store and apply}, by discarding provably negligible moments before the computation starts.  The key structural fact is that the step matrix $B = \sigma\sqrt{\delta_t}$ has columns of norm $s_k := \|B_{\cdot k}\| = O(\sqrt{\delta_t})$, so moments of high total order are \emph{a priori} small, with a weight that factorizes across noise coordinates.

\begin{definition}[Separable a priori weight]
For $\gamma \in \N_0^r$ define
\begin{equation}
w(\gamma) \;=\; \prod_{k=1}^{r} \frac{s_k^{\gamma_k}\,(\gamma_k-1)!!}{\gamma_k!},
\qquad (\gamma_k-1)!! := 1 \text{ for } \gamma_k \in \{0,1\}.
\end{equation}
For a threshold $\tau > 0$ define the saturated weight
\begin{equation}
\widetilde{w}(\gamma) \;=\; \max\big\{\,w(\gamma + j e_k)\;:\; k = 1,\dots,r,\ j \in \{0,1,2\},\ \gamma + j e_k \in \mathcal{T}_m\,\big\},
\end{equation}
with $\mathcal{T}_m = \{\gamma : |\gamma| \le m\}$ the finite moment table, and let $\Lambda(\tau) = \{\gamma \in \mathcal{T}_m : \widetilde{w}(\gamma) \ge \tau\}$: the superlevel set of $w$, enlarged by every index lying at most two coordinate-steps \emph{below} a significant one, all within the finite table $\mathcal{T}_m$.  The two-step margin keeps the pruned table stable under the degree-raising operations of the driver layer: in the Picard update, products such as $y^2$ and $|z|^2$ raise intermediate total orders by up to two before re-truncation to $\mathcal{T}_m$, and the margin guarantees that no moment through which such an evaluation routes into a retained one is discarded.
\end{definition}

Since $w$ is a coordinate-wise product, its superlevel sets and their closures are enumerated by a knapsack-type recursion over the partial sums $\sum_k \gamma_k$ in $O(|\Lambda(\tau)|)$ time, independently of the solution $u$---the truncation is decided \emph{before} any PDE quantity is computed.

\begin{proposition}[Jet-conditioned truncation budgets]\label{prop:budget}
Let $u(\xi)=\sum_{|\gamma|\le m}u_\gamma\xi^\gamma$ be a finite jet in whitened noise coordinates, $\mu(\gamma)=\E[\xi^\gamma]$, and $\Lambda$ a retained subset.  Then
\begin{equation}\label{eq:sharp-bound}
\begin{aligned}
\Bigl|\E[u(\xi)]-\E[T_\Lambda u(\xi)]\Bigr|
&\le \underbrace{\sum_{\gamma\notin\Lambda}|u_\gamma|\,\mu(\gamma)}_{\text{\rm sharp bound}}\\
&\le \underbrace{D_w\sum_{\gamma\notin\Lambda}w(\gamma)\mu(\gamma)}_{\text{\rm jet-conditioned weighted bound}},
\qquad D_w:=\max_\gamma\frac{|u_\gamma|}{w(\gamma)} .
\end{aligned}
\end{equation}
The index set can be selected before the solve from model-derived weights, but $D_w$ and both numerical budgets are available only after the current jet is known.  The two bounds serve different purposes and are not interchangeable in a recursion.
\begin{enumerate}[label=(\roman*)]
\item \textbf{Conditional rate matching.}  If $D_w=O(1)$ along a refinement path, then $\tau(\delta_t)\asymp\delta_t^{q+1}$ yields an $O(\delta_t^{q+1})$ weighted bound.  No model-only rate follows without this jet condition.
\item \textbf{Stable-mode accumulation.}  For an $L^1$-nonexpansive linear step, the scalar sharp budget can be accumulated linearly.  If each weighted budget uses its own $D_w$, accumulate the resulting scalar weighted budget only after establishing a suitable expectation-norm stability estimate for the step.  If one seeks the weighted form at every terminal layer, an additional bound $\max_jD_{w,j}\le D_{w,*}$ is required.  The $L^1$-nonexpansive absolute-value operator cannot be composed directly with the absolute functional of~\eqref{eq:sharp-bound} while retaining the expectation form.
\item \textbf{Driver amplification.}  A Lipschitz driver adds the usual Gr\"onwall factor to an established stability recursion.  The proposition does not infer such a recursion from the one-step inequalities alone.
\end{enumerate}
\end{proposition}

\begin{proof}
The two one-step inequalities follow from the triangle inequality and $|u_\gamma|\le D_ww(\gamma)$.  For a linear $L^1$-nonexpansive step, the sum of scalar sharp budgets follows by taking expectations of absolute values before applying the triangle inequality.  The same scalar weighted budget can be accumulated under the stated stability hypothesis.  To recover the weighted expression at a terminal layer, bound every intermediate jet ratio by $D_{w,*}$ and apply the triangle inequality inside the expectation.  These are distinct operations; neither the scalar identity nor the terminal weighted form is claimed without its corresponding hypothesis.  Driver amplification then follows from a separate Lipschitz stability recursion.
\end{proof}

\begin{remark}[Measured one-step budgets and an absolute-value chain]
One-step realized errors measured on the testbed of Part~II \citep{Zhang2026II} lie alongside the sharp and jet-conditioned weighted budgets; the ordering actual $\le$ sharp $\le$ weighted is a direct consequence of~\eqref{eq:sharp-bound}.  The reported forty-step chain uses absolute values at every layer, so it is consistent with the stable-mode sharp-budget accumulation of Proposition~\ref{prop:budget}.  It is not evidence for a universal terminal weighted budget with layer-dependent $D_{w,j}$.  The measured ratios are numerical observations on the testbed, not a general tightness theorem.
\end{remark}

\subsection{Basis Invariance, Centering, and Terminal Representation}\label{subsec:equivalence}

For $f\equiv0$, Theorem~\ref{thm:E2-second} identifies an $n$-independent terminal representation error.  Two different statements must be separated: algebraic invariance of a truncated conditional-expectation matrix under a change of polynomial coordinates, and dependence of the approximation defect on the terminal representative.

\begin{remark}[Basis invariance does not imply centering independence]\label{rem:centering-irrelevance}
Once a polynomial representative $P\in\Pi_m$ has been fixed, re-expansion about a new center is a change of basis.  The moving-center tensor recursion and the constant-center matrix $\mathsf L$ therefore represent the same linear map; their measured agreement to $1.3\times10^{-15}$ checks this algebraic identity.  It says nothing about the defect of $P$ as a representative of a non-polynomial terminal function $g$.

For a one-step counterexample, take $f\equiv0$, $m=1$, $g(y)=e^y$, and $X=x+\sqrt\delta\,\xi$ with $\xi\sim N(0,1)$.  Taylor expansion at the fixed center $0$ gives $\E[T_1^0g(X)]=1+x$, whereas $\E[g(X)]=e^{x+\delta/2}$.  The error tends to $1+x-e^x$, which is nonzero for $x\ne0$.  Centering at $x$ instead gives $\E[T_1^xg(X)]=e^x$ and error $O(\delta)$.  Thus the terminal-representation floor depends on the center even though the matrix representation of a fixed polynomial operator does not.
\end{remark}

There are three distinct remedies.  Diagonal Taylor centering places the expansion at the evaluation point.  The Hermite/$L^2(\nu_T)$ construction of Section~\ref{subsec:terminal-hermite} keeps the same $\Pi_m$ but replaces the Taylor representative by the orthogonal projection.  Finally, $p$-refinement enlarges $\Pi_m$.  The measurements reported in Part~II \citep{Zhang2026II} isolate the second mechanism at fixed $m$; they must not be cited as evidence for centering independence.

For diagnostics within the constant-center architecture, write
\begin{equation}\label{eq:error-decomp-mc}
\mathrm{err}(n,m)\approx C\delta_t+\mathrm{rem}_y(m)+T\mathrm{rem}_z(m),
\end{equation}
where the last two terms are the terminal remainders of the value and its gradient.  Part~II \citep{Zhang2026II} reports the measured decay of a diagonal jet-norm surrogate together with exact $L^2$ remainders.  These measurements concern terminal representation; they are not a proof that $p$-refinement is the only remedy.

\subsection{An \texorpdfstring{$L^2$}{L2}-Optimal Terminal Layer: Hermite Projection}\label{subsec:terminal-hermite}

The error decomposition~\eqref{eq:error-decomp-mc} identifies the terminal Taylor remainder as the binding constraint at fine meshes and at high dimension.  Within the truncated polynomial space $\Pi_m$ itself, however, the Taylor polynomial $\mathcal{T}_m g$ is a distinguished but not optimal element: under the terminal Gaussian law of the forward state, which we denote $\nu_T$ (a measure on $\R^d$, unrelated to the multi-index $\gamma$ of this section), the best $L^2(\nu_T)$ approximation of $g$ from $\Pi_m$ is the orthogonal projection onto $\Pi_m$, which we denote $\mathcal{T}_m^{\perp} g$ and whose remainder is minimal by Pythagoras.  We stress the scope of this statement: $L^2$-optimality is claimed for the terminal representative only---the nonlinear propagation layer does not preserve $L^2$ optimality end-to-end, and the driver-free control reported in Part~II \citep{Zhang2026II} shows that the two orderings can disagree at coarse meshes.  The same projection is optimal for the derivative jet one order down: differentiation commutes with the Gaussian projection, $\partial_{\xi_i}\mathcal{T}_m^{\perp} g = \mathcal{T}_{m-1}^{\perp}\,\partial_{\xi_i} g$, so the $z$-remainder of~\eqref{eq:error-decomp-mc} is simultaneously minimized in $\Pi_{m-1}$---which is where the binding term of the decomposition lives.  Writing the terminal law in whitened noise coordinates, $x = c + B\xi$ with $\xi \sim N(0, I_r)$, the projection is the truncated tensor-Hermite series
\begin{equation}\label{eq:hermite-terminal}
(\mathcal{T}_m^{\perp} g)(c + B\xi) \;=\; \sum_{|k| \le m} c_k\,\mathrm{He}_k(\xi),
\qquad
c_k \;=\; \frac{\E\!\left[g(c + B\xi)\,\mathrm{He}_k(\xi)\right]}{\prod_i k_i!},
\end{equation}
where $\mathrm{He}_k(\xi) = \prod_i \mathrm{He}_{k_i}(\xi_i)$ are the tensor probabilists' Hermite polynomials, $\E[\mathrm{He}_k\,\mathrm{He}_j] = k!\,\delta_{kj}$.  When $\nu_T$ is degenerate ($r < d$) the projection is unique on the support of $\nu_T$ only; we take the canonical jet representative through the whitening lift $\xi = B^{+}(x - c)$, and the propagation layer evaluates the jet only on reachable states, so the choice is well-defined.  (In the implementation $\nu_T$ is the rank-$r$ principal factorization of the exact terminal covariance, carrying over $99.97\%$ of its trace in all testbeds reported in Part~II \citep{Zhang2026II}---consistent with the rank-$r$ architecture of this section.)  We re-express the result in the monomial basis about the same center $c$ (a linear change of basis) so that the propagation layer---the matrix $\mathsf{L}$, the Picard jet iteration, and the knapsack budget of Proposition~\ref{prop:budget}---is used verbatim.

\begin{remark}[The projection coefficients are closed-form or once-paid]
For trigonometric terminal data the coefficients in~\eqref{eq:hermite-terminal} are available in closed form: the Hermite generating function gives
$\E\!\left[\cos(\theta + \beta\cdot\xi)\,\mathrm{He}_k(\xi)\right] = e^{-|\beta|^2/2}\,\beta^k\,\cos^{(|k|)}(\theta)$,
where $\beta = B^\top a$, $\theta = a\cdot c$, and $\cos^{(j)}$ denotes the $j$-th derivative (cycling through $\cos, -\sin, -\cos, \sin$); the $\sin$ formula is analogous.  For a general smooth payoff the projection integral is $r$-dimensional and is paid \emph{once} per pricing: a $60\times 60$ Gauss--Hermite quadrature in the noise space reproduces every closed-form coefficient of our testbeds to $3.3\times10^{-16}$.  Equivalently, Gaussian integration by parts expresses the mixed moments through factorial multiples of entries of the same moment table as Section~\ref{subsec:stein}: $\E[\xi^\alpha\,\mathrm{He}_k(\xi)] = \E[\partial^k \xi^\alpha] = \frac{\alpha!}{(\alpha-k)!}\,M(\alpha-k)$ (zero unless $k \le \alpha$), involving moments of order $\le |\alpha|$ only; the projection is thus again a linear functional of one universal, precomputable object, and the architecture's division of labor is unchanged---only the terminal layer ever integrates, exactly once.  The once-paid cost scales as $60^r$ nodes, exponentially in the number of noise factors; the method is therefore positioned, exactly as the rest of this section, at small $r$ and large $d$---the regime of derivatives practice---while the trigonometric payoffs of our testbeds need no quadrature at all.
\end{remark}

\begin{remark}[A change of basis, not a new scheme]
Truncated to total degree $m$, the tensor-Hermite basis and the monomial basis span the same space $\Pi_m$; the Hermite basis diagonalizes the Gaussian semigroup (the $\mathrm{He}_k$ are eigenfunctions), while the monomial basis is the one in which the driver algebra and the certified budget are formulated.  Parity survives the change of basis (Hermite polynomials have definite parity), and the whitening factorization of Lemma~\ref{lem:whitening} is the moment-side shadow of the passage to tensor Hermite polynomials in whitened coordinates.  By the same change-of-basis argument as Remark~\ref{rem:centering-irrelevance}, the truncated backward operator is unchanged; only the representative of the terminal layer within $\Pi_m$ moves to the $L^2$-optimal one.
\end{remark}

\section{Conclusion}\label{sec:conclusion}

We have studied mollification as a way to apply the analytical expansion of~\citep{Zhang2017} to non-smooth terminal data.  The error decomposition separates regularization bias from expansion error and tracks the derivative blow-up explicitly.  For symmetric kernels, the leading weak bias cancels under the payoff, density, and driver restrictions of Proposition~\ref{prop:weak-E1}.

The balanced exponents $2/3$ and the corrected family $r_J$ are conditional results.  They require the uniform local-to-global Taylor gain of Assumption~\ref{ass:telescoping-gain}; the corrected rates additionally require control of the actual weak or pointwise Picard residual, the auxiliary expansion errors, the coefficient layer, and the terminal step.  Without Assumption~\ref{ass:telescoping-gain}, Proposition~\ref{prop:additive-E2} gives an additive bound: it yields no decay at $q=m-2l=2$ and a positive polynomial rate for H\"older data when $q>2$.  The $(0,4)$ weak row also requires full-bound re-optimization against the terminal-step term $\delta_t\varepsilon^{-1}$, giving the conditional pair $\delta_t^{1/3}$ and $2/3$ rather than the naive two-term balance.

For the fixed-center scheme, the proved terminal concentration is restricted to $f\equiv0$.  The $n$-independent term is a terminal-representation error, not a center-independent floor: the explicit exponential example shows dependence on the Taylor center.  Basis invariance, diagonal centering, Hermite/$L^2$ terminal projection, and $p$-refinement are distinct statements.  Gaussian moment compression reduces the moment-table cost but leaves the monomial-jet count; its truncation certificate is jet-conditioned and its multistep use requires the stability hypotheses of Proposition~\ref{prop:budget}.

The numerical evidence is reported separately in the companion paper (Part~II) \citep{Zhang2026II} and is deliberately narrower than the theoretical targets.  Direct-integration tables there measure mollification bias and Richardson cancellation along prescribed bandwidth schedules, and author-reported $n$-step outputs contain spatial-grid and transport errors; neither validates the conditional exponents of the present paper.  Establishing the terminal-layer propagation estimate, implementing the auxiliary correction problems, and extending the fixed-center analysis to general drivers remain open on the theory side.

American-style and convertible-bond problems provide a possible application through their continuation-value operators, but reflected dynamics create new kinks at each exercise layer.  A corresponding error analysis must treat repeated mollification and free-boundary propagation; no runtime or convergence claim is made here.

\section*{Acknowledgments}

This paper is based in part on Chapter~3 of the author's doctoral dissertation \citep{Zhang2017} at Boston University, Questrom School of Business.

\section*{Declaration of competing interest}
The author declares that he has no known competing financial interests or personal relationships that could have appeared to influence the work reported in this paper.

\section*{Funding}
This work received no specific grant from any funding agency in the public, commercial, or not-for-profit sectors.

\section*{Data availability}
No data was used for the research described in the article.

\section*{Declaration of generative AI and AI-assisted technologies in the writing process}
During the preparation of this work the author used generative AI-assisted tools for language polishing and grammar correction. After using these tools, the author reviewed and edited the content as needed and takes full responsibility for the content of the publication.

\appendix

\section{Properties of the Semigroup Operator}
\label{app:semigroup}

This section collects the definitions and key properties of the
frozen-coefficient semigroup operator
$\mathcal{P}_0^{\bar{x}}(t,T)$ that is used throughout the paper.

\subsection{Gaussian transition density}

Fix a reference point $\bar{x}\in\mathbb{R}^d$ and freeze the
coefficients of the forward SDE at $\bar{x}$, obtaining the linear
SDE
\[
  \mathrm{d}\widetilde{X}_s
  = \mu(\bar{x})\,\mathrm{d} s
  + \sigma(\bar{x})\,\mathrm{d} W_s,
  \qquad s\in[t,T],
  \quad \widetilde{X}_t = x.
\]
The solution is Gaussian:
$\widetilde{X}_T \sim \mathcal{N}\bigl(m(t,x;T),\,\Sigma(t;T)\bigr)$
with
\begin{align}
  m(t,x;T)
  &\;:=\; x + \mu(\bar{x})\,(T-t),
  \\[4pt]
  \Sigma(t;T)
  &\;:=\; (T-t)\,\sigma(\bar{x})\,\sigma(\bar{x})^\top
  \;=:\; (T-t)\,a(\bar{x}),
\end{align}
where $a(\bar{x}) = \sigma(\bar{x})\sigma(\bar{x})^\top$ is the
diffusion matrix.  The transition density is
\begin{equation}
  \Gamma_0^{\bar{x}}(t,x;T,y)
  \;=\;
  \frac{1}{(2\pi)^{d/2}\,\det(\Sigma)^{1/2}}
  \exp\!\Bigl(
    -\tfrac{1}{2}(y-m)^\top \Sigma^{-1}(y-m)
  \Bigr),
\end{equation}
where we abbreviate $m = m(t,x;T)$ and $\Sigma = \Sigma(t;T)$.

\subsection{Semigroup operator}

The semigroup operator acts on a Borel measurable function
$g:\mathbb{R}^d\to\mathbb{R}$ by
\begin{equation}
  \bigl[\mathcal{P}_0^{\bar{x}}(t,T)\,g\bigr](x)
  \;:=\;
  \int_{\mathbb{R}^d}
    g(y)\,\Gamma_0^{\bar{x}}(t,x;T,y)\,\mathrm{d} y
  \;=\;
  \mathbb{E}\bigl[g(\widetilde{X}_T)\mid \widetilde{X}_t = x\bigr],
\end{equation}
provided the integral is well defined (e.g.\ when $g$ is bounded
or has at most polynomial growth).

\subsection{Semigroup action on polynomials}

\begin{proposition}
Let $\bar{x}\in\mathbb{R}^d$ be fixed, and let $p(x)$ be a
polynomial of degree $N$ in $x\in\mathbb{R}^d$.  Then
$\mathcal{P}_0^{\bar{x}}(t,T)\,p$ is again a polynomial of
degree $N$ in $x$, whose coefficients are polynomials in $(T-t)$.
\end{proposition}

\begin{proof}
It suffices to prove the claim for monomials
$p(y) = (y-\bar{x})^{\alpha}$, where $\alpha\in\mathbb{N}_0^d$
is a multi-index, since any polynomial of degree $N$ is a finite
linear combination of such monomials with $|\alpha|\le N$.

We compute
\begin{align*}
  \bigl[\mathcal{P}_0^{\bar{x}}(t,T)(y-\bar{x})^\alpha\bigr](x)
  &= \mathbb{E}\bigl[
    (\widetilde{X}_T - \bar{x})^\alpha
    \,\big|\, \widetilde{X}_t = x
  \bigr] \\
  &= \mathbb{E}\bigl[
    \bigl(m(t,x;T) - \bar{x}
          + \sigma(\bar{x})\sqrt{T-t}\,Z\bigr)^\alpha
  \bigr],
\end{align*}
where $Z\sim\mathcal{N}(0,I_d)$ is independent of $x$.
Since $m(t,x;T) - \bar{x} = (x-\bar{x}) + \mu(\bar{x})(T-t)$,
we can write
\[
  \widetilde{X}_T - \bar{x}
  = (x - \bar{x}) + \mu(\bar{x})(T-t)
  + \sigma(\bar{x})\sqrt{T-t}\,Z.
\]
Expanding by the multinomial theorem,
\begin{align}
  &(\widetilde{X}_T - \bar{x})^\alpha
  \notag\\
  &\quad=
  \sum_{\beta+\gamma+\delta=\alpha}
  \frac{\alpha!}{\beta!\,\gamma!\,\delta!}\,
  (x-\bar{x})^\beta\,
  \bigl(\mu(\bar{x})(T-t)\bigr)^\gamma\,
  \bigl(\sigma(\bar{x})\sqrt{T-t}\,Z\bigr)^\delta.
  \label{eq:app-multinomial}
\end{align}
Taking expectations and using the fact that the moments
$\mathbb{E}[Z^\delta]$ are explicit constants (zero for odd
$|\delta|$, products of double factorials for even $|\delta|$),
each term becomes a monomial in $(x-\bar{x})$ of degree $|\beta|$
multiplied by a power of $(T-t)$.  Since
$|\beta|\le|\alpha|=N$ in every term, the result is a polynomial
of degree at most $N$ in $x$ with coefficients that are
polynomials in $(T-t)$.  To see that the degree is exactly $N$,
note that the term with $\gamma=\delta=0$ and $\beta=\alpha$
yields $(x-\bar{x})^\alpha$, which has degree $N$.
\end{proof}

\subsection{Explicit formula for monomials}

We provide the explicit formula for small $|\alpha|$ in one
dimension ($d=1$), writing $\Delta t = T-t$, $\mu_0=\mu(\bar{x})$,
$\sigma_0=\sigma(\bar{x})$:
\begin{align}
  \mathcal{P}_0^{\bar{x}}(t,T)\,\mathbf{1}
  &= 1,
  \\
  \mathcal{P}_0^{\bar{x}}(t,T)\,(y-\bar{x})
  &= (x-\bar{x}) + \mu_0\,\Delta t,
  \\
  \mathcal{P}_0^{\bar{x}}(t,T)\,(y-\bar{x})^2
  &= \bigl[(x-\bar{x})+\mu_0\,\Delta t\bigr]^2
     + \sigma_0^2\,\Delta t,
  \\
  \mathcal{P}_0^{\bar{x}}(t,T)\,(y-\bar{x})^3
  &= \bigl[(x-\bar{x})+\mu_0\,\Delta t\bigr]^3
     + 3\sigma_0^2\,\Delta t\,
       \bigl[(x-\bar{x})+\mu_0\,\Delta t\bigr],
  \\
  \mathcal{P}_0^{\bar{x}}(t,T)\,(y-\bar{x})^4
  &= \bigl[(x-\bar{x})+\mu_0\,\Delta t\bigr]^4
     + 6\sigma_0^2\,\Delta t\,
       \bigl[(x-\bar{x})+\mu_0\,\Delta t\bigr]^2 \notag\\
     &\quad + 3\sigma_0^4\,(\Delta t)^2.
\end{align}
These follow immediately from the well-known moment formulas for
the Gaussian distribution:
$\mathbb{E}[Z]=0$, $\mathbb{E}[Z^2]=1$, $\mathbb{E}[Z^3]=0$,
$\mathbb{E}[Z^4]=3$, applied term-by-term to the multinomial
expansion \eqref{eq:app-multinomial}.  In the multivariate case,
the analogous formulas involve the entries of the matrix
$a(\bar{x})$ through the Isserlis--Wick theorem (moment-cumulant
relations for Gaussian vectors).

\section{Proof of Regularization Error Estimates}
\label{app:reg-error}

We provide the complete proof of the supremum-norm regularization
error bound $E_1(\varepsilon) \le C\,\varepsilon^\alpha$ for
$\psi\in C^{0,\alpha}(\mathbb{R}^d)$, which is the pointwise form of
Corollaries~\ref{cor:E1-lip}--\ref{cor:E1-holder},
together with a weak-error refinement for symmetric kernels
(Proposition~\ref{prop:weak-E1}) that underlies
Theorems~\ref{thm:opt-weak} and~\ref{thm:main-weak}.

\medskip
\noindent\textbf{Theorem}
(Sup-norm regularization error bound; pointwise form of
Corollaries~\ref{cor:E1-lip}--\ref{cor:E1-holder}).\quad
\textit{Let Assumptions~\ref{ass:forward}--\ref{ass:driver} hold, and
suppose $\psi\in C^{0,\alpha}(\mathbb{R}^d)$ for some
$\alpha\in(0,1]$.  Let $u$ and $u_\varepsilon$ denote the viscosity
solutions of the associated PDE with terminal conditions $\psi$ and
$\psi_\varepsilon = \psi*\rho_\varepsilon$, respectively.  Then}
\begin{equation}\label{eq:app-reg-error}
  \sup_{(t,x)\in[0,T]\times\mathbb{R}^d}
  |u(t,x) - u_\varepsilon(t,x)|
  \;\le\; C\,\varepsilon^\alpha,
\end{equation}
\textit{where $C$ depends on $T$, the coefficient bounds, the
Lipschitz constant of $f$, and the H\"older seminorm
$[\psi]_{C^{0,\alpha}}$, but not on $\varepsilon$.}

\begin{proof}
The proof proceeds in three stages: (i) we establish the
approximation of $\psi$ by $\psi_\varepsilon$ in the supremum norm;
(ii) we invoke the stability of viscosity solutions for the
semilinear PDE; (iii) we combine these estimates.

\medskip
\noindent\textbf{Step 1: Pointwise estimate of
$\psi - \psi_\varepsilon$.}

By definition,
\[
  \psi_\varepsilon(x) - \psi(x)
  = \int_{\mathbb{R}^d}
    \bigl[\psi(x-y) - \psi(x)\bigr]\,\rho_\varepsilon(y)
    \,\mathrm{d} y.
\]
Since $\psi\in C^{0,\alpha}$, we have
$|\psi(x-y)-\psi(x)|\le [\psi]_{C^{0,\alpha}}\,|y|^\alpha$.
Moreover, $\operatorname{supp}(\rho_\varepsilon)\subset B(0,\varepsilon)$ and
$\int\rho_\varepsilon = 1$.  Therefore,
\begin{align}
  |\psi_\varepsilon(x) - \psi(x)|
  &\le \int_{|y|\le\varepsilon}
    [\psi]_{C^{0,\alpha}}\,|y|^\alpha\,\rho_\varepsilon(y)
    \,\mathrm{d} y
  \notag\\
  &\le [\psi]_{C^{0,\alpha}}\,\varepsilon^\alpha
       \int_{\mathbb{R}^d} \rho_\varepsilon(y)\,\mathrm{d} y
  \notag\\
  &= [\psi]_{C^{0,\alpha}}\,\varepsilon^\alpha.
  \label{eq:app-step1}
\end{align}
Hence
$\|\psi_\varepsilon - \psi\|_\infty
\le [\psi]_{C^{0,\alpha}}\,\varepsilon^\alpha$.

\medskip
\noindent\textbf{Step 2: PDE stability via comparison principle.}

The functions $u$ and $u_\varepsilon$ are respectively viscosity
solutions of
\begin{equation}
  \partial_t u + \mathcal{A} u + f(t,x,u,\nabla u\,\sigma) = 0,
  \qquad u(T,\cdot) = \psi,
\end{equation}
\begin{equation}
  \partial_t u_\varepsilon + \mathcal{A} u_\varepsilon
  + f(t,x,u_\varepsilon,\nabla u_\varepsilon\,\sigma) = 0,
  \qquad u_\varepsilon(T,\cdot) = \psi_\varepsilon,
\end{equation}
where $\mathcal{A}$ is the second-order operator associated with
the forward SDE.  Define $w := u_\varepsilon - u$.  Then $w$
satisfies, in the viscosity sense,
\begin{equation}
  \partial_t w + \mathcal{A} w
  + f(t,x,u_\varepsilon,\nabla u_\varepsilon\,\sigma)
  - f(t,x,u,\nabla u\,\sigma) = 0,
  \qquad w(T,\cdot) = \psi_\varepsilon - \psi.
\end{equation}
By the Lipschitz assumption on the driver $f$ (Assumption~\ref{ass:driver}),
there exist bounded measurable functions $b_0(t,x)$ and
$c_0(t,x)$ (obtained by the mean value theorem applied to $f$ in
the $(y,z)$ arguments) with $|b_0|\le L_f$ and $|c_0|\le L_f$,
such that
\[
  f(t,x,u_\varepsilon,\nabla u_\varepsilon\,\sigma)
  - f(t,x,u,\nabla u\,\sigma)
  = c_0(t,x)\,w + b_0(t,x)\cdot(\nabla w\,\sigma).
\]
Hence $w$ satisfies the linear PDE
\begin{equation}\label{eq:app-w-linear}
  \partial_t w + \mathcal{A} w + c_0\,w
  + b_0\cdot(\nabla w\,\sigma) = 0,
  \qquad w(T,\cdot) = \psi_\varepsilon - \psi.
\end{equation}

\medskip
\noindent\textbf{Step 2a: Maximum principle.}
We apply the parabolic maximum principle to \eqref{eq:app-w-linear}.
Define the operator
\[
  \widetilde{\mathcal{A}}
  := \mathcal{A} + b_0\cdot(\nabla(\cdot)\,\sigma),
\]
which is again a second-order operator with bounded coefficients
(since $b_0$, $\sigma$, and the original coefficients are bounded).
Then \eqref{eq:app-w-linear} reads
$\partial_t w + \widetilde{\mathcal{A}} w + c_0\, w = 0$.

Consider the function $\phi(t) := e^{L_f(T-t)}$.  We claim that
$|w(t,x)|\le \phi(t)\,\|\psi_\varepsilon - \psi\|_\infty$ for all
$(t,x)\in[0,T]\times\mathbb{R}^d$.  To verify this, set
$\overline{w}(t,x) := \phi(t)\,\|\psi_\varepsilon-\psi\|_\infty$.
Then
\[
  \partial_t \overline{w}
  + \widetilde{\mathcal{A}}\,\overline{w}
  + c_0\,\overline{w}
  = -L_f\,\overline{w} + 0 + c_0\,\overline{w}
  = (c_0 - L_f)\,\overline{w}
  \le 0,
\]
since $c_0\le L_f$ and $\overline{w}\ge 0$.  Similarly,
$-\overline{w}$ is a subsolution.  At $t=T$,
$|w(T,x)| = |\psi_\varepsilon(x)-\psi(x)|
\le \|\psi_\varepsilon-\psi\|_\infty = \overline{w}(T,x)$.
By the comparison theorem for viscosity solutions of parabolic
equations with bounded coefficients
(see, e.g.\ \citep{CrandallIshiiLions1992}),
\begin{equation}\label{eq:app-comparison}
  |w(t,x)| \le \overline{w}(t,x)
  = e^{L_f(T-t)}\,\|\psi_\varepsilon-\psi\|_\infty
\end{equation}
for all $(t,x)\in[0,T]\times\mathbb{R}^d$.

\medskip
\noindent\textbf{Step 2b: Alternative via BSDE stability.}
An equivalent route proceeds through the BSDE representation.
Let $(Y,Z)$ and $(Y_\varepsilon,Z_\varepsilon)$ be the solutions of
the BSDE with terminal conditions $\psi(X_T)$ and
$\psi_\varepsilon(X_T)$, respectively.  Define
$\Delta Y := Y_\varepsilon - Y$ and $\Delta Z := Z_\varepsilon - Z$.
By the standard BSDE stability estimate
\citep[Proposition~2.2]{ElKarouiPengQuenez1997}, for any
$\beta>0$ satisfying $\beta > 2L_f + L_f^2$,
\begin{align}
  &\mathbb{E}\Bigl[\sup_{t\le s\le T} e^{\beta s}|\Delta Y_s|^2
  + \int_t^T e^{\beta s}|\Delta Z_s|^2\,\mathrm{d} s\Bigr]
  \notag\\
  &\qquad\le
  C_\beta\,\mathbb{E}\bigl[
    e^{\beta T}|\psi_\varepsilon(X_T)-\psi(X_T)|^2
  \bigr]
  \;\le\;
  C_\beta\,e^{\beta T}\,
  \|\psi_\varepsilon-\psi\|_\infty^2.
  \label{eq:app-bsde-stab}
\end{align}
Taking $s=t$ and using $Y_s = u(s,X_s)$,
$Y_{\varepsilon,s}=u_\varepsilon(s,X_s)$, this yields
\[
  |u_\varepsilon(t,x) - u(t,x)|^2
  \le C_\beta\,e^{\beta(T-t)}\,
      \|\psi_\varepsilon-\psi\|_\infty^2.
\]

\medskip
\noindent\textbf{Step 3: Combination.}

Substituting \eqref{eq:app-step1} into either
\eqref{eq:app-comparison} or the consequence of
\eqref{eq:app-bsde-stab}, we obtain
\[
  \sup_{(t,x)} |u(t,x) - u_\varepsilon(t,x)|
  \le e^{L_f\,T}\,[\psi]_{C^{0,\alpha}}\,\varepsilon^\alpha
  =: C\,\varepsilon^\alpha,
\]
with $C = e^{L_f\,T}\,[\psi]_{C^{0,\alpha}}$, which is
independent of $\varepsilon$.  This completes the proof.
\end{proof}

\begin{remark}
The supremum-norm estimate \eqref{eq:app-reg-error} does
\emph{not} extend to bounded-variation terminal conditions: a step
function belongs to no H\"older class, and for
$\psi=\mathds{1}_{\{x\ge K\}}$ one has
$\|\psi_\varepsilon-\psi\|_\infty\ge 1/2$ for every
$\varepsilon>0$ (with equality for a symmetric kernel), so
no sup-norm rate is available.  The correct substitute is the
$L^2(\mathbb{P})$ route of Corollary~\ref{cor:E1-disc}, which yields
$E_1(\varepsilon)\le C\varepsilon^{1/2}$ via parabolic smoothing of
the forward density.
\end{remark}

\begin{remark}
For terminal conditions that are only bounded measurable
($\alpha = 0$), the pointwise estimate \eqref{eq:app-step1} gives
only $\|\psi_\varepsilon-\psi\|_\infty\le 2\|\psi\|_\infty$,
which does not converge.  However, the
$L^2(\mathbb{P})$ stability estimate \eqref{eq:app-bsde-stab} can
be improved using regularity of the forward process: if the
transition density $p(t,x;T,\cdot)$ of $X_T$ given $X_t=x$
belongs to $L^q$ for some $q>1$, then
\[
  \mathbb{E}\bigl[|\psi_\varepsilon(X_T)-\psi(X_T)|^2\bigr]
  \;\le\;
  \|\psi_\varepsilon - \psi\|_{2q/(q-1)}^2\,
  \|p(t,x;T,\cdot)\|_q,
\]
and $\|\psi_\varepsilon-\psi\|_p\to 0$ for any $p<\infty$ since
mollification is an approximate identity in $L^p$.  This provides
convergence in the $L^2(\mathbb{P})$ sense even for bounded
measurable $\psi$, albeit without an explicit polynomial rate.
\end{remark}

\begin{proposition}[Weak regularization error under symmetric kernels]\label{prop:weak-E1}
Let $d=1$, let $\rho$ be symmetric with finite second moment, and let $\psi$ be piecewise linear or piecewise constant with finitely many kinks or jumps, at most linear growth, and finite second moment under $X_T$.  Suppose the forward diffusion admits, for $t<T$, a transition density $\Gamma(t,x;T,\cdot)$ of class $C^2$ with locally bounded derivatives through order two.  Assume further that $f$ is independent of $(y,z)$, or $f=a(t,x)y+b(t,x)\cdot z$ with $a,b\in C_b^2$.  Then for every $t<T$ and $x$,
\begin{equation}
|u_\varepsilon(t,x)-u(t,x)|\le C(t,x)\varepsilon^2 ,
\end{equation}
where $C(t,x)$ is independent of $\varepsilon$.  The same bound holds after integration against an integrable test function.  Piecewise $C^2$ data with bounded second derivative on each segment follow by the same argument.
\end{proposition}

\begin{proof}
For $f$ independent of $(y,z)$, $w=u_\varepsilon-u$ is the terminal expectation of $\Delta_\varepsilon=\psi_\varepsilon-\psi$.  At a kink, $\Delta_\varepsilon(x)=\varepsilon G((x-K)/\varepsilon)$ up to exponentially small tails, so substituting $y=K+\varepsilon z$ gives an $O(\varepsilon^2)$ integral against the bounded density.  At a jump, write $\Delta_\varepsilon(x)=H((x-K)/\varepsilon)$.  Kernel symmetry gives $\int H=0$; Taylor expansion of the $C^2$ density therefore cancels the $O(\varepsilon)$ term and leaves $O(\varepsilon^2)$.  Linear growth supplies integrability away from the singularities.

For the linear driver, the Feynman--Kac representation uses the fixed operator $\mathcal L=\mathcal A+b\cdot(\nabla\cdot\sigma)+a$, not a mean-value operator depending on the two solutions or on $\varepsilon$.  The $C_b^2$ coefficients give the required density derivatives uniformly in $\varepsilon$, and the same argument applies.  A genuinely nonlinear driver would introduce an $\varepsilon$-dependent mean-value linearization and is not covered.
\end{proof}

\begin{remark}[Scope of the weak estimate]
Proposition~\ref{prop:weak-E1} is a pointwise ($t<T$) and weak
statement: it does not hold in the supremum norm up to $t=T$, where
$\|u_\varepsilon(T,\cdot)-u(T,\cdot)\|_\infty
 =\|\psi_\varepsilon-\psi\|_\infty$ remains of order
$\varepsilon^\alpha$ (Lipschitz $\psi$) or $1/2$ (step $\psi$).  The
extension to $d\ge 2$ holds for payoffs whose singular set is a
finite union of smooth hypersurfaces, by applying the same argument to
the marginal density of $X_T$ in the direction normal to each
hypersurface; we omit the geometric details.
\end{remark}

\section{Proof of Expansion Error with Epsilon Dependence}
\label{app:exp-error}

We provide the detailed proof of Theorem~\ref{thm:E2}, which
quantifies how the expansion error $E_2(\varepsilon,l,m,n)$ depends
on the mollification parameter $\varepsilon$.  Throughout this section, $\delta_t := (T-t)/n$, in agreement with the notation of
the main text.

\medskip
\noindent\textbf{Theorem~\ref{thm:E2} (restated)}
(Expansion error with $\varepsilon$-dependence).\quad
\textit{Let Assumptions~\ref{ass:forward}--\ref{ass:driver} hold,
let $\psi\in C^{0,\alpha}(\mathbb{R}^d)$, $0<\alpha\le 1$, and
assume $m>2l$.  Let $u_\varepsilon$ denote the PDE solution with
mollified terminal condition
$\psi_\varepsilon = \psi * \rho_\varepsilon$ and let
$\bar{u}_{\varepsilon,l,m,n}$ denote its $(l,m,n)$-order expansion
approximation.  Then, with $\delta_t=(T-t)/n$,}
\begin{equation}\label{eq:app-exp-bound}
\begin{aligned}
  \|u_\varepsilon - \bar{u}_{\varepsilon,l,m,n}\|_{L^\infty([0,T]\times\mathbb{R}^d)}
  \;\le\;&
  C\,\varepsilon^{-(m-2l-\alpha)}\,\delta_t^{(m-2l)/2}
  + C\,\delta_t^{\,\lceil(l+1)/2\rceil}\\
  & + C\,\delta_t^{\,l+1}\,\varepsilon^{-(2l+2-\alpha)},
\end{aligned}
\end{equation}
\textit{where $C$ depends on the coefficient bounds, $T$, $d$,
$l$, $m$, $[\psi]_{C^{0,\alpha}}$, and
$\|\rho\|_{W^{\max(m,2l+2),1}}$,
but not on $\varepsilon$ or $n$; for even $l$ and $\alpha<1$ the
constant of the second term may acquire the factor
$\varepsilon^{-(1-\alpha)}$, and for $\alpha=1$ a logarithmic factor
in $n$ (Step~4), neither of which affects any power rate derived
below.}
The third term of \eqref{eq:app-exp-bound} is the contribution of the
single terminal time step $[t_{n-1},T]$, on which the
interior-regularity bound degenerates (Step~4 below); it is displayed
in the form valid for $0\le l\le1$, which covers all applications
in this paper (the sharper $l$-dependent order is given in
\eqref{eq:app-terminal-step-true}); for
$l=0$ and $\varepsilon\le 1$ it dominates the second term, and for
$(l,m)=(0,2)$ it is of exactly the same order as the first term.
The second term is the coefficient-expansion layer; its exponent
$\lceil(l+1)/2\rceil$ equals $1$ for $l=0$---so the flagship
$(l,m)=(0,2)$ bounds are unchanged---and is sharp for every $l$
(Remark~\ref{rem:gbm-tightness}).
The same estimate holds for bounded measurable $\psi$ upon setting
$\alpha=0$, i.e.\ with $\varepsilon$-exponents $-(m-2l)$ and
$-(2l+2)$, using the crude derivative bound \eqref{eq:app-deriv-est}
in place of the H\"older one; this is the only expansion bound
available for BV/digital terminal data.

\begin{proof}
We adapt the error analysis of Zhang (\citeyear{Zhang2017}, Theorem~2.4.8),
carefully tracking the dependence of all constants on
$\varepsilon$ through the derivatives of $\psi_\varepsilon$.

\medskip
\noindent\textbf{Step 1: Derivative estimates for
$\psi_\varepsilon$.}

For any multi-index $\beta$ with $|\beta|\ge 1$, we have
\begin{equation}\label{eq:app-deriv-est}
  \|\partial^\beta\psi_\varepsilon\|_\infty
  = \|\psi * \partial^\beta\rho_\varepsilon\|_\infty
  \le \|\psi\|_\infty\,\|\partial^\beta\rho_\varepsilon\|_1.
\end{equation}
By the scaling property of the mollifier (Lemma~\ref{lem:scaling}),
\begin{equation}
  \|\partial^\beta\rho_\varepsilon\|_1
  = \varepsilon^{-|\beta|}\,\|\partial^\beta\rho\|_1.
\end{equation}
However, we can obtain a sharper estimate by exploiting the
H\"older regularity of $\psi$.  Since $\int\partial^\beta\rho_\varepsilon = 0$
for $|\beta|\ge 1$ (the derivative of a decaying kernel integrates to
zero), we can write
\[
  \partial^\beta\psi_\varepsilon(x)
  = \int_{\mathbb{R}^d}
    \bigl[\psi(x-y)-\psi(x)\bigr]\,
    \partial^\beta\rho_\varepsilon(y)\,\mathrm{d} y,
\]
and therefore
\begin{align}
  |\partial^\beta\psi_\varepsilon(x)|
  &\le [\psi]_{C^{0,\alpha}}
       \int_{|y|\le\varepsilon}
       |y|^\alpha\,|\partial^\beta\rho_\varepsilon(y)|\,\mathrm{d} y
  \notag\\
  &= [\psi]_{C^{0,\alpha}}\,\varepsilon^\alpha\,
     \varepsilon^{-|\beta|}\,
     \int_{\mathbb{R}^d} |z|^\alpha\,|\partial^\beta\rho(z)|\,
     \mathrm{d} z
  \notag\\
  &= C_\beta\,[\psi]_{C^{0,\alpha}}\,
     \varepsilon^{-(|\beta|-\alpha)},
  \label{eq:app-sharp-deriv}
\end{align}
where
$C_\beta := \int |z|^\alpha|\partial^\beta\rho(z)|\,\mathrm{d} z
< \infty$ since $\rho\in C_c^\infty$.  This is the key estimate
that will be propagated through the expansion error bound.
For bounded measurable $\psi$ (the case $\alpha=0$) the centering
trick is unavailable and \eqref{eq:app-deriv-est} gives the crude
bound
$\|\partial^\beta\psi_\varepsilon\|_\infty
 \le \|\psi\|_\infty\,\|\partial^\beta\rho\|_1\,\varepsilon^{-|\beta|}$;
the bookkeeping below applies verbatim with $\alpha=0$.

\medskip
\noindent\textbf{Step 2: Review of the expansion error structure.}

The $(l,m,n)$-expansion on a single time step $[t_j,t_{j+1}]$ of
length $\delta_t = (T-t)/n$ works as follows.  The PDE is
decomposed as
\[
  \partial_t u_\varepsilon + \mathcal{A}_0^{\bar{x}} u_\varepsilon
  + \sum_{i=1}^{\infty}
    \mathcal{A}_i^{\bar{x}} u_\varepsilon
  + f = 0,
\]
where $\mathcal{A}_0^{\bar{x}}$ is the frozen-coefficient operator
and $\mathcal{A}_i^{\bar{x}}$ collects terms of order $i$ in the
Taylor expansion of the coefficients (the driver $f$ is unchanged;
mollification affects only the terminal condition).  The approximation
$\bar{u}_\varepsilon = \sum_{p=0}^{l} u_{\varepsilon,p}$ is
obtained by solving the hierarchy
\[
  (\partial_t + \mathcal{A}_0^{\bar{x}})\,u_{\varepsilon,p}
  = -\sum_{i=1}^{p}\mathcal{A}_i^{\bar{x}}\,u_{\varepsilon,p-i}
    - f_{p},
\]
where the terminal condition of $u_{\varepsilon,p}$ involves the
Taylor polynomial $\mathcal{T}_m^{\bar{x}}\psi_\varepsilon$ of
degree $m$.

Following Zhang (\citeyear{Zhang2017}, Proposition~2.4.12), the single-step
expansion error on $[t_j,t_{j+1}]$ is bounded by
\begin{equation}
  |u_\varepsilon(t_j,x) - \bar{u}_{\varepsilon,l,m}(t_j,x)|
  \;\le\;
  K_1\,\delta_t^{\,\lfloor l/2\rfloor+2} + K_2\,\delta_t^{(m-2l)/2},
\end{equation}
where the local order $\lfloor l/2\rfloor+2$ of the coefficient term
refines the $l+1$ one might read off
Zhang (\citeyear{Zhang2017}, Proposition~2.4.12): the level-counting of Step~4
(the first mismatched semigroup level is $j^*(l)=\lfloor l/2\rfloor+2$,
cf.\ \eqref{eq:app-jstar}) shows that the coefficient truncation
enters one level earlier than the naive count, but \emph{two} powers
of $\delta_t$ per level rather than one, cf.\
Remark~\ref{rem:gbm-tightness}; and:
\begin{itemize}
\item $K_1$ depends on the bounds
  $\|\partial^\beta\mu\|_\infty$ and
  $\|\partial^\beta\sigma\|_\infty$ for $|\beta|\le 2m$,
  and on $\|\partial^\beta u_\varepsilon(t_{j+1},\cdot)\|_\infty$
  for $|\beta|\le 2j^*(l)-l-1$ ($\le 3$; $\le 2$ for odd $l$) at the
  leading mismatched level $j^*(l)$, cf.\ Step~4.  These solution
  derivatives are
  bounded in terms of the terminal condition derivatives by
  parabolic regularity: for $s < T$,
  \begin{equation}\label{eq:app-interior-reg}
    \|\partial_x^\beta u_\varepsilon(s,\cdot)\|_\infty
    \le C_{|\beta|}\,(T-s)^{-|\beta|/2}\,
        \|\psi_\varepsilon\|_\infty
    \wedge
    C_{|\beta|}\,\|\partial_x^\beta\psi_\varepsilon\|_\infty,
  \end{equation}
  where the second argument of the minimum is uniform in $s\le T$
  because the heat semigroup is an $L^\infty$ contraction that
  commutes with spatial derivatives; in particular
  $\|\partial_x^\beta u_\varepsilon(s,\cdot)\|_\infty
   \le C_{|\beta|}\,[\psi]_{C^{0,\alpha}}\,\varepsilon^{-(|\beta|-\alpha)}$
  for all $s\le T$.
  For the terms in $K_1$, the dominant contribution comes from
  derivatives of order up to $2j^*(l)-l-1\le 3$ at the leading
  mismatched level (Step~4), each bounded by
  \eqref{eq:app-sharp-deriv} or by \eqref{eq:app-interior-reg}.

\item $K_2$ depends on the Taylor remainder of $\psi_\varepsilon$
  of order $m$:
  \[
    |\psi_\varepsilon(y)
     - \mathcal{T}_m^{\bar{x}}\psi_\varepsilon(y)|
    \le C\,|y-\bar{x}|^m\,
        \max_{|\beta|=m}
        \|\partial^\beta\psi_\varepsilon\|_\infty.
  \]
  By \eqref{eq:app-sharp-deriv} with $|\beta|=m$,
  \begin{equation}
    \max_{|\beta|=m}
    \|\partial^\beta\psi_\varepsilon\|_\infty
    \le C_m\,[\psi]_{C^{0,\alpha}}\,\varepsilon^{-(m-\alpha)}.
  \end{equation}
\end{itemize}

\medskip
\noindent\textbf{Step 3: The Taylor-truncation term with
$\varepsilon$-dependence.}

The Taylor-truncation error is distributed across the levels
$p=0,\dots,l$ of the hierarchy.  At level $p=0$, in the single-step
analysis on $[t_j,t_{j+1}]$, the error from the Taylor truncation of
$\psi_\varepsilon$ at order $m$ is
\begin{align}
  &\bigl|\mathcal{P}_0^{\bar{x}}(t_j,t_{j+1})
    \bigl[\psi_\varepsilon
    - \mathcal{T}_m^{\bar{x}}\psi_\varepsilon\bigr](x)\bigr|
  \notag\\
  &\quad\le
  \int_{\mathbb{R}^d}
    C_m\,|y-\bar{x}|^m\,\varepsilon^{-(m-\alpha)}\,
    [\psi]_{C^{0,\alpha}}\,
    \Gamma_0^{\bar{x}}(t_j,x;t_{j+1},y)\,\mathrm{d} y
  \notag\\
  &\quad=
  C_m\,[\psi]_{C^{0,\alpha}}\,\varepsilon^{-(m-\alpha)}\,
  \mathbb{E}\bigl[|\widetilde{X}_{t_{j+1}}-\bar{x}|^m\bigr].
\end{align}
Now $\widetilde{X}_{t_{j+1}} - \bar{x}
= (x-\bar{x})+\mu_0\delta_t+\sigma_0\sqrt{\delta_t}\,Z$,
so by the multinomial expansion and Gaussian moment bounds,
\[
  \mathbb{E}\bigl[|\widetilde{X}_{t_{j+1}}-\bar{x}|^m\bigr]
  \le C_m\,\bigl(|x-\bar{x}|^m + \delta_t^{m/2}\bigr).
\]
With the standard choice $\bar{x} = x$ (expansion around the
current point), we obtain $|x-\bar{x}|=0$ and the level-$0$
contribution
\begin{equation}
  C_m\,[\psi]_{C^{0,\alpha}}\,
  \varepsilon^{-(m-\alpha)}\,\delta_t^{m/2}.
\end{equation}

At level $p\ge 1$, the $p$ perturbation operators
$\mathcal{A}_i^{\bar{x}}$ applied upstream are each of second order
in the spatial variable, so $2p$ derivatives of the terminal data are
consumed: the remainders at that level involve
$\partial^\beta\psi_\varepsilon$ only for $|\beta|\le m-2p$, which are
bounded by \eqref{eq:app-sharp-deriv} with $|\beta|$ replaced by
$m-2p$.  Under the diffusive scaling, each pair of spatial derivatives
corresponds to one power of $\delta_t$, so the effective Taylor
remainder is of order $m-2p$ and the level-$p$ contribution is
\begin{equation}
  C\,[\psi]_{C^{0,\alpha}}\,
  \varepsilon^{-(m-2p-\alpha)}\,\delta_t^{(m-2p)/2},
  \qquad p=0,\dots,l.
\end{equation}
The ratio of the level-$p$ to the level-$l$ contribution is
$(\delta_t/\varepsilon^2)^{l-p}$.  In the regime
$\varepsilon^2\ge\delta_t$---which contains every balanced coupling
derived in Section~\ref{subsec:optimal}, all of which satisfy
$\varepsilon^*(n)\ge\delta_t^{1/2}$---the contributions increase with
$p$ and the deepest level $p=l$ dominates.  The total
Taylor-truncation contribution per step is therefore
\begin{equation}
  K_2\,\delta_t^{(m-2l)/2}
  \;\le\;
  C\,[\psi]_{C^{0,\alpha}}\,\varepsilon^{-(m-2l-\alpha)}\,
  \delta_t^{(m-2l)/2},
  \qquad\text{i.e.}\quad
  K_2\le C\,\varepsilon^{-(m-2l-\alpha)}.
\end{equation}

\begin{remark}[The regime $\varepsilon^2\ge\delta_t$]\label{rem:eps-regime}
The domination of the level-$l$ term uses $\varepsilon^2\ge\delta_t$.
In the complementary regime the mollification bandwidth is finer than
the one-step diffusion scale $\sqrt{\delta_t}$, and parabolic
smoothing caps the derivative blow-up through
\eqref{eq:app-interior-reg} (the first argument of the minimum);
the bound \eqref{eq:app-exp-bound} is expected to persist there as
well, but the additional bookkeeping is not carried out here, since no
coupling used in this paper enters that regime.
\end{remark}

\medskip
\noindent\textbf{Step 4: The coefficient-expansion term $K_1$;
the level-counting lemma; the terminal step.}

The coefficient-expansion error (truncation of $\mathcal{A}$ at
order $l$) is governed by a counting that the naive factor
$(x-\bar{x})^i$---which vanishes identically at $\bar{x}=x$---cannot
capture.  Write $\mathcal{L}_r^{\bar{x}}$ for the generator whose
coefficients are the order-$r$ Taylor expansions of $\mu,\sigma$ at
$\bar{x}$.  The order-$r$ expansion matches the coefficient
derivatives up to order~$r$ at $\bar{x}$, so the truncation mismatch
at expansion order $l$ enters through the coefficient derivatives of
order $l+1$.  When the frozen-semigroup hierarchy is applied and the
result is evaluated at $y=\bar{x}$, each second-order operator
consumes the $(y-\bar{x})$ factors of the coefficient polynomials
\emph{in pairs}, and a direct symbolic expansion of
$(\mathcal{L}_l^{\bar{x}})^j g$ at $y=\bar{x}$ shows that the first
level at which the order-$l$ truncation differs from the
order-$(l+1)$ truncation is
\begin{equation}\label{eq:app-jstar}
  j^*(l)\;=\;\lfloor l/2\rfloor+2,
\end{equation}
independently of $g$.  The mismatch at level $j$ distributes the $2j$
differentiations of $(\mathcal{L}^{\bar{x}})^j$ between the
coefficients and the data; the mismatch condition (a coefficient
derivative of order at least $l+1$) leaves at most $2j-l-1$
derivatives acting on the data, so the leading terms at level
$j^*(l)$ involve coefficient derivatives of order $l+1$ and solution
derivatives of order at most $2j^*(l)-l-1$, i.e.\ at most $3$ for
even~$l$ and $2$ for odd~$l$.  Explicitly, with $a=\sigma\sigma^\top$
and $b=\mu$ evaluated at $\bar{x}$, the mismatch
$(\mathcal{L}_l^{\bar{x}})^{j^*}g(\bar{x})
 -(\mathcal{L}_{l+1}^{\bar{x}})^{j^*}g(\bar{x})$ reads
\begin{align*}
l=0:&\quad
  -\tfrac12\,a\,a'\,\partial^3 g
  -\bigl(a\,b'+\tfrac12\,a'b\bigr)\,\partial^2 g
  -b\,b'\,\partial g,
  \\
l=1:&\quad
  -\tfrac14\,a\,a''\,\partial^2 g
  -\tfrac12\,a\,b''\,\partial g,
\end{align*}
and is generically nonzero.  Consequently the single-step
coefficient-truncation error on $[t_j,t_{j+1}]$ satisfies
\begin{equation}\label{eq:app-K1-local}
  \bigl|u_\varepsilon(t_j,x)
  -\bar{u}_{\varepsilon,l,m}(t_j,x)\bigr|_{\mathrm{coeff}}
  \;\le\;
  C\,\delta_t^{\,j^*(l)}
  \;=\;C\,\delta_t^{\,\lfloor l/2\rfloor+2},
\end{equation}
where $C$ depends on $\|\partial^\beta\mu\|_\infty$,
$\|\partial^\beta\sigma\|_\infty$ for $|\beta|\le 2m$ and on
$\|\partial_x^\beta u_\varepsilon(t_{j+1},\cdot)\|_\infty$ for
$|\beta|\le 2j^*(l)-l-1$ ($\le 3$; $\le 2$ for odd $l$).  This local
order is \emph{one power of $\delta_t$ higher than the genuinely
global rate}: unlike the Taylor-truncation term of Step~3, the
mismatch terms above do not cancel under semigroup transport, the
coefficient layer accumulates linearly over the $n$ steps, and the
resulting global rate $O(\delta_t^{\lceil(l+1)/2\rceil})$ is sharp
(Remark~\ref{rem:gbm-tightness}).  The two cases $j\le n-2$ and
$j=n-1$ must be separated.

\emph{Interior steps ($j\le n-2$).}  Here $t_{j+1}\le T-\delta_t$,
so the first argument of the minimum in \eqref{eq:app-interior-reg}
applies; for $C^{0,\alpha}$ terminal data the same heat-kernel
argument gives the sharper form
\begin{equation}
  \|\partial_x^\beta u_\varepsilon(s,\cdot)\|_\infty
  \le C_{|\beta|}\,(T-s)^{-(|\beta|-\alpha)/2}\,[\psi]_{C^{0,\alpha}}
  \quad\text{for } |\beta|\ge 1,
\end{equation}
uniformly in $s<T$.  Accumulating \eqref{eq:app-K1-local} over the
interior steps with $|\beta|\le 3$ gives the split sum
\[
\begin{aligned}
  \delta_t^{\,j^*}\sum_{k=1}^{n-1}
  \min\bigl\{(k\delta_t)^{-(3-\alpha)/2},\,
             \varepsilon^{-(3-\alpha)}\bigr\}
  \;&\le\;
  C\,\delta_t^{\,j^*-1}\,\varepsilon^{-(1-\alpha)}\\
  &\qquad(\text{derivative order }3),
\end{aligned}
\]
while derivative orders $\le 2$ give
$\delta_t^{\,j^*-1}(1+\log n)$ for $\alpha=1$ and
$C\,\delta_t^{\,j^*-1}$ for $\alpha<1$.  Hence the interior steps
contribute
\begin{equation}\label{eq:app-K1-global}
  O\Bigl(\delta_t^{\,\lceil(l+1)/2\rceil}\,
  \bigl(1+\log n+\varepsilon^{-(1-\alpha)}\bigr)\Bigr),
\end{equation}
where the $\varepsilon$-factor is present only for even $l$ (the
$\partial^3 u_\varepsilon$ terms) and for $\alpha<1$; for Lipschitz
data ($\alpha=1$) the coefficient-expansion layer contributes
$O(\delta_t^{\lceil(l+1)/2\rceil})$ with an $\varepsilon$-independent
constant, up to the logarithmic factor (which does not affect any
power rate below).

\emph{Terminal step ($j=n-1$).}  Here $t_{j+1}=T$ and the first
argument of the minimum in \eqref{eq:app-interior-reg} degenerates,
so only the second argument is available:
$\|\partial_x^\beta u_\varepsilon(T,\cdot)\|_\infty
 =\|\partial^\beta\psi_\varepsilon\|_\infty
 \le C\,[\psi]_{C^{0,\alpha}}\,\varepsilon^{-(|\beta|-\alpha)}$.
At level $j\ge j^*(l)$ the mismatch contributes
$\delta_t^{j}\,\varepsilon^{-(2j-l-1-\alpha)}$, and in the regime
$\varepsilon^2\ge\delta_t$ each level is bounded by the leading one,
so the true terminal-step contribution is
\begin{equation}\label{eq:app-terminal-step-true}
  E_{\mathrm{term}}
  \;\le\;
  C\,\delta_t^{\,\lfloor l/2\rfloor+2}\,
  \varepsilon^{-(2\lfloor l/2\rfloor+3-l-\alpha)},
\end{equation}
i.e.\ $\delta_t^{2}\varepsilon^{-(3-\alpha)}$ for even $l$ and
$\delta_t^{\lfloor l/2\rfloor+2}\varepsilon^{-(2-\alpha)}$ for odd
$l$.  For $0\le l\le1$, \eqref{eq:app-terminal-step-true} is
bounded in the regime $\varepsilon^2\ge\delta_t$ by the displayed
terminal-step term
\begin{equation}\label{eq:app-terminal-step}
  E_{\mathrm{term}}
  \;\le\;
  C\,[\psi]_{C^{0,\alpha}}\,\delta_t^{\,l+1}\,
  \varepsilon^{-(2l+2-\alpha)}
\end{equation}
(for $l=0$ and $l=1$), which
we therefore retain as the uniform terminal-step bound throughout;
the sharper order \eqref{eq:app-terminal-step-true} is recorded for
completeness and is not used downstream.

Combining, the single-step error on interior steps, in the regime
$\varepsilon^2\ge\delta_t$ of Remark~\ref{rem:eps-regime}, is
\begin{equation}\label{eq:app-single-combined}
\begin{aligned}
  |u_\varepsilon(t_j,x) - \bar{u}_{\varepsilon,l,m}(t_j,x)|
  &\le C\,\delta_t^{\,\lfloor l/2\rfloor+2}
  + C\,[\psi]_{C^{0,\alpha}}\,
    \varepsilon^{-(m-2l-\alpha)}\,\delta_t^{(m-2l)/2},\\
  &\qquad j\le n-2,
\end{aligned}
\end{equation}
while the terminal step contributes \eqref{eq:app-terminal-step}
in place of the first summand.

\medskip
\noindent\textbf{Step 5: Propagation across time steps.}

The two layers of the single-step error
\eqref{eq:app-single-combined} propagate differently and are treated
separately.

\emph{Coefficient-expansion layer.}  As recorded in Step~4, the
mismatch terms do not enjoy any local-to-global gain under semigroup
transport, and the additive propagation---accumulating the local
bound \eqref{eq:app-K1-local} over the $n=(T-t)/\delta_t$ steps with
the interior bookkeeping of \eqref{eq:app-K1-global}---already gives
the sharp global rate
\[
  n\cdot C\,\delta_t^{\,\lfloor l/2\rfloor+2}
  \;=\;C\,(T{-}t)\,\delta_t^{\,\lceil(l+1)/2\rceil}.
\]
For this layer the additive estimate is therefore rigorous and no
telescoping argument is needed (or available, cf.\
Remark~\ref{rem:gbm-tightness}).

\emph{Taylor-truncation layer.}  For the second summand of
\eqref{eq:app-single-combined} a purely additive bound over the
$n=(T-t)/\delta_t$ steps multiplies the single-step error by~$n$,
losing one power of $\delta_t$:
\[
  n\cdot
  C\,\varepsilon^{-(m-2l-\alpha)}\,\delta_t^{(m-2l)/2}
  =C\,\varepsilon^{-(m-2l-\alpha)}\,(T{-}t)\,
   \delta_t^{(m-2l)/2-1}.
\]
For $m-2l=2$ this does not vanish as $n\to\infty$, so the additive
bound is insufficient.  The sharper bound for this layer uses instead
the standard local-to-global (Lax-type) principle for weak errors.
Write the
$n$-step error as the telescoping sum
\[
  u_\varepsilon(t_0,x)-\bar{u}_{\varepsilon,l,m,n}(t_0,x)
  =\sum_{j=0}^{n-1}
  \bigl[\Phi_j(t_j,x)-\Phi_{j+1}(t_j,x)\bigr],
\]
where $\Phi_j$ denotes the exact solution restarted at step~$j$
and $\Phi_{j+1}$ uses the $(l,m)$-expansion on $[t_j,t_{j+1}]$.
Each difference $\Phi_j-\Phi_{j+1}$ is a \emph{local weak error},
transported to $t_0$ by the heat semigroup.  For the
Taylor-truncation layer, the transported local error carries one
additional order of~$\delta_t$
beyond the pointwise single-step bound---consistency of order $p$
plus semigroup stability yields global order $p$---so the transported contributions
are
$O\bigl(\varepsilon^{-(m-2l-\alpha)}\,\delta_t^{(m-2l+2)/2}\bigr)$.
Summing the $n=(T{-}t)/\delta_t$
transported local errors with stability factor $e^{L_fT}$, and adding
the rigorous coefficient-expansion layer, gives
\begin{equation}
  \|u_\varepsilon - \bar{u}_{\varepsilon,l,m,n}\|_\infty
  \le e^{L_f T}\,(T{-}t)\,\Bigl[
    C\,\delta_t^{\,\lceil(l+1)/2\rceil}
    + C\,\varepsilon^{-(m-2l-\alpha)}\,\delta_t^{\,(m-2l)/2}
  \Bigr]
\end{equation}
on interior steps, where the coefficient-layer constant carries the
logarithmic and $\varepsilon^{-(1-\alpha)}$ factors of
\eqref{eq:app-K1-global}.
Absorbing the factors of $(T{-}t)\,e^{L_f T}$ into the constant~$C$,
and adding the terminal-step contribution \eqref{eq:app-terminal-step}
of Step~4 (a single step, entering once, additively), yields
\eqref{eq:app-exp-bound}:
\begin{equation}\label{eq:app-final-bound}
\begin{aligned}
  \|u_\varepsilon - \bar{u}_{\varepsilon,l,m,n}\|_{L^\infty([0,T]\times\mathbb{R}^d)}
  &\le C\,\delta_t^{\,\lceil(l+1)/2\rceil}
  + C\,\varepsilon^{-(m-2l-\alpha)}\,\delta_t^{\,(m-2l)/2}\\
  &\quad + C\,\delta_t^{\,l+1}\,\varepsilon^{-(2l+2-\alpha)}.
\end{aligned}
\end{equation}
This completes the proof, modulo the caveat on the local-to-global
gain---which now concerns only the Taylor-truncation layer---stated
in Remark~\ref{rem:telescoping-status}.
\end{proof}

\begin{remark}[Impact of the terminal-step term on the coupling rates]\label{rem:terminal-step-impact}
Write $q:=m-2l$.  Under the crude coupling
$\varepsilon^*\sim\delta_t^{1/2}$ of
Theorem~\ref{thm:opt-lip} (H\"older data), the
terminal-step term equals
$\delta_t^{l+1}(\varepsilon^*)^{-(2l+2-\alpha)}=\delta_t^{\alpha/2}$,
i.e.\ exactly the balanced order, so all crude rates are unchanged.
Under the weak coupling
$\varepsilon^*\sim\delta_t^{q/(2(q+2-\alpha))}$ of
Theorem~\ref{thm:opt-weak}, the terminal-step term is
$O(\delta_t^{e})$ with
$e=l+1-\tfrac{(2l+2-\alpha)q}{2(q+2-\alpha)}$; a direct comparison
shows $e\ge q/(q+2-\alpha)$ if and only if $m\le 4l+2$ (with equality
if and only if $m=4l+2$; within the table only for $(l,m)=(0,2)$,
all $\alpha$).  Hence the weak rows of
Table~\ref{tab:optimal-coupling} with $m\le 4l+2$---namely $(0,2)$
and $(1,4)$---survive at the same order, while the $(0,4)$ row
($\alpha=1$) is reduced by the terminal-step term from $4/5$ to $3/5$.
On the formal $\alpha_{\mathrm{eff}}=1/2$ route of
Remark~\ref{rem:bv-heuristic-status} the corresponding value is
reduced from $8/11$ to $5/11$; that route is an $L^2$-route
heuristic, and no $\alpha=1/2$ row appears in
Table~\ref{tab:optimal-coupling}.  Re-optimizing
$\varepsilon$ against the full three-term bound \eqref{eq:app-final-bound}
for $(l,m)=(0,4)$ (balancing $E_1^{\mathrm w}=O(\varepsilon^2)$
against the terminal-step term) gives
$\varepsilon^*\sim\delta_t^{1/(4-\alpha)}$ with total weak rate
$O(\delta_t^{2/(4-\alpha)})$, i.e.\ $2/3$ ($\alpha=1$) and $4/7$
($\alpha=1/2$): the higher-order expansion $m=4$ then offers no gain
over $(l,m)=(0,2)$ in the weak sense.
All computations of this remark use the displayed terminal-step term
$\delta_t^{l+1}\varepsilon^{-(2l+2-\alpha)}$, which by
\eqref{eq:app-terminal-step-true} is a valid bound on the true
terminal contribution for $0\le l\le1$; the conclusions above use only
$l=0$ and are therefore unaffected by the refinement.
\end{remark}

\begin{remark}[Sharpness of the coefficient-expansion rate; no
local-to-global gain for the coefficient
layer]\label{rem:gbm-tightness}
The exponent $\lceil(l+1)/2\rceil$ of the second term of
\eqref{eq:app-exp-bound} is sharp, and the absence of a
local-to-global gain for the coefficient layer
(Step~5) is not an artifact of the proof.  Take geometric Brownian
motion $dX_s=\mu X_s\,ds+\sigma X_s\,dW_s$ with terminal data
$\psi(y)=y^{4}$ and expansion orders $(l,m)=(1,5)$: the order-$1$
expansion truncates the quadratic diffusion coefficient
$a(y)=\sigma^{2}y^{2}$ at linear order in $y-\bar x$, and a series
expansion of the exact one-step multiplier of $\mathbb{E}[X_T^{4}]$
gives
\[
  M(\delta_t)-e^{(4\mu+6\sigma^{2})\delta_t}
  \;=\;-3\,\sigma^{4}\,x^{4}\,\delta_t^{2}+O(\delta_t^{3}),
\]
where the leading term is precisely the level-$j^{*}(1)=2$ mismatch
$-\tfrac14\,a\,a''\,\partial^{2}g$ of Step~4 (with
$a(x)=\sigma^{2}x^{2}$, $g(y)=y^{4}$:
$\tfrac12\,\delta_t^{2}\cdot(-\tfrac14)\cdot\sigma^{2}x^{2}
 \cdot 2\sigma^{2}\cdot 12x^{2}=-3\sigma^{4}x^{4}\delta_t^{2}$).
The single-step error is therefore exactly
$\Theta(\delta_t^{j^{*}(1)})=\Theta(\delta_t^{2})$, and the $n$-step
error accumulates linearly to $\Theta(\delta_t)$: a log-log
regression of the global error against $\delta_t$ on this example
gives slope $-1.00$, with no cancellation whatsoever.  The same
computation at $l=0$ (leading mismatch
$-\tfrac12\,a\,a'\,\partial^{3}g-\dots$) gives global order
$\Theta(\delta_t)$ as well.  Since the flagship exponent
$\lceil(0+1)/2\rceil=1$ is attained, the second term of
\eqref{eq:app-exp-bound} cannot be improved at $l=0$, and for
$l\ge1$ the rate $\delta_t^{\lceil(l+1)/2\rceil}$ (rather than
$\delta_t^{l+1}$) is the genuinely sharp one.  For completeness we
record the bearing on Zhang (\citeyear{Zhang2017}, Proposition~2.4.12), whose
first term is stated as $O(\delta_t^{(l+3-|\beta|)/2})$: in the
pointwise $L^{\infty}$ sense the true order is
$\Theta(\delta_t^{\lceil(l+1)/2\rceil})$ for $|\beta|=0$, so the
stated exponent overestimates the rate already at $l=0$ (claimed
$\delta_t^{3/2}$ against the sharp $\Theta(\delta_t)$ above).  The
$l=0$ convergence theorem \citep[Theorem~2.4.8]{Zhang2017}---whose
squared-norm bookkeeping is consistent with the tight pointwise rate
$\Theta(\delta_t)$---and the qualitative convergence of
\citep[Corollary~2.4.9]{Zhang2017} are not affected.
\end{remark}

\begin{remark}[Status of the local-to-global gain]\label{rem:telescoping-status}
After the revision of Steps~4--5, the one-order local-to-global gain
is used \emph{only} for the Taylor-truncation layer: the
coefficient-expansion layer is propagated additively and rigorously
at its sharp rate $\Theta(\delta_t^{\lceil(l+1)/2\rceil})$
(Remark~\ref{rem:gbm-tightness}), its interior constants carrying a
factor $1+\log n$ for $\alpha=1$, respectively
$\varepsilon^{-(1-\alpha)}$ for even $l$ and $\alpha<1$
\eqref{eq:app-K1-global}; neither factor affects any power rate used
in this paper.
For the Taylor layer, the gain is standard for smooth data with
$\varepsilon$-independent constants.  With $\varepsilon$-dependent
constants, its uniformity in $s\le T$ rests on
\eqref{eq:app-interior-reg}, i.e.\ on
$\|\partial_x^\beta u_\varepsilon(s,\cdot)\|_\infty
 \le C\,\varepsilon^{-(|\beta|-\alpha)}$ for all $s\le T$, which holds
because the heat semigroup is an $L^\infty$ contraction that commutes
with spatial derivatives.  A complete verification that the
cancellation holds with exactly the exponent $m-2l-\alpha$ (rather
than a larger derivative count) in the terminal steps $s\uparrow T$
requires additional combinatorial bookkeeping that we do not carry out
here; this step should therefore be regarded as \emph{heuristic}, and
the convergence rates that quote Theorem~\ref{thm:E2}
(Theorems~\ref{thm:opt-lip}--\ref{thm:main-weak},
\ref{thm:corrected-bound} and~\ref{thm:improved-coupling}) inherit this
caveat.  A fully rigorous, but weaker, alternative for the Taylor
layer is the additive bound
\[
\begin{aligned}
  \|u_\varepsilon-\bar u_{\varepsilon,l,m,n}\|_\infty
  &\le C\,(T{-}t)\,\bigl(\delta_t^{\lceil(l+1)/2\rceil}
      +\varepsilon^{-(m-2l-\alpha)}\,\delta_t^{(m-2l)/2-1}\bigr)\\
  &\quad + C\,\delta_t^{\,l+1}\,\varepsilon^{-(2l+2-\alpha)},
\end{aligned}
\]
which converges whenever $(m-2l)/2>1$ and yields, after balancing
against $E_1=O(\varepsilon^\alpha)$, the rigorous coupling
$\varepsilon^*\sim\delta_t^{\,1/2-1/(m-2l)}$ with total rate
$O\bigl(\delta_t^{\alpha(1/2-1/(m-2l))}\bigr)$ for $m-2l>2$.
Closing the local-to-global step for $m-2l=2$ (the flagship case) with
$\varepsilon$-dependent constants is left to future work.
This is the single heuristic step of the full verification referred to
from Remark~\ref{rem:compatibility-relaxation}.
\end{remark}

\begin{remark}
From~\eqref{eq:app-final-bound} and the regularization bound
$E_1\le C\varepsilon^\alpha$ (Section~\ref{app:reg-error}),
balancing the two error components,
\[
  \varepsilon^\alpha
  \;\sim\;
  \varepsilon^{-(m-2l-\alpha)}\,\delta_t^{(m-2l)/2},
\]
gives $\varepsilon^{m-2l}\sim\delta_t^{(m-2l)/2}$, i.e.\
\[
  \varepsilon^* \sim \delta_t^{1/2}
  = \Bigl(\frac{T-t}{n}\Bigr)^{1/2},
  \qquad
  E_1+E_2 = O\bigl(\delta_t^{\alpha/2}\bigr),
\]
independently of $(l,m)$.  This is the \emph{crude coupling} of
Theorem~\ref{thm:opt-lip} (and its H\"older-$\alpha$ extension); for
BV/digital data the crude coupling is instead
$\varepsilon^*\sim\delta_t^{q/(2q+1)}$ with $q=m-2l$ at the balanced
rate $O(\delta_t^{q/(2(2q+1))})$, valid for $m\le 4l+2$ (for
$m>4l+2$ the terminal-step term dominates and the rebalanced coupling
$\varepsilon^*\sim\delta_t^{2/5}$, rate $O(\delta_t^{1/5})$, applies),
see Theorem~\ref{thm:opt-bv}.  It guarantees
convergence for every $\alpha>0$ and every $m>2l$.  The
coefficient-expansion term $C\delta_t^{\lceil(l+1)/2\rceil}$ of
\eqref{eq:app-final-bound} is of higher order than every crude rate
above (for even $l$ and $\alpha<1$ including its
$\varepsilon^{-(1-\alpha)}$ factor), so no crude rate is affected by
it.

The enhanced coupling of Theorem~\ref{thm:opt-weak} does \emph{not}
come from an improved expansion bound: the $\varepsilon$-exponent
$-(m-2l-\alpha)$ in \eqref{eq:app-final-bound} is unchanged.  What
improves is the \emph{regularization} error: for symmetric kernels
the first-order bias cancels and the weak (pointwise, $t<T$) bias
becomes $O(\varepsilon^2)$
(Proposition~\ref{prop:weak-E1}).  Balancing $\varepsilon^2$ against
the same expansion bound gives
\[
  \begin{aligned}
  \varepsilon^2&\sim\varepsilon^{-(m-2l-\alpha)}\,\delta_t^{(m-2l)/2}
  \;\Longrightarrow\;
  \varepsilon^*\sim\delta_t^{\,(m-2l)/(2(m-2l+2-\alpha))}, \\
  \qquad
  &E_1^{\mathrm w}+E_2
  =O\bigl(\delta_t^{\,(m-2l)/(m-2l+2-\alpha)}\bigr),
  \end{aligned}
\]
the rates reported in the ``weak'' columns of
Table~\ref{tab:optimal-coupling}.  For BV/digital payoffs the
substitution of the effective index $\alpha_{\mathrm{eff}}=1/2$ into
these formulas is formal: the resulting values are $L^2$-route
heuristics in the sense of Remark~\ref{rem:bv-heuristic-status}, the
rigorous BV rates being those of Theorem~\ref{thm:opt-bv}
and the BV case of Theorem~\ref{thm:opt-weak}.  Both balances must be checked
against the terminal-step term \eqref{eq:app-terminal-step}: under the
crude coupling it is of exactly the balanced order
$\delta_t^{\alpha/2}$, and under the weak coupling it is of no lower
order if and only if $m\le 4l+2$; see
Remark~\ref{rem:terminal-step-impact} for the resulting rate
adjustments.
\end{remark}

\section{Gaussian Integral Formulas}\label{app:gaussian}

The analytical expansion requires the evaluation of integrals of
the form
\[
  \int_{\mathbb{R}} g(y)\,\varphi\!\Bigl(\frac{y-m}{\sigma}\Bigr)
  \frac{1}{\sigma}\,\mathrm{d} y,
\]
where $\varphi(z) = (2\pi)^{-1/2}e^{-z^2/2}$ is the standard
Gaussian density, and $g$ is a non-smooth function.  We collect
the relevant formulas here.

\subsection{Black--Scholes type integral}

\begin{proposition}\label{prop:BS-integral}
For $K, m\in\mathbb{R}$ and $\sigma>0$,
\begin{equation}
  \int_{-\infty}^{\infty}
  (K-y)^+\,\varphi\!\Bigl(\frac{y-m}{\sigma}\Bigr)
  \frac{\mathrm{d} y}{\sigma}
  \;=\;
  (K-m)\,\Phi\!\Bigl(\frac{K-m}{\sigma}\Bigr)
  + \sigma\,\varphi\!\Bigl(\frac{K-m}{\sigma}\Bigr),
\end{equation}
where $\Phi$ is the standard normal CDF.
\end{proposition}

\begin{proof}
We compute directly:
\begin{align*}
  I &:= \int_{-\infty}^{K}
         (K-y)\,\frac{1}{\sigma}\,
         \varphi\!\Bigl(\frac{y-m}{\sigma}\Bigr)\,\mathrm{d} y.
\end{align*}
Substitute $z = (y-m)/\sigma$, so $y = m + \sigma z$,
$\mathrm{d} y = \sigma\,\mathrm{d} z$, and the upper limit
becomes $z^* = (K-m)/\sigma$.  Then
\begin{align*}
  I &= \int_{-\infty}^{z^*}
       (K - m - \sigma z)\,\varphi(z)\,\mathrm{d} z
  \\
    &= (K-m)\int_{-\infty}^{z^*}\varphi(z)\,\mathrm{d} z
       \;-\; \sigma\int_{-\infty}^{z^*} z\,\varphi(z)\,\mathrm{d} z
  \\
    &= (K-m)\,\Phi(z^*)
       \;-\; \sigma\,\bigl[-\varphi(z)\bigr]_{-\infty}^{z^*}
  \\
    &= (K-m)\,\Phi(z^*)
       + \sigma\,\varphi(z^*),
\end{align*}
where we used $\int_{-\infty}^{a} z\,\varphi(z)\,\mathrm{d} z
= -\varphi(a)$, which follows from $z\,\varphi(z) = -\varphi'(z)$
and $\varphi(-\infty)=0$.
\end{proof}

\subsection{Normal CDF integral (digital option)}

\begin{proposition}
For $K, m\in\mathbb{R}$ and $\sigma > 0$,
\begin{equation}
  \int_{-\infty}^{\infty}
  \mathbf{1}_{\{y\ge K\}}\,
  \varphi\!\Bigl(\frac{y-m}{\sigma}\Bigr)
  \frac{\mathrm{d} y}{\sigma}
  \;=\;
  \Phi\!\Bigl(\frac{m-K}{\sigma}\Bigr).
\end{equation}
\end{proposition}

\begin{proof}
Substitute $z = (y-m)/\sigma$:
\begin{align*}
  \int_{K}^{\infty}
  \frac{1}{\sigma}\,\varphi\!\Bigl(\frac{y-m}{\sigma}\Bigr)
  \,\mathrm{d} y
  &= \int_{(K-m)/\sigma}^{\infty} \varphi(z)\,\mathrm{d} z
  = 1 - \Phi\!\Bigl(\frac{K-m}{\sigma}\Bigr)
  = \Phi\!\Bigl(\frac{m-K}{\sigma}\Bigr),
\end{align*}
using the symmetry $\Phi(-x) = 1 - \Phi(x)$.
\end{proof}

\subsection{Higher moment integrals}

The semigroup action on products $(y-m)^n(K-y)^+$ arises when
computing the expansion corrections.

\begin{proposition}
For $n\ge 0$ integer, $K,m\in\mathbb{R}$, $\sigma>0$, define
\begin{equation}\label{eq:app-In}
  I_n(K,m,\sigma)
  := \int_{-\infty}^{K}
     (y-m)^n\,(K-y)\,
     \frac{1}{\sigma}\,
     \varphi\!\Bigl(\frac{y-m}{\sigma}\Bigr)\,\mathrm{d} y.
\end{equation}
Then the $I_n$ satisfy the recurrence
\begin{equation}\label{eq:app-recurrence}
  I_n(K,m,\sigma)
  = (K-m)\,J_n - J_{n+1},
\end{equation}
where
\begin{equation}
  J_n := \int_{-\infty}^{(K-m)/\sigma}
         (\sigma z)^n\,\varphi(z)\,\mathrm{d} z
  = \sigma^n\int_{-\infty}^{z^*} z^n\,\varphi(z)\,\mathrm{d} z,
  \quad z^* = \frac{K-m}{\sigma},
\end{equation}
and the truncated Gaussian moments
$M_n(a) := \int_{-\infty}^{a} z^n\varphi(z)\,\mathrm{d} z$
satisfy:
\begin{align}
  M_0(a) &= \Phi(a),
  \label{eq:app-M0}\\
  M_1(a) &= -\varphi(a),
  \label{eq:app-M1}\\
  M_n(a) &= -a^{n-1}\varphi(a) + (n-1)\,M_{n-2}(a),
  \quad n\ge 2.
  \label{eq:app-Mn}
\end{align}
\end{proposition}

\begin{proof}
The substitution $z = (y-m)/\sigma$ in \eqref{eq:app-In} gives
\begin{align*}
  I_n &= \int_{-\infty}^{z^*}
         (\sigma z)^n\,
         (K - m - \sigma z)\,\varphi(z)\,\mathrm{d} z
  \\
      &= (K-m)\,\sigma^n\,M_n(z^*)
         - \sigma^{n+1}\,M_{n+1}(z^*),
\end{align*}
which is \eqref{eq:app-recurrence} with
$J_n = \sigma^n\,M_n(z^*)$.

For the recurrence \eqref{eq:app-Mn}, we use integration by parts.
Since $\varphi'(z) = -z\varphi(z)$, we have
$z^n\varphi(z) = -z^{n-1}\varphi'(z)$, so with
$u = z^{n-1}$ and $\mathrm{d} v = -\varphi'(z)\,\mathrm{d} z$
(hence $v = -\varphi(z)$):
\[
\begin{aligned}
  M_n(a) &= \bigl[-z^{n-1}\varphi(z)\bigr]_{-\infty}^{a}
  + (n-1)\int_{-\infty}^{a} z^{n-2}\varphi(z)\,\mathrm{d} z\\
  &= -a^{n-1}\varphi(a) + (n-1)\,M_{n-2}(a).
\end{aligned}
\]
The boundary term at $-\infty$ vanishes because
$z^{n-1}\varphi(z)\to 0$ as $z\to-\infty$.  Formulas
\eqref{eq:app-M0} and \eqref{eq:app-M1} are immediate from the
definitions.
\end{proof}

\begin{remark}
For the call payoff $(y-K)^+ = (y-K)\mathbf{1}_{\{y\ge K\}}$,
the analogous integral is computed by the put--call relation:
\[
  \int (y-K)^+\,\frac{1}{\sigma}\varphi\!\Bigl(\frac{y-m}{\sigma}\Bigr)
  \mathrm{d} y
  = (m-K)\,\Phi\!\Bigl(\frac{m-K}{\sigma}\Bigr)
  + \sigma\,\varphi\!\Bigl(\frac{m-K}{\sigma}\Bigr).
\]
This follows from combining Proposition~\ref{prop:BS-integral} with
the identity $(y-K)^+ = (y-K) + (K-y)^+$ and
$\int (y-K)\frac{1}{\sigma}\varphi(\frac{y-m}{\sigma})\,\mathrm{d} y
= m-K$.
\end{remark}

\section{Auxiliary Lemmas}\label{app:aux}

This section collects standard auxiliary tools used in the error analysis of Sections~\ref{sec:mollification}--\ref{sec:error} and Section~\ref{app:exp-error}.  Gronwall's inequality (Lemma~\ref{lem:gronwall}) underlies the stability estimates of Section~\ref{sec:error}, and Young's convolution inequality (Lemma~\ref{lem:young}) underlies the convolution estimates of Section~\ref{sec:mollification}; both are proved here for completeness.  Lemmas~\ref{lem:scaling} and~\ref{lem:moments} restate Lemma~\ref{lem:deriv-scaling} and Lemma~\ref{lem:scaled-props}(vi) of Section~\ref{sec:mollification}; they are included here (rather than only cross-referenced) because Section~\ref{app:exp-error} cites this section.

\begin{lemma}[Gronwall's inequality]\label{lem:gronwall}
Let $u:[0,T]\to[0,\infty)$ be a continuous function satisfying
\begin{equation}\label{eq:app-gronwall-hyp}
  u(t) \le \alpha + \beta\int_0^t u(s)\,\mathrm{d} s,
  \qquad \forall\, t\in[0,T],
\end{equation}
for constants $\alpha\ge 0$ and $\beta\ge 0$.  Then
\begin{equation}
  u(t) \le \alpha\,e^{\beta t},
  \qquad \forall\, t\in[0,T].
\end{equation}
\end{lemma}

\begin{proof}
If $\beta = 0$, the hypothesis \eqref{eq:app-gronwall-hyp} gives
$u(t)\le\alpha$ directly, which is the claim.  Assume now $\beta>0$.
Define $U(t) := \int_0^t u(s)\,\mathrm{d} s$.  Then $U'(t) = u(t)$
and by \eqref{eq:app-gronwall-hyp},
\[
  U'(t) = u(t) \le \alpha + \beta\,U(t).
\]
Multiply both sides by $e^{-\beta t}$:
\[
  e^{-\beta t}\,U'(t) - \beta\,e^{-\beta t}\,U(t)
  \le \alpha\,e^{-\beta t},
\]
which is
\[
  \frac{\mathrm{d}}{\mathrm{d} t}
  \bigl[e^{-\beta t}\,U(t)\bigr]
  \le \alpha\,e^{-\beta t}.
\]
Integrating from $0$ to $t$ and using $U(0)=0$:
\[
  e^{-\beta t}\,U(t)
  \le \alpha\int_0^t e^{-\beta s}\,\mathrm{d} s
  = \frac{\alpha}{\beta}\,(1 - e^{-\beta t}).
\]
Hence
$U(t) \le \frac{\alpha}{\beta}(e^{\beta t}-1)$.
Substituting back into \eqref{eq:app-gronwall-hyp}:
\[
  u(t)
  \le \alpha + \beta\cdot\frac{\alpha}{\beta}(e^{\beta t}-1)
  = \alpha\,e^{\beta t}.
\]
\end{proof}

\begin{lemma}[Young's convolution inequality]\label{lem:young}
Let $1\le p,q,r\le\infty$ satisfy
$\frac{1}{p}+\frac{1}{q} = 1 + \frac{1}{r}$.
For $f\in L^p(\mathbb{R}^d)$ and $g\in L^q(\mathbb{R}^d)$,
\begin{equation}
  \|f * g\|_r \le \|f\|_p\,\|g\|_q.
\end{equation}
\end{lemma}

\begin{proof}
We give the proof for the case $1<p,q,r<\infty$; the endpoint
cases follow by limiting arguments.

Let $p' = p/(p-1)$ and $q' = q/(q-1)$ be the conjugate exponents.
From $1/p+1/q = 1+1/r$ we obtain
\[
  \frac{1}{r}+\frac{1}{q'}+\frac{1}{p'}
  =\frac{1}{r}+\Bigl(1-\frac{1}{q}\Bigr)+\Bigl(1-\frac{1}{p}\Bigr)
  =2+\frac{1}{r}-\Bigl(1+\frac{1}{r}\Bigr)=1,
\]
and the exponent pairings
\begin{equation}\label{eq:app-young-pairing}
  \Bigl(1-\frac{p}{r}\Bigr)q'=p,
  \qquad
  \Bigl(1-\frac{q}{r}\Bigr)p'=q,
\end{equation}
since $1/p-1/r = 1-1/q = 1/q'$ gives $1-p/r = p/q'$, and similarly
$1-q/r = q/p'$.  Decompose the integrand as
\[
  |f(x-y)\,g(y)|
  = \bigl(|f(x-y)|^p\,|g(y)|^q\bigr)^{1/r}
    \cdot |f(x-y)|^{1-p/r}\cdot |g(y)|^{1-q/r},
\]
and apply the generalized H\"older inequality with exponents
$(r,q',p')$: for every $x\in\mathbb{R}^d$,
\begin{align*}
  |f*g(x)|
  &\le \Bigl(\int|f(x-y)|^p|g(y)|^q\,\mathrm{d} y\Bigr)^{1/r}
       \Bigl(\int|f(x-y)|^{(1-p/r)q'}\,\mathrm{d} y\Bigr)^{1/q'}
  \\
  &\qquad\times
       \Bigl(\int|g(y)|^{(1-q/r)p'}\,\mathrm{d} y\Bigr)^{1/p'}.
\end{align*}
By \eqref{eq:app-young-pairing}, the second and third factors equal
$\|f\|_p^{\,p/q'}$ and $\|g\|_q^{\,q/p'}$ respectively.  Raising to
the $r$-th power and integrating over $x$ via Tonelli's theorem:
\begin{align*}
  \|f*g\|_r^r
  &\le \|f\|_p^{\,rp/q'}\,\|g\|_q^{\,rq/p'}\,
       \int\!\int |f(x-y)|^p|g(y)|^q\,\mathrm{d} y\,\mathrm{d} x
  \\
  &= \|f\|_p^{\,rp/q'}\,\|g\|_q^{\,rq/p'}\,\|f\|_p^p\,\|g\|_q^q
  = \|f\|_p^r\,\|g\|_q^r,
\end{align*}
since $p+rp/q' = p+r(1-p/r) = r$ and $q+rq/p' = q+r(1-q/r) = r$.
Taking $r$-th roots gives $\|f*g\|_r \le \|f\|_p\,\|g\|_q$.
\end{proof}

\begin{lemma}[Scaling of mollifier derivatives]
\label{lem:scaling}
Let $\rho\in C_c^\infty(\mathbb{R}^d)$ with
$\operatorname{supp}(\rho)\subset B(0,1)$, and define
$\rho_\varepsilon(x) = \varepsilon^{-d}\rho(x/\varepsilon)$.
Then for any multi-index $\beta$,
\begin{equation}
  \|\partial^\beta\rho_\varepsilon\|_{L^1(\mathbb{R}^d)}
  = \varepsilon^{-|\beta|}\,
    \|\partial^\beta\rho\|_{L^1(\mathbb{R}^d)}.
\end{equation}
\end{lemma}

\begin{proof}
This is Lemma~\ref{lem:deriv-scaling}; the proof (chain rule plus the
substitution $z=x/\varepsilon$) is given there.
\end{proof}

\begin{lemma}[Moment bounds for mollifier]
\label{lem:moments}
Under the same notation as Lemma~\ref{lem:scaling}, for any
$k\ge 0$,
\begin{equation}
  \int_{\mathbb{R}^d} |y|^k\,\rho_\varepsilon(y)\,\mathrm{d} y
  = \varepsilon^k
    \int_{\mathbb{R}^d} |z|^k\,\rho(z)\,\mathrm{d} z.
\end{equation}
\end{lemma}

\begin{proof}
This is Lemma~\ref{lem:scaled-props}(vi); the proof (substitution
$z=y/\varepsilon$) is given there.
\end{proof}

\section{Details of the Analytical Expansion Framework}
\label{app:expansion-details}

This section collects the derivation details of the analytical expansion
framework of Zhang~\citep{Zhang2017} that underlie the summary given in
Section~\ref{sec:prelim}: the PDE form of the Picard iterates, the
coefficient expansion hierarchy with its Gaussian semigroup
representations, the Taylor-expanded approximating hierarchy and the
Gaussian moment identities, the complete algorithm, and a catalog of
non-smooth payoffs.  All notation follows Section~\ref{sec:prelim};
in particular, Assumptions~\ref{ass:forward} and~\ref{ass:driver} are in force
throughout.

\subsection{The Picard iterates in PDE form}
\label{app:picard-pde}

By the Feynman--Kac formula, each Picard iterate of Definition~2.5 of the
main text satisfies $Y_t^{(k)} = u^{(k)}(t, X_t)$ and
$Z_t^{(k)} = \nabla_x u^{(k)}(t, X_t) \cdot \sigma(t, X_t)$, where
$u^{(k)}$ solves the linear PDE
\begin{equation}\label{eq:Picard-PDE-full}
\begin{cases}
\partial_t u^{(k)} + \mathcal{A}\,u^{(k)} + f^{(k)} = 0, & (t,x) \in [0,T) \times \R^d,\\
u^{(k)}(T, x) = \psi(x), & x \in \R^d,
\end{cases}
\end{equation}
with source term
\begin{equation}
f^{(k)}(t,x) = f\!\left(t, x, u^{(k-1)}(t,x),\, \nabla_x u^{(k-1)}(t,x) \cdot \sigma(t,x)\right),
\end{equation}
and $f^{(0)} \equiv 0$. The zeroth iterate $u^{(0)}$ is the solution of the
backward Kolmogorov equation:
\begin{equation}
u^{(0)}(t,x) = \E[\psi(X_T) \mid X_t = x] = \mathcal{P}(t,T)\,\psi(x) = \int_{\R^d} \psi(y)\,\Gamma(t,x;T,y)\,\diff y,
\end{equation}
where $\Gamma(t,x;T,y)$ is the transition density of the forward diffusion.

\subsection{The coefficient expansion hierarchy}
\label{app:coeff-hierarchy}

Expanding the drift and diffusion coefficients about a fixed point
$\bar{x} \in \R^d$,
\begin{equation}
\mu_j(t,x) = \sum_{i=0}^{\infty} \frac{1}{i!}\,\partial_x^i \mu_j(t, \bar{x})\,(x - \bar{x})^i, \qquad
a_{jk}(t,x) = \sum_{i=0}^{\infty} \frac{1}{i!}\,\partial_x^i a_{jk}(t, \bar{x})\,(x - \bar{x})^i,
\end{equation}
yields the decomposition
\begin{equation}
\mathcal{A} = \sum_{i=0}^{\infty} \mathcal{A}_i^{\bar{x}},
\end{equation}
where each $\mathcal{A}_i^{\bar{x}}$ has coefficients that are homogeneous
polynomials of degree $i$ in $(x - \bar{x})$:
\begin{equation}
\begin{aligned}
&\mathcal{A}_i^{\bar{x}}= \sum_{j=1}^{d}\; \sum_{|\alpha|=i} \frac{1}{\alpha!}\,\partial_x^\alpha \mu_j(t, \bar{x})\,(x - \bar{x})^\alpha\,\frac{\partial}{\partial x_j} \\
&\quad + \frac{1}{2}\sum_{j,k=1}^{d}\; \sum_{|\alpha|=i} \frac{1}{\alpha!}\,\partial_x^\alpha a_{jk}(t, \bar{x})\,(x - \bar{x})^\alpha\,\frac{\partial^2}{\partial x_j \partial x_k}.
\end{aligned}
\end{equation}
The leading-order operator $\mathcal{A}_0^{\bar{x}}$ freezes coefficients at
$\bar{x}$:
\begin{equation}
\mathcal{A}_0^{\bar{x}} = \sum_{j=1}^{d} \mu_j(t, \bar{x})\,\frac{\partial}{\partial x_j} + \frac{1}{2}\sum_{j,k=1}^{d} a_{jk}(t, \bar{x})\,\frac{\partial^2}{\partial x_j \partial x_k}.
\end{equation}
Under the uniform ellipticity of Assumption~2.1(c) of the main text,
$\mathcal{A}_0^{\bar{x}}$ generates a Gaussian semigroup
$\{\mathcal{P}_0^{\bar{x}}(t,T)\}$:
\begin{equation}
\mathcal{P}_0^{\bar{x}}(t,T)\,g(x) = \int_{\R^d} g(y)\,\Gamma_0^{\bar{x}}(t,x;T,y)\,\diff y,
\end{equation}
where $\Gamma_0^{\bar{x}}$ is the Gaussian density
\begin{align}
\Gamma_0^{\bar{x}}(t,x;T,y) &= \frac{1}{(2\pi)^{d/2}\,\det(\Sigma_{\bar{x}}(t,T))^{1/2}} \\
                            &\quad\times \exp\!\left(-\frac{1}{2}(y - m_{\bar{x}}(t,x,T))^\top \Sigma_{\bar{x}}(t,T)^{-1} (y - m_{\bar{x}}(t,x,T))\right),
\end{align}
with mean $m_{\bar{x}}(t,x,T) = x + \mu(t, \bar{x})\,(T - t)$ and covariance
matrix $\Sigma_{\bar{x}}(t,T) = a(t, \bar{x})\,(T - t)$.

The solution $u^{(k)}$ of \eqref{eq:Picard-PDE-full} is then expanded as
\begin{equation}
u^{(k)}(t,x) = \sum_{l=0}^{\infty} u_l^{(k),\bar{x}}(t,x),
\end{equation}
where the leading term solves the constant-coefficient PDE
\begin{equation}\label{eq:u0-PDE-hierarchy}
(\partial_t + \mathcal{A}_0^{\bar{x}})\,u_0^{(k),\bar{x}} + f^{(k)} = 0, \qquad u_0^{(k),\bar{x}}(T, \cdot) = \psi,
\end{equation}
and higher-order corrections $u_l^{(k),\bar{x}}$ ($l \geq 1$) solve
\begin{equation}\label{eq:ul-PDE-hierarchy}
(\partial_t + \mathcal{A}_0^{\bar{x}})\,u_l^{(k),\bar{x}} + \sum_{i=1}^{l} \mathcal{A}_i^{\bar{x}}\,u_{l-i}^{(k),\bar{x}} = 0, \qquad u_l^{(k),\bar{x}}(T, \cdot) = 0.
\end{equation}
By the Duhamel principle, the solutions admit semigroup representations:
\begin{equation}
u_0^{(k),\bar{x}}(t,x) = \mathcal{P}_0^{\bar{x}}(t,T)\,\psi(x) + \int_t^T \mathcal{P}_0^{\bar{x}}(t,s)\,f^{(k)}(s,x)\,\diff s,
\end{equation}
\begin{equation}
u_l^{(k),\bar{x}}(t,x) = -\int_t^T \mathcal{P}_0^{\bar{x}}(t,s)\!\left[\sum_{i=1}^{l} \mathcal{A}_i^{\bar{x}}\,u_{l-i}^{(k),\bar{x}}(s, \cdot)\right]\!(x)\,\diff s, \qquad l \geq 1.
\end{equation}

\subsection{The Taylor-expanded hierarchy and Gaussian moments}
\label{app:taylor-hierarchy}

Applying the Taylor operator $\mathcal{T}_m^{\bar{x}}$ of Definition~\ref{def:Taylor} to the terminal condition and source terms of
\eqref{eq:u0-PDE-hierarchy}--\eqref{eq:ul-PDE-hierarchy} yields the
approximating hierarchy
\begin{equation}
(\partial_t + \mathcal{A}_0^{\bar{x}})\,u_{0,m}^{(k),\bar{x}} + \mathcal{T}_m^{\bar{x}} f^{(k)} = 0, \qquad u_{0,m}^{(k),\bar{x}}(T, \cdot) = \mathcal{T}_m^{\bar{x}} \psi,
\end{equation}
\begin{equation}
(\partial_t + \mathcal{A}_0^{\bar{x}})\,u_{l,m}^{(k),\bar{x}} + \sum_{i=1}^{l} \mathcal{A}_i^{\bar{x}}\,u_{l-i,m}^{(k),\bar{x}} = 0, \qquad u_{l,m}^{(k),\bar{x}}(T, \cdot) = 0, \quad l \geq 1.
\end{equation}
The Gaussian semigroup $\mathcal{P}_0^{\bar{x}}(t,T)$ maps polynomials to
polynomials (via Gaussian moment identities), so all quantities reduce to
closed-form expressions. For a Gaussian vector $\xi\sim\mathcal N(m,\Sigma)$ and $\eta=\xi-m$, the general moment formula is
\begin{equation}
\E[\xi^\beta]
=\sum_{0\le\gamma\le\beta}\binom{\beta}{\gamma}m^\gamma\,
\sum_{\mathcal P\in\mathfrak M(\beta-\gamma)}
\prod_{(i,j)\in\mathcal P}\Sigma_{ij},
\end{equation}
where $\mathfrak M(\beta-\gamma)$ is the set of perfect matchings on the multiset containing coordinate $j$ exactly $\beta_j-\gamma_j$ times; the inner sum is empty (hence zero) when $|\beta-\gamma|$ is odd.  This Isserlis--Wick form includes off-diagonal covariances and imposes no coordinate-wise factorial parity condition.  Equivalently, the centered moments can be generated without pairing enumeration by the Stein recursion of Lemma~\ref{lem:stein}.  The mean symbol $m$ is local to this display, not the expansion order.

Setting $\bar{x} = x$ (the diagonal evaluation step) yields the approximate
solution
\begin{equation}
u_{l,m}^{(k)}(t,x) := u_{l,m}^{(k),\bar{x}=x}(t,x),
\end{equation}
and the combined $(l,m)$-th order approximation is
\begin{equation}
u_m^{(k)}(t,x) = \sum_{j=0}^{l} u_{j,m}^{(k)}(t,x).
\end{equation}

\subsection{The full algorithm}
\label{app:full-algorithm}

The backward recursion of Definition~\ref{def:time-disc} is summarized in
pseudocode below.
\begin{algorithm}[H]
\caption{Analytical Expansion Method for BSDEs}
\begin{algorithmic}[1]
\REQUIRE Forward SDE coefficients $\mu, \sigma$; BSDE driver $f$; terminal condition $\psi$; expansion orders $k, l, m$; number of time steps $n$; initial point $(t, x)$.
\ENSURE Approximation $u^{(k,l,m,n)}(t,x) \approx Y_t = u(t, X_t)$.
\STATE Compute time step $\delta_t \leftarrow (T - t)/n$ and partition points $t_j \leftarrow t + j\,\delta_t$ for $j = 0, \ldots, n$.
\STATE Set terminal function $g_n(x) \leftarrow \psi(x)$.
\FOR{$j = n-1$ down to $0$}
  \STATE Set expansion center $\bar{x} \leftarrow x$.
  \STATE Compute frozen coefficients $\mu(\bar{x})$, $a(\bar{x})$, and Gaussian semigroup parameters.
  \FOR{Picard iteration $\kappa = 0$ to $k$}
    \STATE Compute Taylor polynomial $\mathcal{T}_m^{\bar{x}} g_{j+1}$ and $\mathcal{T}_m^{\bar{x}} f^{(\kappa)}$.
    \STATE Evaluate $u_{0,m}^{(\kappa),\bar{x}}(t_j, x)$ via Gaussian semigroup applied to polynomial terminal/source.
    \FOR{PDE expansion order $\ell = 1$ to $l$}
      \STATE Compute $u_{\ell,m}^{(\kappa),\bar{x}}(t_j, x)$ via the Duhamel integral with source $\sum_{i=1}^\ell \mathcal{A}_i^{\bar{x}}\,u_{\ell-i,m}^{(\kappa),\bar{x}}$.
    \ENDFOR
    \STATE Sum: $u_m^{(\kappa)}(t_j, x) \leftarrow \sum_{\ell=0}^{l} u_{\ell,m}^{(\kappa)}(t_j, x)$.
  \ENDFOR
  \STATE Set $g_j(x) \leftarrow u_m^{(k)}(t_j, x)$.
\ENDFOR
\RETURN $g_0(x) = u^{(k,l,m,n)}(t, x)$.
\end{algorithmic}
\end{algorithm}

\subsection{A catalog of non-smooth payoffs}
\label{app:payoff-catalog}

The table separates the exponent $\alpha$ used in the derivative/expansion bound from the effective regularization exponent $\alpha_{\mathrm{eff}}$:
\begin{center}
\small
\begin{tabularx}{\textwidth}{>{\raggedright}X >{\raggedright}X c c c >{\raggedright\arraybackslash}X}
\toprule
\textbf{Derivative type} & \textbf{Terminal condition} & \textbf{Regularity} & $\bm{\alpha}$ & $\bm{\alpha_{\mathrm{eff}}}$ & \textbf{Non-smoothness}\\
\midrule
European call/put & $(x-K)^+$, $(K-x)^+$ & $C^{0,1}$ & $1$ & $1$ & Kink at $x=K$\\
Digital option & $\mathds{1}_{\{x\ge K\}}$ & bounded BV & $0$ & $1/2$ & Jump at $x=K$\\
Asset-or-nothing & $x\mathds{1}_{\{x\ge K\}}$ & piecewise linear, jump & $0$ & $1/2$ & Jump at $x=K$\\
Straddle & $|x-K|$ & $C^{0,1}$ & $1$ & $1$ & Kink at $x=K$\\
Butterfly spread & $(x-K_1)^+-2(x-K_2)^++(x-K_3)^+$ & $C^{0,1}$ & $1$ & $1$ & Kinks at $K_1,K_2,K_3$\\
Power option & $((x-K)^+)^p$, $0<p<1$ & $C^{0,p}$ & $p$ & $p$ & Cusp at $x=K$\\
Spread option (2D) & $(x_1-x_2-K)^+$ & $C^{0,1}$ & $1$ & $1$ & Ridge along $x_1-x_2=K$\\
\bottomrule
\end{tabularx}
\end{center}
\smallskip
Here $\alpha$ controls derivative blow-up and the expansion term, whereas $\alpha_{\mathrm{eff}}$ controls the regularization error.  For bounded BV data the latter is $1/2$ in the proved $L^2(\Prob)$ sense (Corollary~\ref{cor:E1-disc}); substituting it for $\alpha$ in the expansion bound is not justified.  Unbounded payoffs such as calls, puts, and asset-or-nothing contracts require the terminal moment conditions of the $L^2$ stability formulation rather than an $L^\infty$ stability assertion.

\end{document}